\documentclass[journal,twocolumn]{IEEEtran}

\usepackage{color}
\definecolor{gray}{rgb}{0.5,0.5,0.5}

\usepackage{amsmath,amsthm}
\usepackage{amssymb,amsfonts}
\usepackage{graphicx}
\usepackage{url}
\usepackage{subfigure}
\newtheorem{theorem}{Theorem}
\newtheorem{assumption}{Assumption}
\newtheorem{lemma}{Lemma}

\newtheorem{remark}{Remark}

\usepackage{cite}

\ifCLASSINFOpdf
\else
\fi

\begin{document}
%
\title{Free Deterministic Equivalents for the Analysis of
MIMO Multiple Access  Channel}

\author{An-An~Lu,~\IEEEmembership{Student Member,~IEEE,}~Xiqi~Gao,~\IEEEmembership{Fellow,~IEEE,}
and ~Chengshan~Xiao,~\IEEEmembership{Fellow,~IEEE}

\thanks{The work of A.-A.~Lu and X.~Q.~Gao was supported in part by National
Natural Science Foundation of China under Grants
61320106003, 61471113 and 61401095, the China High-Tech 863 Plan under Grant Grants 2015AA01A701 and 2014AA01A704, National Science
and Technology Major Project of China under Grant 2014ZX03003006-003, and the Huawei Cooperation Project.
The work of C.~Xiao was supported in part by the U.S. National Science Foundation under Grants ECCS-1231848 and ECCS-1539316. Part of this work was carried out while An-An Lu was visiting Missouri
University of Science and Technology, Rolla, MO. Part of the material in this paper
was presented at the IEEE International Conference on Communications, London, U.K., June 2015.}
\thanks{A.-A.~Lu and X.~Q.~Gao are with the National Mobile Communications Research Laboratory, Southeast University,
Nanjing 210096, China (e-mail: aalu@seu.edu.cn, xqgao@seu.edu.cn).}
\thanks{C.~Xiao is with the Department of Electrical and Computer Engineering,
 Missouri University of Science and Technology, Rolla, MO  65409, USA (
 email: xiaoc@mst.edu).}
 \thanks{Communicated by O.~Simeone, Associate Editor for Communications.}
 \thanks{Copyright \copyright~2014 IEEE. Personal use of this material is permitted. However, permission to use this material for any other purposes must be obtained from the IEEE by sending a request to pubs-permissions@ieee.org.}
 }


\maketitle

\begin{abstract}
In this paper, a free deterministic equivalent is proposed for the capacity analysis of the multi-input multi-output (MIMO) multiple access channel (MAC) with a more general channel model compared to previous works.
Specifically, a MIMO MAC with one base station (BS) equipped with several distributed antenna sets is considered.
Each link between a user and a BS antenna set forms a jointly correlated Rician fading channel.
The analysis is based on operator-valued free probability theory, which broadens the range of applicability of free probability techniques tremendously.
By replacing independent Gaussian random matrices with operator-valued random variables satisfying certain operator-valued freeness relations, the free deterministic equivalent of the considered channel Gram matrix is obtained. The Shannon transform of the free deterministic equivalent is derived, which provides an approximate expression for the ergodic input-output mutual information of the channel.
The sum-rate capacity achieving input covariance matrices are also derived based on the approximate ergodic input-output mutual information.
The free deterministic equivalent results are easy to compute, and simulation results show that these approximations are numerically accurate and computationally efficient.
\end{abstract}


%
\IEEEpeerreviewmaketitle

\section{Introduction}
\IEEEPARstart{F}{or} the development of next generation communication systems, massive multiple-input multiple-output (MIMO) technology has been widely investigated during the last few years \cite{larsson2014massive, haider2014cellular, yongpeng2015secure, lulow2016, You16Channel, overview_LSAS_2016}. Massive MIMO systems provide huge capacity enhancement by employing hundreds of antennas at a base station (BS).
The co-location of so many antennas on a single BS is a major challenge in realizing massive MIMO, whereas  dividing the BS antennas into distributed antenna sets (ASs) provides an alternative solution \cite{truong2013viability}.
In most massive MIMO literature, it is assumed that each user equipment (UE) is equipped with a single-antenna. Since multiple antenna UEs are already used in practical systems, it would be of both theoretical and practical interest to investigate the capacity of massive MIMO with distributed ASs and multiple antenna users.

In \cite{zhang2013capacity},  Zhang \textit{et al.} investigated the capacity of a MIMO multiple access channel (MAC) with distributed sets of correlated antennas. The results of \cite{zhang2013capacity} can be applied to a massive MIMO uplink with distributed ASs and multiple antenna UEs directly. The channel between a user and an AS in \cite{zhang2013capacity} is assumed to be a Kronecker correlated MIMO channel \cite{kermoal2002stochastic} with line-of-sight (LOS) components. In \cite{oestges2006validity}, Oestges concluded that the validity of the Kronecker model decreases as the array size increases. Thus, we consider in this paper a MIMO MAC with a more general channel model than that in \cite{zhang2013capacity}. More precisely, we consider also distributed ASs and multiple antenna UEs, but assume that each link between a user and an AS forms a jointly correlated Rician fading channel \cite{weichselberger2006stochastic,gao:statistical}. If the BS antennas become co-located, then the considered channel model reduces to that in \cite{wen2011on}. To the best of our knowledge, a capacity analysis for such MIMO MACs has not been addressed to date.

For the MIMO MAC under consideration, an exact capacity analysis is difficult and might be unsolvable when the number of antennas grows large. In this paper, we aim at deriving an approximate capacity expression. Deterministic equivalents \cite{couillet2011random}, which have been addressed extensively, are successful methods to derive the approximate capacity for various MIMO channels. These deterministic equivalent approaches fall into four main categories: the Bai and Silverstein method\cite{couillet2011deterministic,couillet2012random,wen2013deterministic}, the Gaussian method\cite{hachem2008new,dupuy2011capacity,zhang2013capacity}, the replica method\cite{taricco2008asymptotic,wen2011on} and free probability theory\cite{far2008slow,speicher2012free}.

The Bai and Silverstein method has been applied to various MIMO MACs. Couillet \textit{et al.} \cite{couillet2011deterministic} used it to investigate the capacity of a MIMO MAC with separately correlated channels.
Combining it with the generalized Lindeberg principle \cite{korada2011applications}, Wen \textit{et al.}   \cite{wen2013deterministic} derived the ergodic input-output mutual information of a MIMO MAC where the channel matrix consists of correlated non-Gaussian entries. In the Bai and Silverstein method, one needs to ``guess'' the deterministic equivalent of the Stieltjes transform. This limits its applicability since the deterministic equivalents of some involved models might be hard to ``guess'' \cite{couillet2011random}. By using an integration by parts formula and the Nash-Poincare inequality, the Gaussian method is able to derive directly the deterministic equivalents and can be applied to random matrices with involved correlations.
It is particularly suited to random matrices with Gaussian entries.
Combined with the Lindeberg principle, the Gaussian method can be used to treat random matrices with non-Gaussian entries as in \cite{zhang2013capacity}.

The replica method developed in statistical physics \cite{edwards1975theory} is a widely used approach in wireless communications. It has also been applied to the MIMO MAC. Wen \textit{et al.} \cite{wen2011on} used it to investigate the sum-rate of multiuser MIMO uplink channels with jointly correlated Rician fading. Free probability theory \cite{voiculescu1997free}
provides a better way to understand the asymptotic behavior of large dimensional random matrices. It was first applied to wireless communications by Evans and Tse to investigate the multiuser wireless communication systems \cite{evans2000large}.

The Bai and Silverstein method and the Gaussian method are very flexible. Both of them have been used to handle
deterministic equivalents for advanced Haar models \cite{couillet2012random,pastur2011eigenvalue}. Although its validity has not yet been proved \cite{couillet2011random}, the replica method is also a powerful tool. Meanwhile, the applicability of free probability theory is commonly considered very limited as it can  be only applied to large random matrices with unitarily invariant properties, such as standard Gaussian matrices and Haar unitary matrices.

The domain of applicability of free probability techniques can be broadened tremendously by operator-valued free probability theory \cite{voiculescu1985symmetries, nica2002operator}, which is a more general version of free probability theory and allows one to deal with random matrices with correlated entries \cite{far2008slow}.
In \cite{far2008slow}, Far \textit{et al.} first used operator-valued free probability theory in wireless communications to study slow-fading MIMO systems with nonseparable correlation. The results of \cite{far2008slow} were then used by Pan \textit{et al.} to study the approximate capacity of uplink network MIMO systems \cite{pan2013capacity} and the asymptotic spectral efficiency of uplink MIMO-CDMA systems over arbitrarily spatially correlated Rayleigh fading channels \cite{pan2013asymptotic}. Quaternionic free probability used in \cite{muller2012channel} by M\"{u}ller and Cakmak can be seen as a particular kind of operator-valued free probability\cite{nica2008free}.

In \cite{speicher2012free}, Speicher and Vargas provided the free deterministic equivalent method to derive the deterministic equivalents under the operator-valued free probability framework. A free deterministic equivalent of a random matrix is a non-commutative random variable or an operator-valued random variable,
and the difference between the distribution of the latter and the expected distribution of the random matrix goes to zero in the large dimension limit. They viewed the considered random matrix as a polynomial in
several matrices, and obtained its free deterministic equivalent by replacing the matrices with operator-valued random variables satisfying certain freeness relations. They observed that the Cauchy transform of the free deterministic equivalent is actually the solution to the iterative deterministic equivalent equation derived by the Bai and Silverstein method or the Gaussian method. Using the free deterministic equivalent approach, they recovered the deterministic equivalent results for the advanced Haar model from \cite{couillet2010deterministic}.

Motivated by the results from \cite{speicher2012free}, we propose a free deterministic equivalent for the capacity analysis of the general channel model considered in this paper. The method of free deterministic equivalents provides a relatively formalized methodology to obtain the deterministic equivalent of the Cauchy transform.
By replacing independent Gaussian matrices with random matrices that are composed of non-commutative random variables and satisfying certain operator-valued freeness relations, we obtain the free deterministic equivalent of the channel Gram matrix.
The Cauchy transform of the free deterministic equivalent is easy to derive by using operator-valued free probability techniques, and is asymptotically the same as that of the channel Gram matrix.
Then, we compute the approximate Shannon transform of the channel Gram matrix and the approximate ergodic input-output mutual information of the channel. Furthermore, we derive the sum-rate capacity achieving input covariance matrices based on the approximate ergodic input-output mutual information.

Our considered channel model reduces to that in \cite{zhang2013capacity} when the channel between a user and an AS is a Kronecker correlated MIMO channel, and to the channel model in \cite{wen2011on} when there is one AS at the BS. In this paper, we will show that the results of \cite{zhang2013capacity} and \cite{wen2011on} can be recovered by using the free deterministic equivalent method. Since many existing channel models are special cases of the channel models in \cite{zhang2013capacity} and \cite{wen2011on}, we will also be able to provide a new approach to derive the deterministic equivalent results for them.

The rest of this article is organized as follows.
The preliminaries and problem formulation are presented in Section II.
The main results are provided in Section III. Simulations are contained in Section IV. The conclusion is drawn in Section V. A tutorial on free probability theory and operator-valued free probability theory is presented in Appendix A, where the free deterministic equivalents used in this paper are also introduced and a rigorous mathematical justification
of the free deterministic equivalents is provided. Proofs of Lemmas and Theorems are provided in Appendices B to G.

\textit{Notations:}
Throughout this paper, uppercase boldface letters and lowercase boldface letters are used for matrices and vectors, respectively. The superscripts $(\cdot)^*$, $(\cdot)^T$ and $(\cdot)^H$ denote the conjugate, transpose and conjugate transpose operations, respectively. The notation ${\mathbb E}\{\cdot\}$ denotes the mathematical expectation operator. In some cases, where it is not clear from the context, we will employ subscripts to emphasize the definition. The notation $g \circ f$ represents the composite function $g(f(x))$. We use $\mathbf{A} \odot \mathbf{B}$ to denote the Hadamard product of two matrices $\mathbf{A}$ and $\mathbf{B}$ of the same dimensions. The $N \times N$ identity matrix is denoted by $\mathbf{I}_N$. The $N \times N$ and $N \times M$ zero matrices are denoted by $\mathbf{0}_N$ and $\mathbf{0}_{N\times M}$. We use $[\mathbf{A}]_{ij}$ to denote the $(i,j)$-th entry of the matrix $\mathbf{A}$. The operators ${\rm{tr}}(\cdot)$ and $\det(\cdot)$ represent the matrix trace and determinant, respectively. ${\rm {diag}}(\mathbf{x})$ denotes a diagonal matrix with $\mathbf{x}$ along its main diagonal. $\Re(\mathbf{W})$ and $\Im(\mathbf{W})$  denote $\frac{1}{2}(\mathbf{W}+\mathbf{W}^H)$ and $\frac{1}{2i}(\mathbf{W}-\mathbf{W}^H)$, respectively. $\mathbf{D}_N(\mathbb{C})$ denotes the algebra of $N \times N$ diagonal matrices with elements in the complex field $\mathbb{C}$. Finally,  we denote by $\mathbf{M}_{N}(\mathbb{C})$ the algebra of $N \times N$ complex matrices and by $\mathbf{M}_{N \times M}(\mathbb{C})$ the algebra of $N \times M$ complex matrices.

\section{Preliminaries and Problem Formulation}
In this section, we first present the definitions of the Shannon transform and the Cauchy transform, and introduce the free deterministic equivalent method with a simple channel model, while our rigorous mathematical justification
of the free deterministic equivalents is provided in Appendix A. Then, we present the general model of the MIMO MAC considered in this work, followed by the problem formulation.
\vspace{-1em}
\subsection{Shannon Transform and Cauchy Transform}

Let $\mathbf{H}$ be an $N \times M$ random matrix and ${\mathbf{B}_N}$ denote the Gram matrix $\mathbf{H}\mathbf{H}^H$.
Let $F_{\mathbf{B}_N}(\lambda)$ denote the expected cumulative distribution of the eigenvalues of
${\mathbf{B}_N}$. The Shannon transform $\mathcal{V}_{\mathbf{B}_N}(x)$ is defined as \cite{tulino2004random}
    \begin{equation}
         \mathcal{V}_{\mathbf{B}_N}(x) =  \int_{0}^{\infty}\log(1+\frac{1}{x}\lambda)dF_{\mathbf{B}_N}(\lambda).
    \end{equation}
Let $\mu$ be a probability measure on $\mathbb{R}$ and $\mathbb{C}^+$ denote the set
\begin{equation}
\left\{z \in \mathbb{C}:\Im(z) > 0\right\}. \nonumber
\end{equation} The Cauchy transform $G_{\mu}(z)$ for $z\in\mathbb{C}^+$
is defined by \cite{nica2006lectures}
    \begin{eqnarray}
        {G}_{\mu}(z) = \int_{0}^{\infty}\frac{1}{z-\lambda}d\mu(\lambda).
    \end{eqnarray}
Let $G_{\mathbf{B}_N}(z)$ denote the Cauchy transform for $F_{\mathbf{B}_N}(\lambda)$. Then, we have $G_{\mathbf{B}_N}(z)=\frac{1}{N}{\mathbb{E}}\{{\rm{tr}}((z\mathbf{I}_N-\mathbf{B}_N)^{-1})\}.$
The relation between the Cauchy transform $G_{\mathbf{B}_N}(z)$ and the Shannon transform $\mathcal{V}_{\mathbf{B}_N}(x)$ can be expressed as \cite{tulino2004random}
    \begin{equation}
        \mathcal{V}_{\mathbf{B}_N}(x) = \int_{x}^{+\infty}\left(\frac{1}{z}+G_{\mathbf{B}_N}(-z)\right)dz.
        \label{eq:relation_between_Cauchy_transform_and_shannon_transform}
    \end{equation}
Differentiating both sides of \eqref{eq:relation_between_Cauchy_transform_and_shannon_transform} with respect to $x$, we obtain
    \begin{equation}
        \frac{d\mathcal{V}_{\mathbf{B}_N}(x)}{dx} = -x^{-1}-G_{\mathbf{B}_N}(-x).
        \label{eq:relation_between_Shannon_transform_and_Cauchy_transform}
    \end{equation}
Thus, if we are able to find a function whose derivative with respect to $x$ is $-x^{-1}-G_{\mathbf{B}_N}(-x)$, then we can obtain $\mathcal{V}_{\mathbf{B}_N}(x)$.
In conclusion, if the Cauchy transform $G_{\mathbf{B}_N}(x)$ is known, then the Shannon transform $\mathcal{V}_{\mathbf{B}_N}(x)$ can
be immediately obtained by applying \eqref{eq:relation_between_Shannon_transform_and_Cauchy_transform}.

\subsection{Free Deterministic Equivalent Method}
In this subsection, we introduce the free deterministic equivalent method, which can be used to
derive the approximation of $G_{\mathbf{B}_N}(z)$.
The associated definitions, such as that of free independence, circular elements, R-cyclic matrices and semicircular elements over $\mathbf{D}_n(\mathbb{C})$,  are provided in Appendix
\ref{sec:Free Probability and Operator-valued Free Probability}.

The term free deterministic equivalent was coined by Speicher and Vargas in \cite{speicher2012free}. The considered random matrix in \cite{speicher2012free} was viewed as a polynomial in several deterministic matrices and several independent random matrices. The free deterministic equivalent of the considered random matrix was then obtained by replacing the matrices with operator-valued random variables satisfying certain freeness relations.
Moreover, the difference between the Cauchy transform of the free deterministic equivalent and that of the considered random matrix goes to zero in the large dimension limit.

However, the method in \cite{speicher2012free} only showed how to obtain the free deterministic equivalents for the
case where the random matrices are standard Gaussian matrices and Haar unitary matrices.
A method similar to that in \cite{speicher2012free} was presented by Speicher in \cite{Speicher2008what}, which
appeared earlier than \cite{speicher2012free}.
The method in \cite{Speicher2008what} showed that the random matrix with independent Gaussian entries having different variances can be replaced
by the random matrix with free (semi)circular elements having different variances.
But, it only considered a very simple case, and the replacement process had no rigorous mathematical proof.
Moreover, the free deterministic equivalents were not mentioned in \cite{Speicher2008what}.

In this paper, we introduce in Appendix \ref{sec:Free Deterministic Equivalent} the free deterministic equivalents for the case where all the matrices are square and
have the same size, and the random matrices are Hermitian and composed of independent Gaussian entries with different variances. Similarly to \cite{speicher2012free}, the free deterministic equivalent of a polynomial in matrices is defined.
The replacement process used is that in \cite{Speicher2008what}. Moreover,
a rigorous mathematical justification of the free deterministic equivalents we introduce is also provided in Appendix \ref{sec:Free Deterministic Equivalent} and Appendix \ref{New_asymptotic_freeness_results}.

\newcounter{tempequationcounter}
\begin{figure*}[!t]
\normalsize
\setcounter{tempequationcounter}{\value{equation}}
\begin{eqnarray}
\setcounter{equation}{19}
        &&\!\!\!\!\!\!\!\!\!\!\!\!\!\!\!\!\left(
            \begin{array}{cccccc}
                  z\mathcal{G}_{\boldsymbol{\mathcal{B}_N}}^{\mathcal{D}_N}(z^2\mathbf{I}_{N})  & \mathbf{0}  \\
                  \mathbf{0}   & z\mathcal{G}_{\boldsymbol{\mathcal{H}}^H\boldsymbol{\mathcal{H}}}^{\mathcal{D}_M}(z^2\mathbf{I}_{M})   \\
            \end{array}
        \right) \nonumber \\
        &&= E_{\mathcal{D}_n}\left\{
        \left(
            \begin{array}{cccccc}
               z\mathbf{I}_N - z\eta_{\mathcal{D}_N}(\mathcal{G}_{\boldsymbol{\mathcal{H}}^H \boldsymbol{\mathcal{H}}}^{\mathcal{D}_M}(z^2\mathbf{I}_{M}))   &  -\overline{\mathbf{H}}  \\
               -\overline{\mathbf{H}}{}^H      &  z\mathbf{I}_M - z\eta_{\mathcal{D}_M}(\mathcal{G}_{\boldsymbol{\mathcal{B}}_N}^{\mathcal{D}_N}(z^2\mathbf{I}_{N})) \\
            \end{array}
        \right)^{-1}\right\}
    \label{eq:floatingequation_tmp1}
    \end{eqnarray}
\setcounter{equation}{\value{tempequationcounter}}
\hrulefill
\end{figure*}

In \cite{Speicher2008what}, the deterministic equivalent results of \cite{hachem2007deterministic} were rederived.
But the description in \cite{Speicher2008what} is not easy to follow.
To show how the introduced free deterministic equivalents can be used to derive the approximation of the Cauchy transform $G_{\mathbf{B}_N}(z)$, we
use the channel model in \cite{hachem2007deterministic} as a toy example and
restate the method used in \cite{Speicher2008what} as follows.

The channel matrix $\mathbf{H}$ in \cite{hachem2007deterministic} consists of an $N \times M$ deterministic matrix $\overline{\mathbf{H}}$ and an $N \times M$ random matrix $\widetilde{\mathbf{H}}$, \textit{i.e.},  $\mathbf{H}=\overline{\mathbf{H}}+\widetilde{\mathbf{H}}$. The entries of $\widetilde{\mathbf{H}}$ are independent zero mean complex Gaussian random variables with variances $\mathbb{E}\{[\widetilde{\mathbf{H}}]_{ij}[\widetilde{\mathbf{H}}]_{ij}^*\}=\frac{1}{N}\sigma_{ij}^2$.

Let $n$ denote $N+M$, $\mathcal{P}$ denote the algebra of complex random variables and $\mathbf{M}_n(\mathcal{P})$ denote the algebra of $n \times n$ complex random matrices. We define $\mathbb{E}_{\mathcal{D}_n}:\mathbf{M}_n(\mathcal{P}) \rightarrow \mathbf{D}_n(\mathbb{C})$ by
    \begin{eqnarray}
        & &\!\!\!\!\!\!\!\!\!\!\!\!{\mathbb E}_{\mathcal{D}_n}\left\{\left(
                    \begin{array}{ccccc}
                      X_{11}     & X_{12} & \cdots & X_{1n} \\
                      X_{21}    & X_{22} & \ldots & X_{2n} \\
                      \vdots  &  \vdots & \ddots & \vdots \\
                      X_{n1}    & X_{n2} & \ldots & X_{nn} \\
                    \end{array}
        \right)\right\}
        \nonumber \\
        &&=\left(
                    \begin{array}{cccc}
                      {\mathbb E}\{X_{11}\}        & 0        & \cdots & 0 \\
                       0       & {\mathbb E}\{X_{22}\}        & \cdots & 0 \\
                      \vdots            & \vdots   & \ddots & \vdots \\
                      0    & 0  & \ldots & {\mathbb E}\{X_{nn}\} \\
                    \end{array}
        \right)
    \end{eqnarray}
where each $X_{ij}$ is a complex random variable. Hereafter, we use the notations $\mathcal{M}_n:=\mathbf{M}_n(\mathbb{C})$ and $\mathcal{D}_n:=\mathbf{D}_n(\mathbb{C})$ for brevity.

Let $\mathbf{X}$ be an $n \times n$ matrix defined by \cite{far2008slow}
    \begin{eqnarray}
        {\mathbf{X}} = \left(
                    \begin{array}{cc}
                      \mathbf{0}_{N}              & {\mathbf{H}}  \\
                      {\mathbf{H}}^H    & \mathbf{0}_{M}  \\
                    \end{array}
        \right).
        \label{eq:definition_of_matrix_bold_captial_x}
    \end{eqnarray}
The matrix $\mathbf{X}$ is even, \textit{i.e.}, all the odd moments of $\mathbf{X}$ are zeros, and
    \begin{eqnarray}
        {\mathbf{X}}^2 = \left(\begin{array}{cc}
                      {\mathbf{H}}{\mathbf{H}}^H              & \mathbf{0}_{N\times M}   \\
                         \mathbf{0}_{M \times N}            & {\mathbf{H}}^H{\mathbf{H}}  \\
                    \end{array}
        \right).
        \label{eq:mathbf_x_square}
    \end{eqnarray}
Let $\boldsymbol{\Delta}_n\in \mathcal{D}_n$ be a diagonal matrix with $\Im(\boldsymbol{\Delta}_n) \succ 0$. The $\mathcal{D}_n$-valued Cauchy transform $\mathcal{G}_{\mathbf{X}}^{\mathcal{D}_n}(\boldsymbol{\Delta}_n)$ is given by
    \begin{eqnarray}
        \mathcal{G}_{\mathbf{X}}^{\mathcal{D}_n}(\boldsymbol{\Delta}_n) = {\mathbb E}_{\mathcal{D}_n}\{(\boldsymbol{\Delta}_n-\mathbf{X})^{-1}\}.
    \end{eqnarray}
When $\boldsymbol{\Delta}_n=z\mathbf{I}_{n}$ and $z \in \mathbb{C}^+$, we have that
    \begin{eqnarray}
        &&     \!\!\!\!\!\!\!\!\!\!\!\!\!\!\mathcal{G}_{\mathbf{X}}^{\mathcal{D}_n}(z\mathbf{I}_{n})
        \nonumber \\
        &&     \!\!\!\!\!\!\!\!\!\!=  {\mathbb E}_{\mathcal{D}_n} \!\!\left\{(z\mathbf{I}_{n}-\mathbf{X})^{-1}\right\}
        \nonumber \\
         &&    \!\!\!\!\!\!\!\!\!\!=
        {\mathbb E}_{\mathcal{D}_n} \!\!\left\{ \!\! \left(\!\!\begin{array}{cc}
           \!\!z(z^2\mathbf{I}_{N}-\mathbf{H}\mathbf{H}^H)^{-1}  \!\! & \!\!{\mathbf{H}}(z^2\mathbf{I}_{M}-\mathbf{H}^H\mathbf{H})^{-1} \!\! \\
           \!\!{\mathbf{H}}^H(z^2\mathbf{I}_{N}-\mathbf{H}\mathbf{H}^H)^{-1}  \!\!& \!\!z(z^2\mathbf{I}_{M}-\mathbf{H}^H\mathbf{H})^{-1}  \!\!\\
        \end{array}\right)\!\!\right\} \nonumber \\
        \label{eq:diagonal_matrix_valued_cauchy_transform_of_X}
    \end{eqnarray}
where the second equality is due to the block matrix inversion formula \cite{petersen2008matrix}.
From \eqref{eq:mathbf_x_square} and \eqref{eq:diagonal_matrix_valued_cauchy_transform_of_X},  we obtain
    \begin{equation}
        \mathcal{G}_{\mathbf{X}}^{\mathcal{D}_n}(z\mathbf{I}_{n}) = z\mathcal{G}_{\mathbf{X}^{2}}^{\mathcal{D}_n}(z^2\mathbf{I}_{n})
        \label{eq:realtion_of_Cauchy_transform_of X_and_X2}
    \end{equation}
for each $z,z^2 \in \mathbb{C}^+$.
Furthermore, we write $\mathcal{G}_{\mathbf{X}^{2}}^{\mathcal{D}_n}(z\mathbf{I}_{n})$ as
    \begin{equation}
        \mathcal{G}_{\mathbf{X}^2}^{\mathcal{D}_n}(z\mathbf{I}_{n})=\left(
                    \begin{array}{cccccc}
                      \mathcal{G}_{\mathbf{B}_N}^{\mathcal{D}_N}(z\mathbf{I}_{N})      & \mathbf{0}  \\
                       \mathbf{0}       & \mathcal{G}_{{\mathbf{H}}^H{\mathbf{H}}}^{\mathcal{D}_M}(z\mathbf{I}_{M})   \\
                    \end{array}
        \right)
        \label{eq:Cauchy_transform_of_Hsquare_in_detail}
    \end{equation}
where
\begin{IEEEeqnarray}{Rl}
&\mathcal{G}_{\mathbf{B}_N}^{\mathcal{D}_N}(z\mathbf{I}_{N}) = {\mathbb E}_{\mathcal{D}_N}\{(z\mathbf{I}_{N}-\mathbf{B}_N)^{-1}\}\nonumber \\
&\mathcal{G}_{{\mathbf{H}}^H{\mathbf{H}}}^{\mathcal{D}_M}(z\mathbf{I}_M) = {\mathbb E}_{\mathcal{D}_M}\{(z\mathbf{I}_{M}-{\mathbf{H}}^H{\mathbf{H}})^{-1}\}.\nonumber
\end{IEEEeqnarray}
Since $G_{\mathbf{B}_N}(z)=\frac{1}{N}{\rm{tr}}(\mathcal{G}_{\mathbf{B}_N}^{\mathcal{D}_N}(z\mathbf{I}_{N}))$, we have related the calculation of ${G}_{\mathbf{B}_N}(z)$ with that of $\mathcal{G}_{\mathbf{X}}^{\mathcal{D}_n}(z\mathbf{I}_{n})$.

We define $\overline{\mathbf{X}}$ and $\widetilde{\mathbf{X}}$ by
    \begin{eqnarray}
        \overline{\mathbf{X}} = \left(
                    \begin{array}{cc}
                      \mathbf{0}_{N}              & \overline{\mathbf{H}}  \\
                      \overline{\mathbf{H}}{}^H    & \mathbf{0}_{M}  \\
                    \end{array}
        \right)
        \label{eq:definition_of_matrix_overline_bold_captial_x}
    \end{eqnarray}
and
    \begin{eqnarray}
        \widetilde{\mathbf{X}} = \left(
                    \begin{array}{cc}
                      \mathbf{0}_{N}              & \widetilde{\mathbf{H}}  \\
                      \widetilde{\mathbf{H}}^H    & \mathbf{0}_{M}  \\
                    \end{array}
        \right).
        \label{eq:definition_of_matrix_widetilde_bold_captial_x}
    \end{eqnarray}
Then, we have that $\mathbf{X} = \overline{\mathbf{X}} + \widetilde{\mathbf{X}}$.

The free deterministic equivalent of $\mathbf{X}$ is constructed as follows.
Let $\mathcal{A}$ be a unital algebra, $(\mathcal{A},\phi)$ be a non-commutative probability space and
$\widetilde{\boldsymbol{\mathcal{H}}}$ denote an $N \times M$ matrix with entries from $\mathcal{A}$.
The entries $[\widetilde{\boldsymbol{\mathcal{H}}}]_{ij}
\in \mathcal{A}$ are freely independent centered circular elements with variances $\phi([\widetilde{\boldsymbol{\mathcal{H}}}]_{ij}[\widetilde{\boldsymbol{\mathcal{H}}}]_{ij}^*)=\frac{1}{N}\sigma_{ij}^2$.
Let $\boldsymbol{\mathcal{H}}$ denote $\overline{\mathbf{H}}+\widetilde{\boldsymbol{\mathcal{H}}}$, $\widetilde{\boldsymbol{\mathcal{X}}}$ denote
    \begin{eqnarray}
        \widetilde{\boldsymbol{\mathcal{X}}}=\left(
                    \begin{array}{ccccc}
                      \mathbf{0}   & \widetilde{\boldsymbol{\mathcal{H}}}   \\
                      \widetilde{\boldsymbol{\mathcal{H}}}^H     &   \mathbf{0} \\
                    \end{array}
        \right)
        \label{eq:definition_of_matrix_widetilde_cal_captial_x}
    \end{eqnarray}
and
$\boldsymbol{\mathcal{X}}$ denote
    \begin{eqnarray}
        \boldsymbol{\mathcal{X}}=\left(
                    \begin{array}{ccccc}
                      \mathbf{0}   & \boldsymbol{\mathcal{H}}   \\
                      \boldsymbol{\mathcal{H}}^H     &   \mathbf{0} \\
                    \end{array}
        \right).
        \label{eq:definition_of_matrix_cal_captial_x}
    \end{eqnarray}
It follows that $\boldsymbol{\mathcal{X}} = \overline{\mathbf{X}} +  \widetilde{\boldsymbol{\mathcal{X}}}$.
The matrix $\boldsymbol{\mathcal{X}}$ is the free deterministic equivalent of $\mathbf{X}$.

We define $E_{\mathcal{D}_n}:\mathbf{M}_n(\mathcal{A}) \rightarrow \mathcal{D}_n$ by
    \begin{eqnarray}
        & &\!\!\!\!\!\!\!\!\!\!\!\!E_{\mathcal{D}_n}\left\{\left(
                    \begin{array}{ccccc}
                      x_{11}     & x_{12} & \cdots & x_{1n} \\
                      x_{21}    & x_{22} & \ldots & x_{2n} \\
                      \vdots  &  \vdots & \ddots & \vdots \\
                      x_{n1}    & x_{n2} & \ldots & x_{nn} \\
                    \end{array}
        \right)\right\}
        \nonumber \\
        &&=\left(
                    \begin{array}{cccc}
                      \phi(x_{11})        & 0        & \cdots & 0 \\
                       0       & \phi(x_{22})        & \cdots & 0 \\
                      \vdots            & \vdots   & \ddots & \vdots \\
                      0    & 0  & \ldots & \phi(x_{nn}) \\
                    \end{array}
        \right)
        \label{eq:definition_of_E_sub_mathcalD_n}
    \end{eqnarray}
where each $x_{ij}$ is a non-commutative random variable from $(\mathcal{A},\phi)$. Then, $(\mathbf{M}_{n}(\mathcal{A}), E_{\mathcal{D}_n})$ is a $\mathcal{D}_n$-valued probability space.

From the discussion of the free deterministic equivalents provided in Appendix \ref{sec:Free Deterministic Equivalent}, we have that
$\mathcal{G}^{\mathcal{D}_n}_{\boldsymbol{\mathcal{X}}}(z\mathbf{I}_n)$ and
$\mathcal{G}^{\mathcal{D}_n}_{\mathbf{X}}(z\mathbf{I}_n)$ are asymptotically the same. Let $\boldsymbol{\mathcal{B}}_N$ denote  $\boldsymbol{\mathcal{H}}\boldsymbol{\mathcal{H}}^H$.
The relation between $\mathcal{G}_{\boldsymbol{\mathcal{X}}}^{\mathcal{D}_n}(z\mathbf{I}_n)$ and $\mathcal{G}_{\boldsymbol{\mathcal{B}}_N}^{\mathcal{D}_N}(z\mathbf{I}_N)$ is the same as that between $\mathcal{G}_{\mathbf{X}}^{\mathcal{D}_n}(z\mathbf{I}_n)$ and $\mathcal{G}_{\mathbf{B}_N}^{\mathcal{D}_N}(z\mathbf{I}_N)$.
Thus, we also have that $\mathcal{G}_{\boldsymbol{\mathcal{B}}_N}^{\mathcal{D}_N}(z\mathbf{I}_N)$ and
$\mathcal{G}_{\mathbf{B}_N}^{\mathcal{D}_N}(z\mathbf{I}_N)$ are asymptotically the same
and $G_{\boldsymbol{\mathcal{B}}_N}(z)$ is the deterministic equivalent of $G_{\mathbf{B}_N}(z)$.
For convenience, we also call $\boldsymbol{\mathcal{B}}_N$ the free deterministic equivalent of $\mathbf{B}_N$.
In the following, we derive
 the Cauchy transform
$G_{\boldsymbol{\mathcal{B}}_N}(z)$ by using operator-valued free probability techniques.

Since its elements on and above the diagonal are freely independent, we have that $\widetilde{\boldsymbol{\mathcal{X}}}$ is an R-cyclic matrix.
From Theorem 8.2 of \cite{nica2002r}, we then have that $\overline{\boldsymbol{\mathbf{X}}}$ and $\widetilde{\boldsymbol{\mathcal{X}}}$ are free over $\mathcal{D}_n$.
The $\mathcal{D}_n$-valued Cauchy transform of the sum of two $\mathcal{D}_n$-valued free random variables is given by \eqref{eq:operator_cauchy_transform_of_sum_of_free_varaible} in Appendix \ref{sec:Free Probability and Operator-valued Free Probability}. Applying \eqref{eq:operator_cauchy_transform_of_sum_of_free_varaible}, we have that
    \begin{eqnarray}
        \!\!\!\!\mathcal{G}_{\boldsymbol{\mathcal{X}}}^{\mathcal{D}_n}(z\mathbf{I}_n)
        \!\!\!\!&=&\!\!\!\! \mathcal{G}_{\overline{\boldsymbol{\mathbf{X}}}}^{\mathcal{D}_n}\!\!\left(z\mathbf{I}_n - \mathcal{R}_{\widetilde{\boldsymbol{\mathcal{X}}}}^{\mathcal{D}_n}\!\! \left(\mathcal{G}_{\boldsymbol{\mathcal{X}}}^{\mathcal{D}_n}(z\mathbf{I}_n)\right)\right)
        \nonumber \\
        \!\!\!\!&=&\!\!\!\! E_{\mathcal{D}_n}\!\!\left\{\!\!\left(z\mathbf{I}_n - \mathcal{R}_{\widetilde{\boldsymbol{\mathcal{X}}}}^{\mathcal{D}_n}\!\! \left(\mathcal{G}_{\boldsymbol{\mathcal{X}}}^{\mathcal{D}_n}(z\mathbf{I}_n)\right)-\overline{\boldsymbol{\mathbf{X}}}\right)^{-1}\!\!\right\}
        \label{eq:Cauchy_transform_of_the_sum_in_example}
    \end{eqnarray}
where $\mathcal{R}_{\widetilde{\boldsymbol{\mathcal{X}}}}^{\mathcal{D}_n} $ is the $\mathcal{D}_n$-valued R-transform of $\boldsymbol{\mathcal{X}}$.

Let $\eta_{\mathcal{D}_n}(\mathbf{C})$ denote $E_{\mathcal{D}_n}\{\widetilde{\boldsymbol{\mathcal{X}}}\mathbf{C}\widetilde{\boldsymbol{\mathcal{X}}}\}$, where $\mathbf{C} \in \mathcal{D}_n$. From Theorem 7.2 of \cite{nica2002r}, we obtain that
$\widetilde{\boldsymbol{\mathcal{X}}}$ is semicircular over $\mathcal{D}_n$, and thus its $\mathcal{D}_n$-valued R-transform
is given by
\begin{equation}
\mathcal{R}_{\widetilde{\boldsymbol{\mathcal{X}}}}^{\mathcal{D}_n}(\mathbf{C})=\eta_{\mathcal{D}_n}(\mathbf{C}).
\label{eq:digonal_matrix_valued_R_transform_of_X}
\end{equation}
From \eqref{eq:Cauchy_transform_of_the_sum_in_example} and the counterparts of \eqref{eq:realtion_of_Cauchy_transform_of X_and_X2} and \eqref{eq:Cauchy_transform_of_Hsquare_in_detail} for $\mathcal{G}_{\boldsymbol{\mathcal{X}}}^{\mathcal{D}_n}(z\mathbf{I}_n)$ and $\mathcal{G}_{\boldsymbol{\mathcal{X}}^2}^{\mathcal{D}_n}(z\mathbf{I}_n)$, we obtain equation
\eqref{eq:floatingequation_tmp1}
at the top of this page.
\setcounter{equation}{19}
\begin{figure*}[!t]
\normalsize
\setcounter{tempequationcounter}{\value{equation}}
    \begin{eqnarray}
        &&\!\!\!\!z\mathcal{G}_{\boldsymbol{\mathcal{B}}_N}^{\mathcal{D}_N}(z\mathbf{I}_{N})
        = E_{\mathcal{D}_N}\left\{\left(\mathbf{I}_N - \eta_{\mathcal{D}_N}(\mathcal{G}_{\boldsymbol{\mathcal{H}}^H\boldsymbol{\mathcal{H}}}^{\mathcal{D}_M}(z\mathbf{I}_{M}))
        -\overline{\mathbf{H}}\left(z\mathbf{I}_M-z\eta_{\mathcal{D}_M}(\mathcal{G}_{\boldsymbol{\mathcal{B}}_N}^{\mathcal{D}_N} (z\mathbf{I}_{N}))\right)^{-1}\overline{\mathbf{H}}{}^H\right)^{-1}\right\} \label{eq:floatingequation_tmp2}
        \\
        &&\!\!\!\!z\mathcal{G}_{\boldsymbol{\mathcal{H}}^H\boldsymbol{\mathcal{H}}}^{\mathcal{D}_M}(z\mathbf{I}_{M})
        =E_{\mathcal{D}_M}\left\{\left(\mathbf{I}_M - \eta_{\mathcal{D}_M}(\mathcal{G}_{\boldsymbol{\mathcal{B}}_N}^{\mathcal{D}_N}(z\mathbf{I}_{N}))
        -\overline{\mathbf{H}}{}^H\left(z\mathbf{I}_N-z\eta_{\mathcal{D}_N}(\mathcal{G}_{\boldsymbol{\mathcal{H}}^H \boldsymbol{\mathcal{H}}}^{\mathcal{D}_M}(z\mathbf{I}_{M}))\right)^{-1}\overline{\mathbf{H}}\right)^{-1}\right\}
        \label{eq:floatingequation_tmp3}
    \end{eqnarray}
\addtocounter{tempequationcounter}{2}
\setcounter{equation}{\value{tempequationcounter}}
\hrulefill
\end{figure*}
Furthermore, we obtain equations \eqref{eq:floatingequation_tmp2} and \eqref{eq:floatingequation_tmp3}
at the top of the following page, where
\begin{IEEEeqnarray}{Rl}
&\eta_{\mathcal{D}_N}(\mathbf{C}_1) =  E_{\mathcal{D}_N}\{\widetilde{\boldsymbol{\mathcal{H}}}\mathbf{C}_1\widetilde{\boldsymbol{\mathcal{H}}}{}^H\}, \mathbf{C}_1 \in \mathcal{D}_M \nonumber \\
&\eta_{\mathcal{D}_M}(\mathbf{C}_2) = E_{\mathcal{D}_M}\{\widetilde{\boldsymbol{\mathcal{H}}}{}^H\mathbf{C}_2\widetilde{\boldsymbol{\mathcal{H}}}\}, \mathbf{C}_2 \in \mathcal{D}_N.\nonumber
\end{IEEEeqnarray}
Equations (20) and (21) are equivalent to the
ones provided by Theorem 2.4 of \cite{hachem2007deterministic}.
Finally, the Cauchy transform $G_{\boldsymbol{\mathcal{B}}_N}(z)$ is obtained by $G_{\boldsymbol{\mathcal{B}}_N}(z)=\frac{1}{N}{\rm{tr}}(\mathcal{G}_{\boldsymbol{\mathcal{B}}_N}^{\mathcal{D}_N}(z\mathbf{I}_{N}))$.

In conclusion, the free deterministic equivalent method provides a way to
derive the approximation of the Cauchy transform $G_{\mathbf{B}_N}(z)$.
The fundamental step is to construct the free deterministic equivalent $\boldsymbol{\mathcal{B}}_N$ of $\mathbf{B}_N$.
After the construction, the Cauchy transform
$G_{\boldsymbol{\mathcal{B}}_N}(z)$ can be derived by using operator-valued free probability techniques.
Moreover, $G_{\boldsymbol{\mathcal{B}}_N}(z)$ is the deterministic equivalent of $G_{\mathbf{B}_N}(z)$.

\subsection{General Channel Model of MIMO MAC}
We consider a frequency-flat fading MIMO MAC channel with one BS and $K$ UEs. The BS antennas are divided into $L$ distributed ASs. The $l$-th AS is equipped with $N_l$ antennas. The $k$-th UE is equipped with $M_k$ antennas. Furthermore, we assume $\sum\nolimits_{l=1}^L N_l=N$ and $\sum\nolimits_{k=1}^K M_k=M$. Let $\mathbf{x}_k$ denote the $M_k \times 1$ transmitted vector of the $k$-th UE. The covariance matrices of $\mathbf{x}_k$ are given by
    \begin{equation}
        {\mathbb{E}}\{\mathbf{x}_k\mathbf{x}_{k'}^H\}=\left\{\begin{array}{cc}
                                           \frac{P_k}{M_k}\mathbf{Q}_k, &{\rm if}~ k=k'\\
                                           \mathbf{0}, &~~{\rm otherwise}
                                         \end{array}\right.
    \end{equation}
where $P_k$ is the total transmitted power of the $k$-th UE, and $\mathbf{Q}_k$ is an $M_k \times M_k$ positive semidefinite matrix with the constraint ${\rm{tr}}(\mathbf{Q}_k)\leq M_k$. The received signal $\mathbf{y}$ for a single symbol interval can be written as
    \begin{equation}
        \mathbf{y} = \sum\limits_{k=1}^{K}\mathbf{H}_k\mathbf{x}_k + \mathbf{z}
    \end{equation}
where $\mathbf{H}_k$ is the $N \times M_k$ channel matrix between the BS and the $k$-th UE, and $\mathbf{z}$ is a
complex Gaussian noise vector distributed as $\mathcal{CN}(0,\sigma_z^2\mathbf{I}_N)$. The channel matrix $\mathbf{H}_k$ is normalized as
    \begin{equation}
        \mathbb{E}\{{\rm{tr}}(\mathbf{H}_k\mathbf{H}_k^H)\}=\frac{NM_k}{M}.
        \label{eq:constraint_of_channel_matrix}
    \end{equation}
Furthermore, $\mathbf{H}_k$ has the following structure
\begin{equation}
\mathbf{H}_k=\overline{\mathbf{H}}_k + \widetilde{\mathbf{H}}_k
\label{eq:channel_matrix_one}
\end{equation}
where $\overline{\mathbf{H}}_k$ and $\widetilde{\mathbf{H}}_k$ are defined by
\begin{IEEEeqnarray}{Rl}
 &\overline{\mathbf{H}}_k=\left(\vphantom{\widetilde{\mathbf{H}}} \overline{\mathbf{H}}{}_{1k}^T~\overline{\mathbf{H}}{}_{2k}^T~\cdots~\overline{\mathbf{H}}{}_{Lk}^T\right)^T
 \label{eq:channel_matrix_two}
 \\
 &\widetilde{\mathbf{H}}_k=\left(\widetilde{\mathbf{H}}_{1k}^T~\widetilde{\mathbf{H}}_{2k}^T~\cdots~\widetilde{\mathbf{H}}_{Lk}^T\right)^T.
 \label{eq:channel_matrix_three}
\end{IEEEeqnarray}
Each $\overline{\mathbf{H}}_{lk}$ is an $N_l \times M_k$ deterministic matrix, and each $\widetilde{\mathbf{H}}_{lk}$ is a jointly correlated channel matrix defined by \cite{weichselberger2006stochastic,gao:statistical}
    \begin{equation}
        \widetilde{\mathbf{H}}_{lk}=\mathbf{U}_{lk}(\mathbf{M}_{lk}\odot{\mathbf{W}}_{lk})\mathbf{V}_{lk}^H
        \label{eq:channel_matrix_four}
    \end{equation}
where $\mathbf{U}_{lk}$ and $\mathbf{V}_{lk}$ are deterministic unitary matrices, $\mathbf{M}_{lk}$ is an $N_{l}\times M_{k}$ deterministic matrix with nonnegative elements, and $\mathbf{W}_{lk}$ is a complex Gaussian random matrix with independent and identically distributed (i.i.d.), zero mean and unit variance entries.
The jointly correlated channel model not only accounts for the correlation at both link ends, but also
characterizes their mutual dependence.
It provides a more adequate  model for realistic massive MIMO channels since the validity of the widely used Kronecker model decreases as the number of antennas increases.
Furthermore, the justification of using the jointly correlated channel model for massive MIMO channels has been
provided in  \cite{sunbeam, you2015pilot, adhikary2013joint}.
We assume that the channel matrices of different links are independent in this paper, \textit{i.e.}, when $k\neq m$ or $j\neq n$, we have that
\begin{IEEEeqnarray}{Rl}
   &{\mathbb E}\left\{\widetilde{\mathbf{H}}_{kj}\mathbf{C}_{jn}\widetilde{\mathbf{H}}_{mn}^H\right\} =  \mathbf{0}_{N_k \times N_m} \\
   &{\mathbb E}\left\{\widetilde{\mathbf{H}}_{kj}^H{\widetilde{\mathbf{C}}}_{km}\widetilde{\mathbf{H}}_{mn}\right\}  = \mathbf{0}_{M_j \times M_n}
\end{IEEEeqnarray}
where $\mathbf{C}_{jn} \in \mathbf{M}_{M_j\times M_n}(\mathbb{C})$ and $\widetilde{\mathbf{C}}_{km} \in \mathbf{M}_{N_k\times N_m}(\mathbb{C})$. Let $\widetilde{\mathbf{W}}_{lk}$ denote $\mathbf{M}_{lk}\odot{\mathbf{W}}_{lk}$. We define $\mathbf{G}_{lk}$ as $\mathbf{G}_{lk}=\mathbf{M}_{lk}\odot\mathbf{M}_{lk}$. The parameterized one-sided correlation matrix ${\tilde{\eta}}_k(\mathbf{C}_k)$ is given by
    \begin{IEEEeqnarray}{Rl}
          {\tilde{\eta}}_k(\mathbf{C}_k) =& {\mathbb E}\left\{\widetilde{\mathbf{H}}_k\mathbf{C}_k\widetilde{\mathbf{H}}_k^H\right\}
          \nonumber \\
          =& {\rm{diag}}\left(\mathbf{U}_{1k}\widetilde{\mathbf{\Pi}}_{1k}(\mathbf{C}_k)\mathbf{U}_{1k}^H, \mathbf{U}_{2k}\widetilde{\mathbf{\Pi}}_{2k}(\mathbf{C}_k)\mathbf{U}_{2k}^H, \right.
          \nonumber \\
          &~~~~~~~~~~~~~~~~~~~~\left.\cdots, \mathbf{U}_{Lk}\widetilde{\mathbf{\Pi}}_{Lk}(\mathbf{C}_k)\mathbf{U}_{Lk}^H\right)
          \label{eq:eta_function}
    \end{IEEEeqnarray}
where $\mathbf{C}_k \in \mathcal{M}_{M_k}$, and $\widetilde{\mathbf{\Pi}}_{lk}(\mathbf{C}_k)$ is an $N_l \times N_l$ diagonal matrix valued function with the diagonal entries obtained by
\begin{equation}
\left[\widetilde{\mathbf{\Pi}}_{lk}(\mathbf{C}_k)\right]_{ii}=\sum\limits_{j=1}^{M_k}\left[{\mathbf{G}}_{lk}\right]_{ij}\left[\mathbf{V}_{lk}^H\mathbf{C}_k\mathbf{V}_{lk}\right]_{jj}.
\label{eq:eta_function_component}
\end{equation}
Similarly, the other parameterized one-sided correlation matrix ${\eta}_{k}(\widetilde{\mathbf{C}})$ is expressed as
\begin{equation}
   {\eta}_{k}(\widetilde{\mathbf{C}}) = {\mathbb E}\left\{\widetilde{\mathbf{H}}_k^H{\widetilde{\mathbf{C}}}\widetilde{\mathbf{H}}_k\right\} =
   \sum\limits_{l=1}^{L}\mathbf{V}_{lk}{\mathbf{\Pi}}_{lk}(\langle{\widetilde{\mathbf{C}}}\rangle_l)\mathbf{V}_{lk}^H
  \label{eq:eta_function_of_widetilde}
  \end{equation}
where $\widetilde{\mathbf{C}} \in \mathcal{M}_{N}$, the notation $\langle{\widetilde{\mathbf{C}}}\rangle_l$
denotes the $N_l \times N_l$ diagonal block of $\widetilde{\mathbf{C}}$, \textit{i.e.}, the submatrix of ${\widetilde {\mathbf{C}}}$ obtained by extracting the entries of the rows and
columns with indices from $\sum\nolimits_{i=1}^{l-1}N_i + 1$ to $\sum\nolimits_{i=1}^{l}N_i$,
  and ${\mathbf{\Pi}}_{lk}(\langle{\widetilde{\mathbf{C}}}\rangle_l)$ is an $M_k \times M_k$ diagonal matrix valued function with the diagonal entries computed by
\begin{equation}
\left[{\mathbf{\Pi}}_{lk}(\langle{\widetilde{\mathbf{C}}}\rangle_l)\right]_{ii}=\sum\limits_{j=1}^{N_l}\left[{\mathbf{G}}_{lk}\right]_{ji}\left[\mathbf{U}_{lk}^H\langle{\widetilde{\mathbf{C}}}\rangle_l\mathbf{U}_{lk}\right]_{jj}.
\label{eq:eta_function_of_widetilde_component}
\end{equation}

The channel model described above is suitable for describing cellular systems employing cooperative
multipoint (CoMP) processing \cite{jungnickel2014role}, and also conforms with the framework of cloud radio access
networks (C-RANs) \cite{liu2014joint}.
Moreover, it
embraces many existing channel models as special cases.  When $L=1$,
the MIMO MAC in \cite{wen2011on} is described.
Let $\mathbf{J}_{lk}$ be an $N_l \times M_k$ matrix of all $1$s, $\boldsymbol{\Lambda}_{r,lk}$ be an $N_l \times N_l$
diagonal matrix with positive entries and $\boldsymbol{\Lambda}_{t,lk}$ be an $M_k \times M_k$ diagonal matrix with positive entries.
Set $\mathbf{M}_{lk}=\boldsymbol{\Lambda}_{r,lk}^{1/2}\mathbf{J}_{lk}\boldsymbol{\Lambda}_{t,lk}^{1/2}$.
Then, we obtain $\widetilde{\mathbf{H}}_{lk}=\mathbf{U}_{lk}(\boldsymbol{\Lambda}_{r,lk}^{1/2}\mathbf{J}_{lk}\boldsymbol{\Lambda}_{t,lk}^{1/2}\odot{\mathbf{W}}_{lk})\mathbf{V}_{lk}^H=\mathbf{U}_{lk}\boldsymbol{\Lambda}_{r,lk}^{1/2}(\mathbf{J}_{lk}\odot{\mathbf{W}}_{lk})\boldsymbol{\Lambda}_{t,lk}^{1/2}\mathbf{V}_{lk}^H $ \cite{hom1994topics}.
Thus, each $\widetilde{\mathbf{H}}_{lk}$ reduces to
the Kronecker model, and the considered channel model reduces to that in \cite{zhang2013capacity}. Many channel models are already
included in the channel models of \cite{zhang2013capacity} and \cite{wen2011on}.
See the references for more details.

\subsection{Problem Formulation}
Let $\mathbf{H}$ denote $[\mathbf{H}_1  ~ \mathbf{H}_2 ~  \cdots ~ \mathbf{H}_K]$.
In this paper, we are interested in computing the ergodic input-output mutual information of the channel
$\mathbf{H}$ and deriving the sum-rate capacity achieving input covariance matrices.
In particular, we consider the large-system regime where $L$ and $K$ are fixed but $N_l$ and $M_k$ go to infinity with ratios $\frac{M_k}{N_l} = \beta_{lk}$
such that
\begin{equation}
0 < \min\limits_{l,k}\liminf\limits_{N}\beta_{lk} < \max\limits_{l,k}\limsup\limits_{N}\beta_{lk} < \infty.
\label{eq:antenna_size_limiting_regime}
\end{equation}

We first consider the problem of computing the ergodic input-output mutual information.
For simplicity, we assume $\frac{P_k}{M_k}\mathbf{Q}_k=\mathbf{I}_{M_k}$.
The results for general precoders can then be obtained by replacing
$\mathbf{H}_k$ with $\sqrt{\frac{P_k}{M_k}}\mathbf{H}_k\mathbf{Q}_k^{\frac{1}{2}}$.
Let $\mathcal{I}_{\mathbf{B}_N}(\sigma_z^2)$ denote the ergodic input-output mutual information of the channel ${\mathbf{H}}$ and $\mathbf{B}_N$ denote the channel Gram matrix ${\mathbf{H}}{\mathbf{H}}^H$. Under the assumption that the transmitted vector is a Gaussian random vector having an identity covariance matrix and the receiver at the BS has perfect channel state information (CSI), $\mathcal{I}_{\mathbf{B}_N}(\sigma_z^2)$ is given by \cite{Goldsmith03capacitylimits}
    \begin{equation}
          {\mathcal{I}_{\mathbf{B}_N}(\sigma_z^2)} = {\mathbb{E}}\left\{\log\det(\mathbf{I}_N+\frac{1}{\sigma_z^2}\mathbf{B}_N)\right\}.
          \label{mutual_information}
    \end{equation}
Furthermore, we have $\mathcal{I}_{\mathbf{B}_N}(\sigma_z^2) = N\mathcal{V}_{\mathbf{B}_N}(\sigma_z^2)$.
For the considered channel model, an exact expression of $\mathcal{I}_{\mathbf{B}_N}(\sigma_z^2)$ is intractable.
Instead, our goal is to find an approximation of $\mathcal{I}_{\mathbf{B}_N}(\sigma_z^2)$.
From Section II-A and Section II-B, we know that the Shannon transform $\mathcal{V}_{\mathbf{B}_N}(\sigma_z^2)$ can
be obtained from the Cauchy transform $G_{\mathbf{B}_N}(z)$ and the free deterministic equivalent method
can be used to derive the approximation of $G_{\mathbf{B}_N}(z)$. Thus, the problem becomes to construct the free deterministic equivalent
$\boldsymbol{\mathcal{B}}_N$ of $\mathbf{B}_N$, and to derive the Cauchy transform $G_{\boldsymbol{\mathcal{B}}_N}(z)$ and the Shannon transform $\mathcal{V}_{\boldsymbol{\mathcal{B}}_N}(x)$. This problem will be treated in Sections III-A to III-C.

To derive the sum-rate capacity achieving input covariance matrices, we then consider the problem of maximizing the ergodic input-output mutual information $\mathcal{I}_{\mathbf{B}_N}(\sigma_z^2)$.
Since $\mathcal{I}_{\mathbf{B}_N}(\sigma_z^2)=N\mathcal{V}_{\mathbf{B}_N}(\sigma_z^2)$, the problem can be formulated as
    \begin{equation}
        (\mathbf{Q}_{1}^{\diamond},\mathbf{Q}_{2}^{\diamond},\cdots,\mathbf{Q}_{K}^{\diamond}) =\mathop{\arg\max}\limits_{(\mathbf{Q}_{1},\cdots,\mathbf{Q}_{K})\in\mathbb{Q}} \mathcal{V}_{\mathbf{B}_N}(\sigma_z^2)
        \label{eq:optimization_of_information}
    \end{equation}
where the constraint set $\mathbb{Q}$ is defined by
    \begin{IEEEeqnarray}{Rl}
        \!\!\!\!\!\!\!\!\mathbb{Q}=\{(\mathbf{Q}_{1},\mathbf{Q}_{2},\cdots,\mathbf{Q}_{K}):{\rm{tr}}(\mathbf{Q}_k)\leq M_k,\mathbf{Q}_k\succeq 0,\forall k\}.
    \end{IEEEeqnarray}
We assume that the UEs have no CSI, and that each $\mathbf{Q}_k$ is fed back from the BS to the $k$-th UE. Moreover, we assume that all $\mathbf{Q}_k$ are computed from the deterministic matrices $\overline{\mathbf{H}}_{lk},\mathbf{G}_{lk},\mathbf{U}_{lk}$ and $\mathbf{V}_{lk},{1\leq l \leq L, 1\leq k \leq K}$.

Since $\mathcal{I}_{\mathbf{B}_N}(\sigma_z^2)$ is an expected value of the input-output mutual information, the optimization problem in \eqref{eq:optimization_of_information} is a stochastic programming problem. As mentioned in \cite{zhang2013capacity} and \cite{wen2013deterministic}, it is also a convex optimization problem, and thus can be solved by using approaches based on convex optimization with Monte-Carlo methods \cite{boyd2009convex}. More specifically, it can be solved by the Vu-Paulraj algorithm \cite{vu2005capacity}, which was developed from the barrier method \cite{boyd2009convex} with the gradients and Hessians provided by Monte-Carlo methods.

However, the computational complexity of the aforementioned method is very high \cite{zhang2013capacity}. Thus, new approaches are needed. Since the approximation $\mathcal{V}_{\boldsymbol{\mathcal{B}}_N}(\sigma_z^2)$ of
$\mathcal{V}_{\mathbf{B}_N}(\sigma_z^2)$ will be obtained, we can use it as the objective function. Thus, the optimization problem can be reformulated as
\begin{equation}
(\mathbf{Q}_{1}^{\star},\mathbf{Q}_{2}^{\star},\cdots,\mathbf{Q}_{K}^{\star})=\mathop{\arg\max}\limits_{(\mathbf{Q}_{1},\cdots,\mathbf{Q}_{K})\in\mathbb{Q}} \mathcal{V}_{\boldsymbol{\mathcal{B}}_N}(\sigma_z^2).
\end{equation}
The above problem will be solved in Section \ref{sec:Sum_rate_Capacity_Achieving_Input_Covariance_Matrices}.

\section{Main Results}

In this section, we present the free deterministic equivalent of $\mathbf{B}_N$, the deterministic equivalents of the Cauchy transform $G_{\mathbf{B}_N}(z)$ and the Shannon transform $\mathcal{V}_{\mathbf{B}_N}(x)$.
We also present the results for the problem of maximizing the approximate ergodic input-output mutual information $N\mathcal{V}_{\boldsymbol{\mathcal{B}}_N}(\sigma_z^2)$.

Let $\overline{\mathbf{H}}=[\overline{\mathbf{H}}_1 ~ \overline{\mathbf{H}}_2~   \cdots  ~\overline{\mathbf{H}}_K]$ and $\widetilde{\mathbf{H}}=[\widetilde{\mathbf{H}}_1~   \widetilde{\mathbf{H}}_2~   \cdots ~ \widetilde{\mathbf{H}}_K]$. We define
$\mathbf{X}$, $\overline{\mathbf{X}}$ and $\widetilde{\mathbf{X}}$ as in \eqref{eq:definition_of_matrix_bold_captial_x}, \eqref{eq:definition_of_matrix_overline_bold_captial_x} and \eqref{eq:definition_of_matrix_widetilde_bold_captial_x}, respectively.
\subsection{Free Deterministic Equivalent of ${\mathbf{B}}_N$ }
In \cite{benaych2009rectangular}, independent rectangular random matrices are found to be asymptotically free
over a subalgebra when they are embedded in a larger square matrix space.
Motivated by this, we embed $\widetilde{\mathbf{H}}_{lk}$ in the larger matrix space $\mathbf{M}_{N \times M}(\mathcal{P})$.
Let $\widehat{\mathbf{H}}_{lk}$ be the $N \times M$  matrix defined by
    \begin{equation}
         \!\!\widehat{\mathbf{H}}_{lk}= [\mathbf{0}_{N \times {M_{1}}}\cdots \mathbf{0}_{N \times {M_{k-1}}}\check{\mathbf{H}}_{lk}~
        \mathbf{0}_{N \times {M_{k+1}} }  \cdots \mathbf{0}_{N \times {M_K} }]
    \end{equation}
where $\check{\mathbf{H}}_{lk}$ is defined by
\begin{equation}
        \check{\mathbf{H}}_{lk}=[\mathbf{0}_{N_1\times M_k}^T\cdots\mathbf{0}_{N_{l-1}\times M_k}^T\widetilde{\mathbf{H}}_{lk}^T ~{\mathbf{0}_{N_{l+1}\times M_k}^T}\cdots\mathbf{0}_{N_L\times M_k}^T]^T.
\end{equation}
Then, $\widetilde{\mathbf{X}}$ can be rewritten as
\begin{equation}
\widetilde{\mathbf{X}} = \sum\limits_{k=1}^{K}\sum\limits_{l=1}^{L}\widehat{\mathbf{X}}_{lk}
\end{equation}
where $\widehat{\mathbf{X}}_{lk}$ is defined by
\begin{equation}
\widehat{\mathbf{X}}_{lk} = \left(
            \begin{array}{cc}
              \mathbf{0}_{N}              & \widehat{\mathbf{H}}_{lk}  \\
              \widehat{\mathbf{H}}_{lk}^H    & \mathbf{0}_{M}  \\
            \end{array}
\right).
\end{equation}
Recall that $\widetilde{\mathbf{H}}_{lk}={\mathbf{U}}_{lk}\widetilde{\mathbf{W}}_{lk}{\mathbf{V}}_{lk}^H$.
Inspired by \cite{far2008slow},
we rewrite $\widehat{\mathbf{X}}_{lk}$ as
\begin{equation}
\widehat{\mathbf{X}}_{lk}
= {\mathbf{A}}_{lk}\mathbf{Y}_{lk}{\mathbf{A}}_{lk}^H
\end{equation}
where $\mathbf{Y}_{lk}$ and ${\mathbf{A}}_{lk}$ are defined by
\begin{equation}
\mathbf{Y}_{lk}
= \left(
            \begin{array}{cc}
               \mathbf{0}_{N}              &  \widehat{\mathbf{W}}_{lk}                 \\
                \widehat{\mathbf{W}}_{lk}^H  &  \mathbf{0}_{M}    \\
            \end{array}
\right)
\end{equation}
and
\begin{equation}
{\mathbf{A}}_{lk}
= \left(
            \begin{array}{cc}
               {\widehat{\mathbf{U}}}_{lk}  & \mathbf{0}_{N\times M}                           \\
                \mathbf{0}_{M\times N}   &  {\widehat{\mathbf{V}}}_{lk}    \\
            \end{array}
\right)
\end{equation}
where
\begin{IEEEeqnarray}{Rl}
&\widehat{\mathbf{W}}_{lk}=[\mathbf{0}_{N \times {M_{1}}}\!\cdots\mathbf{0}_{N \times {M_{k-1}}}   \check{\mathbf{W}}_{lk}~\mathbf{0}_{N \times {M_{k+1}} } \! \cdots \mathbf{0}_{N \times {M_K} }] \IEEEnonumber
\\
\\
&\check{\mathbf{W}}_{lk}=[\mathbf{0}_{N_1\times M_k}^T\!\cdots\mathbf{0}_{N_{l-1}\times M_k}^T\widetilde{\mathbf{W}}_{lk}^T~{\mathbf{0}_{N_{l+1}\times M_k}^T}\!\cdots\mathbf{0}_{N_L\times M_k}^T]^T \IEEEnonumber
\\
\end{IEEEeqnarray}
and
\begin{IEEEeqnarray}{Rl}
&\!\!\!\!\!\!\widehat{\mathbf{U}}_{lk}={\rm{diag}}( {\mathbf{0}}_{N_1},\cdots,{\mathbf{0}}_{N_{l-1}},{\mathbf{U}}_{lk},{\mathbf{0}}_{N_{l+1}},\cdots,{\mathbf{0}}_{N_{L}})
\\
&\!\!\!\!\!\!\widehat{\mathbf{V}}_{lk}={\rm{diag}}(
{\mathbf{0}}_{M_1},\cdots,{\mathbf{0}}_{M_{k-1}},{\mathbf{V}}_{lk},{\mathbf{0}}_{M_{k+1}},\cdots,{\mathbf{0}}_{M_{K}}).
\end{IEEEeqnarray}

The free deterministic equivalents of $\mathbf{X}$ and $\mathbf{B}_N$ are constructed as follows.
Let $\mathcal{A}$ be a unital algebra, $(\mathcal{A},\phi)$ be a non-commutative probability space and $\boldsymbol{\mathcal{Y}}_{11}, \cdots, \boldsymbol{\mathcal{Y}}_{LK} \in \mathbf{M}_{n}(\mathcal{A})$ be a family of selfadjoint matrices. The entries $[\boldsymbol{\mathcal{Y}}_{lk}]_{ii}$ are centered semicircular elements, and the entries $[\boldsymbol{\mathcal{Y}}_{lk}]_{ij}, i \neq j$, are centered circular elements. The variance of the entry $[\boldsymbol{\mathcal{Y}}_{lk}]_{ij}$ is given by $\phi([\boldsymbol{\mathcal{Y}}_{lk}]_{ij}[\boldsymbol{\mathcal{Y}}_{lk}]_{ij}^*) = \mathbb{E}\{[\mathbf{Y}_{lk}]_{ij}[\mathbf{Y}_{lk}]_{ij}^*\}$. Moreover, the entries on and above the diagonal
of $\boldsymbol{\mathcal{Y}}_{lk}$ are free, and the entries from different $\boldsymbol{\mathcal{Y}}_{lk}$ are also
free. Thus, we also have $\phi([\boldsymbol{\mathcal{Y}}_{lk}]_{ij}[\boldsymbol{\mathcal{Y}}_{pq}]_{rs}) = \mathbb{E}\{[\mathbf{Y}_{lk}]_{ij}[\mathbf{Y}_{pq}]_{rs}\}$, where $lk \neq pq$, $1 \leq l, p \leq L$, $1 \leq k, q \leq K$ and $1 \leq i, j, r, s \leq n $.

\vspace{0.1em}
Let $\widetilde{\boldsymbol{\mathcal{X}}}$ denote $\sum_{k=1}^{K} \sum_{l=1}^{L} \widehat{\boldsymbol{\mathcal{X}}}_{lk}$, where $\widehat{\boldsymbol{\mathcal{X}}}_{lk}=\mathbf{A}_{lk}\boldsymbol{\mathcal{Y}}_{lk}\mathbf{A}_{lk}^H$.
Based on the definitions of $\boldsymbol{\mathcal{Y}}_{lk}$, we have that both the $N \times N$ upper-left block matrix and the $M \times M$ lower-right block matrix of $\widetilde{\boldsymbol{\mathcal{X}}}$ are equal to zero matrices.
Thus,  $\widetilde{\boldsymbol{\mathcal{X}}}$ can be rewritten as \eqref{eq:definition_of_matrix_widetilde_cal_captial_x}, where $\widetilde{\boldsymbol{\mathcal{H}}}$
denotes the $N \times M$ upper-right block matrix of $\widetilde{\boldsymbol{\mathcal{X}}}$.
For fixed $n$, we define the map $E:\mathbf{M}_{n}(\mathcal{A}) \rightarrow \mathcal{M}_n$  by $[E\{\boldsymbol{\mathcal{Y}}_{lk}\}]_{ij} = \phi([\boldsymbol{\mathcal{Y}}_{lk}]_{ij})$.
Then, we have that
\begin{equation}
E\{\widetilde{\boldsymbol{\mathcal{X}}}\mathbf{C}_n\widetilde{\boldsymbol{\mathcal{X}}}\}
=\mathbb{E}\{\widetilde{\boldsymbol{\mathbf{X}}}\mathbf{C}_n\widetilde{\boldsymbol{\mathbf{X}}}\}
\nonumber
\end{equation}
where $\mathbf{C}_n \in \mathcal{M}_n$. Let $\boldsymbol{\mathcal{H}}$ denote $\overline{\mathbf{H}}+\widetilde{\boldsymbol{\mathcal{H}}}$ and $\boldsymbol{\mathcal{B}}_N$ denote $\boldsymbol{\mathcal{H}}\boldsymbol{\mathcal{H}}^H$.
Finally, we define $\boldsymbol{\mathcal{X}}$ as in \eqref{eq:definition_of_matrix_cal_captial_x}.
The matrices $\boldsymbol{\mathcal{X}}$ and $\boldsymbol{\mathcal{B}}_N$ are the free deterministic equivalents of $\mathbf{X}$ and $\mathbf{B}_N$ under the following assumptions.

\begin{assumption}
\label{assump:temp1}
The entries $[M\mathbf{G}_{lk}]_{ij}$ are uniformly bounded.
\end{assumption}
Let $\psi_{lk}[n]:\mathcal{D}_n \rightarrow \mathcal{D}_n$ be defined by
$\psi_{lk}[n](\mathbf{\Delta}_n)=\mathbb{E}_{\mathcal{D}_n}\{\mathbf{Y}_{lk}\mathbf{\Delta}_n\mathbf{Y}_{lk}\}$,
where $\mathbf{\Delta}_n  \in \mathcal{D}_n$. We define $i_n: \mathcal{D}_n \rightarrow L^{\infty}[0, 1]$ by $i_n({\rm{diag}}(d_1,d_2,\cdots,d_n))=\sum_{j=1}^nd_j \chi_{[\frac{j-1}{n},\frac{j}{n}]}$, where
$\chi_{U}$ is the characteristic function of the set $U$.
\begin{assumption}
There exist maps $\psi_{lk}: L^{\infty}[0, 1] \rightarrow L^{\infty}[0, 1]$ such that whenever $i_n(\mathbf{\Delta}_n) \rightarrow d \in L^{\infty}[0, 1]$ in norm, then also $\lim_{n\rightarrow \infty}\psi_{lk}[n](\mathbf{\Delta}_n) = \psi_{lk}(d)$.
\label{assump:variance_operator_valued_limit_1}
\end{assumption}
\begin{assumption}
\label{assump:temp2}
The spectral norms of $\overline{\mathbf{H}}_k\overline{\mathbf{H}}{}_k^H$ are uniformly bounded in $N$.
\end{assumption}

To rigorously show the relation between
$\mathcal{G}^{\mathcal{D}_n}_{\mathbf{X}}(z\mathbf{I}_n)$ and $\mathcal{G}^{\mathcal{D}_n}_{\boldsymbol{\mathcal{X}}}(z\mathbf{I}_n)$, we present the following theorem.
\begin{theorem}
\label{th:coro_of_determinstic_matrices_gaussian_matrices_asymptotic_operator_valued_free}
Let $\mathcal{E}_n$ denote the algebra of $n \times n$ diagonal matrices with uniformly bounded entries and $\mathcal{N}_n$ denote the algebra generated by $\mathbf{A}_{11}, \cdots, \mathbf{A}_{LK}$, $\overline{\mathbf{X}}$ and $\mathcal{E}_n$. Let $m$ be a positive integer and $\mathbf{C}_0, \mathbf{C}_1,  \cdots, \mathbf{C}_m \in \mathcal{N}_n$ be a family of $n \times n$ deterministic matrices. Assume that Assumptions \ref{assump:temp1}
and \ref{assump:temp2} hold.
Then,
\begin{IEEEeqnarray}{Rl}
        &\lim\limits_{n \rightarrow \infty} i_n (\mathbb{E}_{\mathcal{D}_n} \{\mathbf{C}_{0}\mathbf{Y}\!_{p_1q_1}\!\mathbf{C}_{1}\mathbf{Y}\!_{p_2q_2}\!\mathbf{C}_{2} \cdots\mathbf{Y}\!_{p_mq_m}\!\mathbf{C}_{m}\}
         \IEEEnonumber \\
        & ~~~~~-E_{\mathcal{D}_n}\{ \mathbf{C}_{0}\boldsymbol{\mathcal{Y}}\!_{p_1q_1}\!\mathbf{C}_{1} \boldsymbol{\mathcal{Y}}\!_{p_2q_2}\!\mathbf{C}_{2}\cdots\boldsymbol{\mathcal{Y}}\!_{p_mq_m}\! \mathbf{C}_{m}\}) = 0_{L^{\infty}[0, 1]}
        \IEEEnonumber \\
\end{IEEEeqnarray}
where $1 \leq p_1,\cdots, p_m \leq L$, $1 \leq q_1,\cdots, q_m \leq K$ and the definition of $E_{\mathcal{D}_n}\{\cdot\}$ is given in \eqref{eq:definition_of_E_sub_mathcalD_n}. Furthermore, if Assumption \ref{assump:variance_operator_valued_limit_1} also holds, then $\mathbf{Y}_{11}, \cdots, \mathbf{Y}_{LK}$, $\mathcal{N}_n$ are asymptotically free over $L^{\infty}[0, 1]$.
\end{theorem}
\begin{IEEEproof}
From \eqref{eq:antenna_size_limiting_regime} and Assumption 1, we obtain that the entries $[n\mathbf{G}_{lk}]_{ij}$ are uniformly bounded. According to Assumption \ref{assump:temp2}, the spectral norm of $\overline{\mathbf{X}}$ is uniformly bounded in $n$.
Furthermore, the matrices $\mathbf{A}_{lk}$ have unit spectral norm.
Thus, this theorem can be seen as a corollary of Theorem \ref{th:determinstic_matrices_gaussian_matrices_asymptotic_operator_valued_free} in Appendix \ref{New_asymptotic_freeness_results}.
\end{IEEEproof}

Theorem \ref{th:coro_of_determinstic_matrices_gaussian_matrices_asymptotic_operator_valued_free} implies that $\boldsymbol{\mathcal{X}} $ and $\mathbf{X} $ have the same
asymptotic $L^{\infty}[0, 1]$-valued distribution.
This further indicates that $\mathcal{G}_{\mathbf{X}}^{\mathcal{D}_n}(z\mathbf{I}_{n})$ and $\mathcal{G}_{\boldsymbol{\mathcal{X}}}^{\mathcal{D}_n}(z\mathbf{I}_{n})$ are the same in the limit, \textit{i.e.},
\begin{equation}
\lim\limits_{n \rightarrow \infty} i_n \left(\mathcal{G}_{\mathbf{X}}^{\mathcal{D}_n}(z\mathbf{I}_n) - \mathcal{G}_{\boldsymbol{\mathcal{X}}}^{\mathcal{D}_n}(z\mathbf{I}_n)\right) = 0_{L^{\infty}[0, 1]}.
\label{eq:asymptotic_equivalent_of_diagonal_cauchy_transform}
\end{equation}
Following a derivation similar to that of \eqref{eq:realtion_of_Cauchy_transform_of X_and_X2},  we have that
    \begin{equation}
        \mathcal{G}_{\boldsymbol{\mathcal{X}}}^{\mathcal{D}_n}(z\mathbf{I}_n)  = z\mathcal{G}_{\boldsymbol{\mathcal{X}}^{2}}^{\mathcal{D}_n}(z^2\mathbf{I}_n)
        \label{eq:realtion_of_Cauchy_transform_of X_and_X2_over_diagonal_matrix}
    \end{equation}
where $z,z^2 \in \mathbb{C}^+$.
According to \eqref{eq:realtion_of_Cauchy_transform_of X_and_X2},  \eqref{eq:asymptotic_equivalent_of_diagonal_cauchy_transform} and \eqref{eq:realtion_of_Cauchy_transform_of X_and_X2_over_diagonal_matrix}, we have
\begin{equation}
\lim\limits_{n \rightarrow \infty}i_n \left(\mathcal{G}_{\mathbf{X}^2}^{\mathcal{D}_n}(z\mathbf{I}_n) - \mathcal{G}_{\boldsymbol{\mathcal{X}}^2}^{\mathcal{D}_n}(z\mathbf{I}_n)\right) = 0_{L^{\infty}[0, 1]}.
\label{eq:realtion_of_Cauchy_transform_of X2_and_CalX2_over_diagonal_matrix}
\end{equation}
Furthermore, from \eqref{eq:Cauchy_transform_of_Hsquare_in_detail} and its counterpart for $\mathcal{G}_{\boldsymbol{\mathcal{X}}^2}^{\mathcal{D}_n}(z\mathbf{I}_{n})$ we obtain
\begin{equation}
\lim\limits_{N \rightarrow \infty}i_N \left(\mathcal{G}_{\mathbf{B}_N}^{\mathcal{D}_N}(z\mathbf{I}_N) - \mathcal{G}_{\boldsymbol{\mathcal{B}}_N}^{\mathcal{D}_N}(z\mathbf{I}_N)\right) = 0_{L^{\infty}[0, 1]}.
\label{eq:deterministric_equivalent_of_cauchy_transform}
\end{equation}
Since
\begin{equation}
G_{\boldsymbol{\mathcal{B}}_N}(z) = \frac{1}{N}{\rm{tr}}(\mathcal{G}_{\boldsymbol{\mathcal{B}}_N}^{\mathcal{D}_N}(z\mathbf{I}_N)) \nonumber
\end{equation}
and
\begin{equation}
G_{\boldsymbol{\mathbf{B}}_N}(z) = \frac{1}{N}{\rm{tr}}(\mathcal{G}_{\boldsymbol{\mathbf{B}}_N}^{\mathcal{D}_N}(z\mathbf{I}_N)) \nonumber
\end{equation}
we have that $G_{\boldsymbol{\mathcal{B}}_N}(z)$ is the deterministic equivalent of $G_{\boldsymbol{\mathbf{B}}_N}(z)$.

\subsection{Deterministic Equivalent of $G_{\mathbf{B}_N}(z)$}
The calculation of $G_{\boldsymbol{\mathcal{B}}_N}(z)$ can be much easier than that of $G_{\boldsymbol{\mathbf{B}}_N}(z)$ by using operator-valued free probability techniques.
Let $\mathcal{G}_{\boldsymbol{\mathcal{B}}_N}^{\mathcal{M}_N}(z\mathbf{I}_N) = E\{(z\mathbf{I}_{N} - \boldsymbol{\mathcal{B}}_N)^{-1}\}$.
Since $\mathcal{G}_{\boldsymbol{\mathcal{B}}_N}^{\mathcal{D}_N}(z\mathbf{I}_{N}) = E_{\mathcal{D}_N}\{\mathcal{G}_{\boldsymbol{\mathcal{B}}_N}^{\mathcal{M}_N}(z\mathbf{I}_{N})\}$, where $E_{\mathcal{D}_N}\{\cdot\}$ is defined according to \eqref{eq:definition_of_E_sub_mathcalD_n}, we can obtain $G_{\boldsymbol{\mathcal{B}}_N}(z)$ from
$\mathcal{G}_{\boldsymbol{\mathcal{B}}_N}^{\mathcal{M}_N}(z\mathbf{I}_{N})$. We denote by $\mathcal{D}$ the algebra of the form
    \begin{equation}
        \mathcal{D}=\left(
                    \begin{array}{ccccc}
                      \mathcal{M}_N     & \mathbf{0} & \cdots & \mathbf{0} \\
                      \mathbf{0}    & \mathcal{M}_{M_1} & \ldots & \mathbf{0} \\
                      \vdots  &  \vdots & \ddots & \vdots \\
                      \mathbf{0}    & \mathbf{0} & \ldots & \mathcal{M}_{M_K} \\
                    \end{array}
        \right).
    \end{equation}
We define the conditional expectation $E_{\mathcal{D}}: \mathbf{M}_n(\mathcal{A}) \rightarrow \mathcal{D}$ by
    \begin{eqnarray}
        E_{\mathcal{D}}\left\{\left(
                    \begin{array}{ccccc}
                      \boldsymbol{\mathcal{C}}_{11}       & \boldsymbol{\mathcal{C}}_{12}       & \cdots & \boldsymbol{\mathcal{C}}_{1(K+1)} \\
                      \boldsymbol{\mathcal{C}}_{21}       & \boldsymbol{\mathcal{C}}_{22}       & \ldots & \boldsymbol{\mathcal{C}}_{2(K+1)} \\
                      \vdots  &  \vdots & \ddots & \vdots \\
                      \boldsymbol{\mathcal{C}}_{(K+1)1}   & \boldsymbol{\mathcal{C}}_{(K+1)2}   & \ldots & \boldsymbol{\mathcal{C}}_{(K+1)(K+1)} \\
                    \end{array}
        \right)\right\}& &   \nonumber \\
        =\left(
                    \begin{array}{ccccc}
                      E\{\boldsymbol{\mathcal{C}}_{11}\}       & \mathbf{0}       & \cdots &  \mathbf{0} \\
                       \mathbf{0}       & E\{\boldsymbol{\mathcal{C}}_{22}\}       & \ldots &  \mathbf{0} \\
                      \vdots  &  \vdots & \ddots & \vdots \\
                       \mathbf{0}   &  \mathbf{0}   & \ldots & E\{\boldsymbol{\mathcal{C}}_{(K+1)(K+1)}\} \\
                    \end{array}
                    \right)& &
    \end{eqnarray}
where $\boldsymbol{\mathcal{C}}_{11} \in \mathbf{M}_N(\mathcal{A})$, and $\boldsymbol{\mathcal{C}}_{kk} \in \mathbf{M}_{M_{k-1}}(\mathcal{A})$ for $k=2,3,\cdots,K+1$.
Then, we can write $\mathcal{G}_{\boldsymbol{\mathcal{X}}^2}^{\mathcal{D}}(z\mathbf{I}_n)$ for $z \in \mathbb{C}^+$ as
    \begin{eqnarray}
         \!\!\!\!\mathcal{G}_{\boldsymbol{\mathcal{X}}^2}^{\mathcal{D}}(z\mathbf{I}_n)
        \!\!\!\!&=&\!\!\!\!E_{\mathcal{D}}\left\{(z\mathbf{I}_n - \boldsymbol{\mathcal{X}}^2)^{-1}\right\}
        \nonumber \\
        \!\!\!\!&=&\!\!\!\!\left(
                    \begin{array}{ccccc}
                      \mathcal{G}_{\boldsymbol{\mathcal{B}}_N}^{\mathcal{M}_N}(z\mathbf{I}_N)      & \mathbf{0} & \cdots & \mathbf{0} \\
                      \mathbf{0}    & \mathcal{G}_1(z)  & \ldots & \mathbf{0} \\
                      \vdots  & \vdots &  \ddots & \vdots \\
                      \mathbf{0}    & \mathbf{0} &  \ldots & \mathcal{G}_K(z) \\
                    \end{array}
        \right)
        \label{eq:Cauchy_transform_of_Hfreesquare_in_detail}
    \end{eqnarray}
where $\mathcal{G}_k(z)$ denotes $(E\{(z\mathbf{I}_{M} - \boldsymbol{\mathcal{H}}^H\boldsymbol{\mathcal{H}})^{-1}\})_k$ for $k=1, \cdots, K$, and $(\mathbf{A})_k$ denotes the submatrix of $\mathbf{A}$ obtained by extracting the entries of the rows and columns with indices from $\sum\nolimits_{i=1}^{k-1}M_i+1$ to $\sum\nolimits_{i=1}^{k}M_i$. Thus, we can obtain
$\mathcal{G}_{\boldsymbol{\mathcal{B}}_N}^{\mathcal{M}_N}(z\mathbf{I}_N)$ from $\mathcal{G}_{\boldsymbol{\mathcal{X}}^2}^{\mathcal{D}}(z\mathbf{I}_n)$, which is further related to $\mathcal{G}_{\boldsymbol{\mathcal{X}}}^{\mathcal{D}}(z\mathbf{I}_n)$.

\begin{lemma}
    $\widetilde{\boldsymbol{\mathcal{X}}}$ is semicircular over $\mathcal{D}$. Furthermore, $\widetilde{\boldsymbol{\mathcal{X}}}$ and $\mathcal{M}_n$ are free over $\mathcal{D}$.
    \label{lm:semicircular_lemma}
\end{lemma}
\begin{IEEEproof}
    The proof is given in Appendix \ref{sec:proof_of_semicircular_lemma}.
\end{IEEEproof}

Since $\overline{\boldsymbol{\mathbf{X}}} \in \mathcal{M}_n$, we have that $\widetilde{\boldsymbol{\mathcal{X}}}$ and $\overline{\boldsymbol{\mathbf{X}}}$ are free over $\mathcal{D}$. Recall that $\boldsymbol{\mathcal{X}} = \overline{\mathbf{X}}+ \widetilde{\boldsymbol{\mathcal{X}}}$.  Then, $\mathcal{G}_{\boldsymbol{\mathcal{X}}}^{\mathcal{D}}(z\mathbf{I}_n)$ and $\mathcal{G}_{\boldsymbol{\mathcal{X}}^2}^{\mathcal{D}}(z\mathbf{I}_n)$ can be derived. Moreover, we obtain  $\mathcal{G}_{\boldsymbol{\mathcal{B}}_N}^{\mathcal{M}_N}(z\mathbf{I}_N)$ as shown in the following theorem.
\begin{theorem}
\label{th:cauchy_transform}
The $\mathcal{M}_N$-valued Cauchy transform $\mathcal{G}_{\boldsymbol{\mathcal{B}}_N}^{\mathcal{M}_N}(z\mathbf{I}_N)$ for $z \in \mathbb{C}^+$ satisfies
    \begin{IEEEeqnarray}{Rl}
        &{\Tilde{{\boldsymbol{\Phi}}}}(z) = \mathbf{I}_N - \sum\limits_{k=1}^{K}{\tilde{\eta}}_{k} (\mathcal{G}_k(z))
        \label{eq:cauchy_transform_tilde_phi}\\
        &{{\boldsymbol{\Phi}}}(z) = {\rm{diag}}\left(\mathbf{I}_{M_1} - \eta_{1} (\mathcal{G}_{\boldsymbol{\mathcal{B}}_N}^{\mathcal{M}_N}(z\mathbf{I}_N)),\right.
        \IEEEnonumber \\
        &~~~~~~~~~~~~~~~~~~~~\mathbf{I}_{M_2} - \eta_{2} (\mathcal{G}_{\boldsymbol{\mathcal{B}}_N}^{\mathcal{M}_N}(z\mathbf{I}_N)),\cdots,  \IEEEnonumber \\
        & \left.~~~~~~~~~~~~~~~~~~~~~~~\mathbf{I}_{M_K} - \eta_{K} (\mathcal{G}_{\boldsymbol{\mathcal{B}}_N}^{\mathcal{M}_N}(z\mathbf{I}_N))\right)
        \label{eq:cauchy_transform_phi}\\
        &\mathcal{G}_{\boldsymbol{\mathcal{B}}_N}^{\mathcal{M}_N}(z\mathbf{I}_N)
        = \left(z{\Tilde{{\boldsymbol{\Phi}}}}(z)
        -  \overline{\mathbf{H}}{{\boldsymbol{\Phi}}}(z)^{-1}\overline{\mathbf{H}}{}^H\right)^{-1}
        \label{eq:cauchy_transform_of_mathcal_BN} \\
        &\mathcal{G}_k(z)
        = \left(\left(z{{\boldsymbol{\Phi}}}(z)
        - \overline{\mathbf{H}}{}^H{\Tilde{{\boldsymbol{\Phi}}}}(z)^{-1} \overline{\mathbf{H}}\right)^{-1}\right)_k.
        \label{eq:cauchy_transform_of_mathcal_HkHkH}
    \end{IEEEeqnarray}
Furthermore, there exists a unique solution of $\mathcal{G}_{\boldsymbol{\mathcal{B}}_N}^{\mathcal{M}_N}(z\mathbf{I}_N) \in \mathbb{H}_{-}(\mathcal{M}_N) := \{b \in \mathcal{M}_N: \Im(b) \prec 0\}$   for each $z \in \mathbb{C}^+$, and the solution is obtained by iterating (59)-(62).
The Cauchy transform $G_{\boldsymbol{\mathcal{B}}_N}(z)$ is given by
    \begin{equation}
        G_{\boldsymbol{\mathcal{B}}_N}(z) = \frac{1}{N}{\rm{tr}}(\mathcal{G}_{\boldsymbol{\mathcal{B}}_N}^{\mathcal{M}_N}(z\mathbf{I}_N)).
        \label{eq:cauchy_transform_of_BN}
    \end{equation}
\end{theorem}
\begin{IEEEproof}
The proof is given in Appendix {\ref{sec:proof_of_cauchy_theorem}}.
\end{IEEEproof}

In massive MIMO systems, $N_l$ can go to a very large value. In this case, $\mathbf{U}_{lk}$ can be assumed to be independent of $k$, \textit{i.e.}, $\mathbf{U}_{{l1}}=\mathbf{U}_{{l2}}=\cdots=\mathbf{U}_{{lK}}$, under some antenna configurations \cite{noh2014pilot,zhou2006experimental,you2015pilot}. When uniform linear arrays (ULAs) are employed in all ASs and $N_l$ grows very large, each $\mathbf{U}_{lk}$ is closely approximated by a discrete Fourier transform (DFT) matrix \cite{noh2014pilot,zhou2006experimental}. In \cite{you2015pilot}, a more general BS antenna configuration is considered, and it is shown that the eigenvector matrices of the channel covariance matrices at the BS for different users tend to be the same as the number of antennas increases.

Under the assumption $\mathbf{U}_{l1}=\mathbf{U}_{l2}=\cdots=\mathbf{U}_{lK}$, we can obtain simpler results. For brevity, we denote all $\mathbf{U}_{lk}$ by $\mathbf{U}_{l}$. Consider the Rayleigh channel case, \textit{i.e.}, $\overline{\mathbf{H}}=\mathbf{0}$. Let ${\widetilde{\boldsymbol{\Lambda}}}_l(z)$ denote $(\mathbf{I}_{N_l} - \sum_{k=1}^{K}\widetilde{\mathbf{\Pi}}_{lk}(\mathcal{G}_k(z)))^{-1}$. Then, \eqref{eq:cauchy_transform_tilde_phi} becomes
    \begin{IEEEeqnarray}{Rl}
      {\Tilde{{\boldsymbol{\Phi}}}}(z)
      =& {\rm{diag}}\left(\mathbf{U}_{1}({\widetilde{\boldsymbol{\Lambda}}}_1(z))^{-1}\mathbf{U}_{1}^H, \mathbf{U}_{2}({\widetilde{\boldsymbol{\Lambda}}}_2(z))^{-1}\mathbf{U}_{2}^H,  \right.
      \nonumber \\
      &~~~~~~~~~~~~~~~~~~~\left. \cdots,\mathbf{U}_{L}({\widetilde{\boldsymbol{\Lambda}}}_L(z))^{-1}\mathbf{U}_{L}^H\right).
    \end{IEEEeqnarray}
Furthermore, \eqref{eq:cauchy_transform_of_mathcal_BN} and \eqref{eq:cauchy_transform_of_mathcal_HkHkH} become
    \begin{IEEEeqnarray}{Rl}
        &\mathcal{G}_{\boldsymbol{\mathcal{B}}_N}^{\mathcal{M}_N}(z\mathbf{I}_N) =z^{-1}{\rm{diag}}\left(\mathbf{U}_{1}{\widetilde{\boldsymbol{\Lambda}}}_1(z)\mathbf{U}_{1}^H, \mathbf{U}_{2}{\widetilde{\boldsymbol{\Lambda}}}_2(z)\mathbf{U}_{2}^H, \right.
        \nonumber \\
        &~~~~~~~~~~~~~~~~~~~~~~~~~~~~~~~~~~~~~~\left.
         \cdots, \mathbf{U}_{L}{\widetilde{\boldsymbol{\Lambda}}}_L(z)\mathbf{U}_{L}^H\right)
        \\
        &\mathcal{G}_k(z) = z^{-1}\left(\mathbf{I}_{M_k} - \eta_{k} (\mathcal{G}_{\boldsymbol{\mathcal{B}}_N}^{\mathcal{M}_N}(z\mathbf{I}_N))\right)^{-1}.
    \end{IEEEeqnarray}
From \eqref{eq:eta_function_of_widetilde} and \eqref{eq:eta_function_of_widetilde_component}, we have that
    \begin{equation}
        \eta_{k}(\mathcal{G}_{\boldsymbol{\mathcal{B}}_N}^{\mathcal{M}_N}(z\mathbf{I}_N))
        =  \sum\limits_{l=1}^{L}\mathbf{V}_{lk}{\mathbf{\Pi}}_{lk} (\mathbf{U}_{l}{\widetilde{\boldsymbol{\Lambda}}}_l(z)\mathbf{U}_{l}^H)\mathbf{V}_{lk}^H
    \end{equation}
where ${\mathbf{\Pi}}_{lk}(\mathbf{U}_{l}{\widetilde{\boldsymbol{\Lambda}}}_l(z)\mathbf{U}_{l}^H)$ is an $M_k \times M_k$ diagonal matrix valued function with the diagonal entries computed by
    \begin{eqnarray}
        \left[{\mathbf{\Pi}}_{lk}(\mathbf{U}_{l}{\widetilde{\boldsymbol{\Lambda}}}_l(z)\mathbf{U}_{l}^H)\right]_{ii}
        \!\!\!\!&=& \!\!\!\!\sum\limits_{j=1}^{N_l}\left[{\mathbf{G}}_{lk}\right]_{ji} \!\left[\mathbf{U}_{{l}}^H\mathbf{U}_{l}{\widetilde{\boldsymbol{\Lambda}}}_l(z)\mathbf{U}_{l}^H\mathbf{U}_{{l}}\right]_{jj} \nonumber \\
        \!\!\!\!&=& \!\!\!\! \sum\limits_{j=1}^{N_l}\left[{\mathbf{G}}_{lk}\right]_{ji}\!\left[{\widetilde{\boldsymbol{\Lambda}}}_l(z)\right]_{jj}.
    \end{eqnarray}
Thus, $\mathbf{U}_{1}, \mathbf{U}_{2}, \cdots, \mathbf{U}_{L}$ can be omitted in the iteration process, and hence (59)-(63) reduce to
    \begin{IEEEeqnarray}{Rl}
        &\left[{\widetilde{\boldsymbol{\Lambda}}}_l(z)\right]_{ii} =  \left(1 - \sum\limits_{k=1}^{K}\left[\widetilde{\mathbf{\Pi}}_{lk}(\mathcal{G}_k(z))\right]_{ii}\right)^{-1}
        \label{eq:reduce_formula_lambda_tilde_diagonal} \\
        &\widetilde{\boldsymbol{\Lambda}}(z)={\rm{diag}}\left({\widetilde{\boldsymbol{\Lambda}}}_1(z), {\widetilde{\boldsymbol{\Lambda}}}_2(z), \cdots, {\widetilde{\boldsymbol{\Lambda}}}_L(z)\right) \\
        &\mathcal{G}_k(z) = \left(z\mathbf{I}_{M_k} - \eta_{k}({\widetilde{\boldsymbol{\Lambda}}}(z))\right)^{-1}
        \label{eq:tmp1}
           \\
        &{G}_{\boldsymbol{\mathcal{B}}_N}(z) =   z^{-1}\frac{1}{N}\sum\limits_{l=1}^{L}\sum\limits_{i=1}^{N_l}
        \left[{\widetilde{\boldsymbol{\Lambda}}}_l(z)\right]_{ii}
    \label{eq:cauchy_transform_of_BN_large_mimo_1}
    \end{IEEEeqnarray}
where the diagonal entries of ${\mathbf{\Pi}}_{lk}(\langle{{\widetilde{\boldsymbol{\Lambda}}}(z)}\rangle_l)$, which
is needed in the computation of $\eta_{k} ({\widetilde{\boldsymbol{\Lambda}}}(z))$, are now redefined by
    \begin{equation}
        \left[{\mathbf{\Pi}}_{lk}(\langle{{\widetilde{\boldsymbol{\Lambda}}}(z)}\rangle_l)\right]_{ii} =\sum\limits_{j=1}^{N_l}\left[{\mathbf{G}}_{lk}\right]_{ji}\left[{\widetilde{\boldsymbol{\Lambda}}}_l(z)\right]_{jj}.
    \end{equation}
Furthermore, the matrix inversion in \eqref{eq:cauchy_transform_of_mathcal_BN} has been avoided.
When $L=1$, we have that
    \begin{IEEEeqnarray}{Rl}
        &\eta_{k}({\widetilde{\boldsymbol{\Lambda}}}(z)) = \mathbf{V}_{1k}{\mathbf{\Pi}}_{1k}({\widetilde{\boldsymbol{\Lambda}}}_1(z))\mathbf{V}_{1k}^H
        \\
        &\mathcal{G}_k(z) = \mathbf{V}_{1k}\left(z\mathbf{I}_{M_k} - {\mathbf{\Pi}}_{1k}({\widetilde{\boldsymbol{\Lambda}}}_1(z))\right)^{-1}\mathbf{V}_{1k}^H.
    \end{IEEEeqnarray}
Let ${\boldsymbol{\Lambda}}_k(z)$ denote $(z\mathbf{I}_{M_k} - {\mathbf{\Pi}}_{1k}({\widetilde{\boldsymbol{\Lambda}}}_1(z)))^{-1}$. From \eqref{eq:eta_function_component}, we then obtain
    \begin{eqnarray}
        \left[\widetilde{\mathbf{\Pi}}_{1k}(\mathcal{G}_k(z))\right]_{ii} \!\!\!\!&=&\!\!\!\!\sum\limits_{j=1}^{M_k}\left[{\mathbf{G}}_{1k}\right]_{ij} \left[\mathbf{V}_{1k}^H\mathbf{V}_{1k}{\boldsymbol{\Lambda}}_k(z)\mathbf{V}_{1k}^H\mathbf{V}_{1k}\right]_{jj}
        \nonumber \\
        \!\!\!\!&=&\!\!\!\!\sum\limits_{j=1}^{M_k}\left[{\mathbf{G}}_{1k}\right]_{ij}\left[{\boldsymbol{\Lambda}}_k(z)\right]_{jj}.
    \end{eqnarray}
Thus, we can further omit $\mathbf{V}_{11},\mathbf{V}_{12},\cdots,\mathbf{V}_{1K}$ in the iteration process. We redefine
$\widetilde{\mathbf{\Pi}}_{k}({\boldsymbol{\Lambda}}_k(z))$ by
    \begin{equation}
        \left[\widetilde{\mathbf{\Pi}}_{k}({\boldsymbol{\Lambda}}_k(z))\right]_{ii} = \sum\limits_{j=1}^{M_k}\left[{\mathbf{G}}_{1k}\right]_{ij}\left[{\boldsymbol{\Lambda}}_k(z)\right]_{jj}.
    \end{equation}
Equations (59)-(63) can be further reduced to
    \begin{IEEEeqnarray}{Rl}
        &\left[{\widetilde{\boldsymbol{\Lambda}}}_1(z)\right]_{ii} =  \left(1 - \sum\limits_{k=1}^{K}\left[\widetilde{\mathbf{\Pi}}_{k} ({\boldsymbol{\Lambda}}_k(z))\right]_{ii}\right)^{-1} \\
        &\left[{\boldsymbol{\Lambda}}_k(z)\right]_{ii}  = \left(z - \left[{\mathbf{\Pi}}_{k}({\widetilde{\boldsymbol{\Lambda}}}_1(z))\right]_{ii}\right)^{-1}   \\
        &G_{\boldsymbol{\mathcal{B}}_N}(z) =   z^{-1}\frac{1}{N}\sum\limits_{i=1}^{N}
        \left[{\widetilde{\boldsymbol{\Lambda}}}_1(z)\right]_{ii}.
    \end{IEEEeqnarray}
In this case, all matrix inversions have been avoided. Since $\mathbf{U}_1$ and $\mathbf{V}_{11}, \mathbf{V}_{12}, \cdots, \mathbf{V}_{1K}$ have been omitted in the iteration process, we have that the distribution of $\boldsymbol{\mathcal{B}}_N$ depends only on $\{{\mathbf{G}}_{1k}\}$.

Consider now the Rician channel case, \textit{i.e.}, $\overline{\mathbf{H}}\neq \mathbf{0}$. If $\overline{\mathbf{H}}$ has some special structures, we can still obtain simpler results. Let $L=1$ and $\overline{\mathbf{H}}_{1k} = \mathbf{U}_{1}\mathbf{\Sigma}_{1k}\mathbf{V}_{1k}^H$, where $\mathbf{\Sigma}_{1k}$ is an $N\times M_k$ deterministic matrix with at most one nonzero element in each row and each column. In this case, we have that
    \begin{IEEEeqnarray}{Rl}
        &\!\!\!\!\overline{\mathbf{H}}{{\boldsymbol{\Phi}}}(z)^{-1}\overline{\mathbf{H}}{}^H = \mathbf{U}_{1}\left(\sum\limits_{k=1}^K \mathbf{\Sigma}_{1k}\mathbf{V}_{1k}^H\left(\mathbf{I}_{M_k}-
        \vphantom{\mathcal{G}_{\boldsymbol{\mathcal{B}}_N}^{\mathcal{M}_N}}
        \right.\right.
        \nonumber \\
        &~~~~~~~~~~~~~~~~~~ \left.\left.
        \eta_{k}(\mathcal{G}_{\boldsymbol{\mathcal{B}}_N}^{\mathcal{M}_N}(z\mathbf{I}_N))\right)^{-1}\!\!
         \vphantom{\left(\sum\limits_{k=1}^K \mathbf{\Sigma}_{1k}\mathbf{V}_{1k}^H\left(\mathbf{I}_{M_k}-
        \right.\right.}
        \mathbf{V}_{1k}\mathbf{\Sigma}_{1k}^H\right)\mathbf{U}_{1}^H
        \\
        &\!\!\!\!\eta_{k}(\mathcal{G}_{\boldsymbol{\mathcal{B}}_N}^{\mathcal{M}_N}(z\mathbf{I}_N)) = \mathbf{V}_{1k}{\mathbf{\Pi}}_{1k}({\widetilde{\boldsymbol{\Lambda}}}_1(z))\mathbf{V}_{1k}^H
        \\
        &\!\!\!\!{\Tilde{{\boldsymbol{\Phi}}}}(z)
        = \mathbf{U}_{1}({\widetilde{\boldsymbol{\Lambda}}}_1(z))^{-1}\mathbf{U}_{1}^H.
    \end{IEEEeqnarray}
Recall from \eqref{eq:cauchy_transform_of_mathcal_BN} that
\begin{equation}
\mathcal{G}_{\boldsymbol{\mathcal{B}}_N}^{\mathcal{M}_N}(z\mathbf{I}_N) = (z{\Tilde{{\boldsymbol{\Phi}}}}(z)
-  \overline{\mathbf{H}}{{\boldsymbol{\Phi}}}(z)^{-1}\overline{\mathbf{H}}{}^H)^{-1}.
\nonumber
\end{equation}
The matrix inversion in \eqref{eq:cauchy_transform_of_mathcal_BN} can still be avoided, and the distribution of $\boldsymbol{\mathcal{B}}_N$ also does not vary with $\mathbf{U}_{1}$. However,  the matrix inversion in \eqref{eq:cauchy_transform_of_mathcal_BN} can not be avoided even with the assumption $\overline{\mathbf{H}}_{lk}=\mathbf{U}_{{l}}\mathbf{\Sigma}_{lk}\mathbf{V}_{lk}^H$ when $L\neq 1$.

\subsection{Deterministic Equivalent of $\mathcal{V}_{\boldsymbol{\mathbf{B}}_N}(x)$}
In this subsection, we derive the Shannon transform $\mathcal{V}_{{\boldsymbol{\mathcal{B}}}_N}(x)$ from the Cauchy transform $G_{{\boldsymbol{\mathcal{B}}}_N}(z)$.

According to \eqref{eq:deterministric_equivalent_of_cauchy_transform}, we have that
\begin{equation}
\lim\limits_{N \rightarrow \infty} \mathcal{V}_{\boldsymbol{\mathcal{B}}_N}(x) - \mathcal{V}_{{\boldsymbol{\mathbf{B}}}_N}(x) = 0.
\end{equation}
Thus, $\mathcal{V}_{\boldsymbol{\mathcal{B}}_N}(x)$ is the deterministic equivalent of $\mathcal{V}_{{\boldsymbol{\mathbf{B}}}_N}(x)$.
To derive $\mathcal{V}_{\boldsymbol{\mathcal{B}}_N}(x)$, we introduce the following two lemmas.

\begin{lemma}
\label{lm:shannon_theorem_lemma_1}
{Let $\mathbf{E}_k(x)$ denote} $-x\mathcal{G}_k(-x)$ and {$\mathbf{A}(x)$ denote} $({\Tilde{{\boldsymbol{\Phi}}}}(-x)
+ x^{-1}\overline{\mathbf{H}}{{\boldsymbol{\Phi}}}(-x)^{-1} \overline{\mathbf{H}}{}^H)^{-1}$, we have that
\begin{IEEEeqnarray}{Rl}
&\!\!\!\!\!\!\!\!\!\!\!\!-{\rm{tr}}\left(x^{-1}\overline{\mathbf{H}}{}^H\mathbf{A}(x)\overline{\mathbf{H}}\frac{d{{\boldsymbol{\Phi}}}(-x)^{-1}}{dx}\right) \nonumber \\
&= \sum\limits_{k=1}^{K}{\rm{tr}}\left(\left({{\boldsymbol{\Phi}}}_k(-x)^{-1}-\mathbf{E}_k(x)\right)\frac{d{{\boldsymbol{\Phi}}}_k(-x)}{dx}\right)
\label{eq:shannon_theorem_lemma_1}
\end{IEEEeqnarray}
where ${{\boldsymbol{\Phi}}}_k(-x) = \mathbf{I}_{M_k} - \eta_{k} (\mathcal{G}_{\boldsymbol{\mathcal{B}}_N}^{\mathcal{M}_N}(-x\mathbf{I}_N))$.
\end{lemma}
\begin{IEEEproof}
The proof is given in Appendix {\ref{sec:proof_of_shannon_theorem_lemma_1}}
\end{IEEEproof}

\begin{lemma}
\label{lm:shannon_theorem_lemma_2}
\begin{IEEEeqnarray}{Rl}
&\!\!\!\!\!\!\!\!\!\!\!\!{\rm{tr}}\left(\frac{d(x^{-1}\mathbf{A}(x))}{dx}\left({\Tilde{{\boldsymbol{\Phi}}}}(-x)- \mathbf{I}_N\right)\right)
\nonumber \\
&= \sum\limits_{k=1}^{K}{\rm{tr}}\left(\frac{d{{\boldsymbol{\Phi}}}_k(-x)}{dx}x^{-1}\mathbf{E}_k(x)\right).
\label{eq:shannon_theorem_lemma_2}
\end{IEEEeqnarray}
\end{lemma}
\begin{IEEEproof}
The proof is given in Appendix {\ref{sec:proof_of_shannon_theorem_lemma_2}}.
\end{IEEEproof}

Using the above two lemmas and a technique similar to that in \cite{hachem2007deterministic}, we obtain the following theorem.
\begin{theorem}
\label{th:shannon_theorem}
The Shannon transform $\mathcal{V}_{\boldsymbol{\mathcal{B}}_N}(x)$ of $\boldsymbol{\mathcal{B}}_N$ satisfies
    \begin{IEEEeqnarray}{Rl}
        \!\!\!\!\mathcal{V}_{\boldsymbol{\mathcal{B}}_N}(x)
        = & \log\det\left({\Tilde{{\boldsymbol{\Phi}}}}(-x) + x^{-1}\overline{\mathbf{H}}{{\boldsymbol{\Phi}}}(-x)^{-1} \overline{\mathbf{H}}{}^H\right)
        \nonumber \\
        & +\log\det\left({{\boldsymbol{\Phi}}}(-x)\right)
        \nonumber \\
        & ~- {\rm{tr}}\left(x\sum\limits_{k=1}^{K}{{\eta}}_{k} (\mathcal{G}_{\boldsymbol{\mathcal{B}}_N}^{\mathcal{M}_N}(-x\mathbf{I}_N))\mathcal{G}_k(-x) \right)
        \label{eq:theorem_2_present_1}
    \end{IEEEeqnarray}
or equivalently
    \begin{IEEEeqnarray}{Rl}
         \!\!\!\!\mathcal{V}_{\boldsymbol{\mathcal{B}}_N}(x)
    =& \log\det\left({{{\boldsymbol{\Phi}}}}(-x)
        + x^{-1}\overline{\mathbf{H}}{}^H\Tilde{{\boldsymbol{\Phi}}}(-x)^{-1} \overline{\mathbf{H}}\right)
        \nonumber \\
        & +\log\det(\Tilde{{\boldsymbol{\Phi}}}(-x))
        \nonumber \\
        & ~- {\rm{tr}} \left(x\sum\limits_{k=1}^{K}{\tilde{\eta}}_{k} (\mathcal{G}_k(-x))\mathcal{G}_{\boldsymbol{\mathcal{B}}_N}^{\mathcal{M}_N}(-x\mathbf{I}_N)\right).
        \label{eq:theorem_2_present_2}
    \end{IEEEeqnarray}
\end{theorem}
\begin{IEEEproof}
The proof of \eqref{eq:theorem_2_present_1} is given in Appendix {\ref{sec:proof_of_shannon_theorem}}. Equation  \eqref{eq:theorem_2_present_2} can be obtained from \eqref{eq:theorem_2_present_1} easily, and thus its proof is omitted for brevity.
\end{IEEEproof}

\begin{remark}
From Theorems \ref{th:cauchy_transform} and \ref{th:shannon_theorem}, we observe that the deterministic equivalent $\mathcal{V}_{\boldsymbol{\mathcal{B}}_N}(\sigma_z^2)$
is totally determined by  the parameterized one-sided correlation matrices $ {\tilde{\eta}}_{k}(\mathbf{C}_k)$ and ${\eta}_{k}({\widetilde{\mathbf{C}}})$.
 In \cite{zhang2013capacity}, each sub-channel matrix $\widetilde{\mathbf{H}}_{lk}$ reduces to $\mathbf{R}_{lk}^{\frac{1}{2}}\mathbf{W}_{lk}\mathbf{T}_{lk}^{\frac{1}{2}}$, where $\mathbf{R}_{lk}$ and $\mathbf{T}_{lk}$ are deterministic positive definite matrices.
In this case, ${\tilde{\eta}}_k(\mathbf{C}_k)$ becomes
\begin{IEEEeqnarray}{Rl}
  {\tilde{\eta}}_k(\mathbf{C}_k) = & {\rm{diag}}\left(\mathbf{R}_{1k}{\rm{tr}}\left(\mathbf{T}_{1k}\mathbf{C}_k\right),\mathbf{R}_{2k}{\rm{tr}}\left(\mathbf{T}_{2k}\mathbf{C}_k\right),  \right.
  \nonumber \\
  &~~~~~~~~~~~~~~~~~~~~~~
  \left.\cdots,\mathbf{R}_{Lk}{\rm{tr}}\left(\mathbf{T}_{Lk}\mathbf{C}_k\right)\right)
\end{IEEEeqnarray}
and $\eta_k(\widetilde{\mathbf{C}})$ becomes
\begin{equation}
  \eta_k(\widetilde{\mathbf{C}}) = \sum\limits_{l=1}^{L}\mathbf{T}_{lk}{\rm{tr}}(\mathbf{R}_{lk}\langle{\widetilde{\mathbf{C}}}\rangle_l).
\end{equation}
Let $e_{lk}={\rm{tr}}(\mathbf{R}_{lk}\langle{\mathcal{G}_{\boldsymbol{\mathcal{B}}_N}^{\mathcal{M}_N}(-x\mathbf{I}_N)}\rangle_l)$ and
$\tilde{e}_{lk}={\rm{tr}}(\mathbf{T}_{lk}\mathcal{G}_{k}(-x))$. Then, it is easy to show that the deterministic equivalent  $\mathcal{V}_{\boldsymbol{\mathcal{B}}_N}(\sigma_z^2)$ provided by
 \eqref{eq:theorem_2_present_1} or \eqref{eq:theorem_2_present_2} reduces to that provided by Theorem $2$ of \cite{zhang2013capacity} when $\widetilde{\mathbf{H}}_{lk}$ reduces to $\mathbf{R}_{lk}^{\frac{1}{2}}\mathbf{W}_{lk}\mathbf{T}_{lk}^{\frac{1}{2}}$.
\end{remark}

We now summarize the method to compute the deterministic equivalent $\mathcal{V}_{\boldsymbol{\mathcal{B}}_N}(\sigma_z^2)$ of the Shannon transform $\mathcal{V}_{\boldsymbol{\mathbf{B}}_N}(\sigma_z^2)$ as follows: First, initialize $\mathcal{G}_{\boldsymbol{\mathcal{B}}_N}^{\mathcal{M}_N}(-\sigma_z^2\mathbf{I}_N)$ with $\mathbf{I}_N$ and $\mathcal{G}_k(-\sigma_z^2)$ with $\mathbf{I}_{M_k}$. Second, iterate (59)-(62)  until the desired tolerances of $\mathcal{G}_{\boldsymbol{\mathcal{B}}_N}^{\mathcal{M}_N}(-\sigma_z^2\mathbf{I}_N)$ and $\mathcal{G}_k(-\sigma_z^2)$ are satisfied. Third, obtain the deterministic equivalent $\mathcal{V}_{\boldsymbol{\mathcal{B}}_N}(\sigma_z^2)$ by \eqref{eq:theorem_2_present_1} or \eqref{eq:theorem_2_present_2}.

When $N_l$ goes to a very large value, we can also obtain simpler results from Theorem {\ref{th:shannon_theorem}} under some scenarios.
 Consider $\overline{\mathbf{H}}=\mathbf{0}$. Let $\tilde{\lambda}_{li}(x)=1 - \sum_{k=1}^{K}[\widetilde{\mathbf{\Pi}}_{lk}(\mathcal{G}_k(-x))]_{ii}$. We can rewrite $\mathcal{V}_{\boldsymbol{\mathcal{B}}_N}(x)$ as
  \begin{eqnarray}
  &&\!\!\!\!\!\!\!\!\!\!\!\!\!\!\!\!\!\!\!\!\!\!\!\!\mathcal{V}_{\boldsymbol{\mathcal{B}}_N}(x)  = \sum\limits_{i=1}^{K}\log\det(\mathbf{I}_{M_k} - \eta_{k}( {\widetilde{\boldsymbol{\Lambda}}}(-x)))
  \nonumber \\
  &&
  ~+\sum\limits_{l=1}^{L}\sum\limits_{i=1}^{N_l}\log(\tilde{\lambda}_{li}(x))
  + \sum\limits_{l=1}^{L}\sum\limits_{i=1}^{N_l}\frac{1-\tilde{\lambda}_{li}(x)}{\tilde{\lambda}_{li}(x)}.
  \label{eq:Shannon_theorem_remark_1}
\end{eqnarray}
When $L=1$, \eqref{eq:Shannon_theorem_remark_1} further reduces to
   \begin{eqnarray}
  &&\mathcal{V}_{\boldsymbol{\mathcal{B}}_N}(x)  = \sum\limits_{k=1}^{K}\sum\limits_{i=1}^{M_k}\log({\lambda}_{ki}(x))
  +\sum\limits_{i=1}^{N}\log(\tilde{\lambda}_{1i}(x)) \nonumber \\
  &&~~~~~~~~~~~~
  + \sum\limits_{i=1}^{N}\frac{1-\tilde{\lambda}_{1i}(x)}{\tilde{\lambda}_{1i}(x)}
\end{eqnarray}
where ${\lambda}_{ki}(x)$ denotes $1 - [{\mathbf{\Pi}}_{k}({\widetilde{\boldsymbol{\Lambda}}}(-x))]_{ii}$. In the case of $L=1$ and $\overline{\mathbf{H}}_{1k}=\mathbf{U}_{1}\mathbf{\Sigma}_{1k}\mathbf{V}_{1k}$, similar results
to \eqref{eq:Shannon_theorem_remark_1} can still be obtained and are omitted here for brevity.

\subsection{Sum-rate Capacity Achieving Input Covariance Matrices}
\label{sec:Sum_rate_Capacity_Achieving_Input_Covariance_Matrices}
In this subsection, we consider the optimization problem
\begin{equation}
(\mathbf{Q}_{1}^{\star},\mathbf{Q}_{2}^{\star},\cdots,\mathbf{Q}_{K}^{\star})=\mathop{\arg\max}\limits_{(\mathbf{Q}_{1},\cdots,\mathbf{Q}_{K})\in\mathbb{Q}} \mathcal{V}_{\boldsymbol{\mathcal{B}}_N}(\sigma_z^2).
\label{eq:optimization_problem_of_deterministic_equivalent}
\end{equation}
In the previous section, we have obtained the expression of $\mathcal{V}_{\boldsymbol{\mathcal{B}}_N}(x)$  when assuming $\frac{P_k}{M_k}\mathbf{Q}_k=\mathbf{I}_{M_k}$. The results for  general $\mathbf{Q}_k$'s are obtained by replacing the matrices $\overline{\mathbf{H}}_k$ with
 $\sqrt{\frac{P_k}{M_k}}\overline{\mathbf{H}}_k\mathbf{Q}_k^{\frac{1}{2}}$ and $\widetilde{\mathbf{H}}_k$ with $\sqrt{\frac{P_k}{M_k}}\widetilde{\mathbf{H}}_k\mathbf{Q}_k^{\frac{1}{2}}$.
Let ${\tilde{\eta}}_{Q,k}(\mathbf{C}_k)$ and $\eta_{Q,k}({\widetilde{\mathbf{C}}})$ be defined by
    \begin{IEEEeqnarray}{Rl}
          &\!\!\!\!\!\!{\tilde{\eta}}_{Q,k}(\mathbf{C}_k)=\frac{P_k}{M_k}{\rm{diag}}\left(\mathbf{U}_{1k}\widetilde{\mathbf{\Pi}}_{1k} \left(\mathbf{Q}_k^{\frac{1}{2}}\mathbf{C}_k\mathbf{Q}_k^{\frac{1}{2}}\right)\mathbf{U}_{1k}^H, \right.
          \nonumber \\
          & ~~~~~~~~~~~~~~~~~~\mathbf{U}_{2k}\widetilde{\mathbf{\Pi}}_{2k}\left(\mathbf{Q}_k^{\frac{1}{2}}\mathbf{C}_k\mathbf{Q}_k^{\frac{1}{2}}\right) \mathbf{U}_{2k}^H, \cdots,
          \nonumber \\
          &\left.~~~~~~~~~~~~~~~~~~~~~~ \mathbf{U}_{Lk}\widetilde{\mathbf{\Pi}}_{Lk}\left(\mathbf{Q}_k^{\frac{1}{2}}\mathbf{C}_k\mathbf{Q}_k^{\frac{1}{2}}\right) \mathbf{U}_{Lk}^H\right)
          \label{eq:etaQ_function}
    \end{IEEEeqnarray}
and
    \begin{eqnarray}
      \eta_{Q,k}({\widetilde{\mathbf{C}}})
     =\frac{P_k}{M_k}\sum\limits_{l=1}^{L}\mathbf{Q}_k^{\frac{1}{2}}\mathbf{V}_{lk}{\mathbf{\Pi}}_{lk}(\langle{\widetilde{\mathbf{C}}}\rangle_l)\mathbf{V}_{lk}^H\mathbf{Q}_k^{\frac{1}{2}}.
    \label{eq:etaQ_function_of_widetilde}
    \end{eqnarray}
The right-hand sides (RHSs) of \eqref{eq:etaQ_function} and \eqref{eq:etaQ_function_of_widetilde} are obtained by
replacing $\widetilde{\mathbf{H}}_k$ with $\sqrt{\frac{P_k}{M_k}}\widetilde{\mathbf{H}}_k\mathbf{Q}_k^{\frac{1}{2}}$ in
\eqref{eq:eta_function} and \eqref{eq:eta_function_of_widetilde}, respectively.
Let $\overline{\mathbf{S}}$ denote
$[\sqrt{\tfrac{P_1}{M_1}}\overline{\mathbf{H}}_1~\sqrt{\tfrac{P_2}{M_2}}\overline{\mathbf{H}}_2~\cdots~\sqrt{\tfrac{P_K}{M_K}}\overline{\mathbf{H}}_K]
$ and $\mathbf{Q}={\rm{diag}}(\mathbf{Q}_1,\mathbf{Q}_2,\cdots,\mathbf{Q}_K)$.
Then, \eqref{eq:theorem_2_present_1} becomes
\begin{eqnarray}
  \mathcal{V}_{\boldsymbol{\mathcal{B}}_N}(x)
  \!\!\!\!&=&\!\!\!\! \log\det\left(\mathbf{I}_M+\mathbf{\Gamma}\mathbf{Q}\right)
  +\log\det(\Tilde{{\boldsymbol{\Phi}}}(-x))
  \nonumber \\
  &&- {\rm{tr}}\left(x\sum\limits_{k=1}^{K}{\tilde{\eta}}_{Q,k} (\mathcal{G}_k(-x))\mathcal{G}_{\boldsymbol{\mathcal{B}}_N}^{\mathcal{M}_N}(-x)\right)
\label{eq:shannon_transform_general_Q}
\end{eqnarray}
with the following notations
\begin{IEEEeqnarray}{Rl}
&\mathbf{\Gamma}
={\rm{diag}}\left(-\eta_{1} (\mathcal{G}_{\boldsymbol{\mathcal{B}}_N}^{\mathcal{M}_N}(-x\mathbf{I}_N)),-\eta_{2} (\mathcal{G}_{\boldsymbol{\mathcal{B}}_N}^{\mathcal{M}_N}(-x\mathbf{I}_N)), \cdots,\right.
\nonumber \\
&~~~~~~~~~~~~\left.-\eta_{K}(\mathcal{G}_{\boldsymbol{\mathcal{B}}_N}^{\mathcal{M}_N}(-x\mathbf{I}_N))\right) + x^{-1}\overline{\mathbf{S}}{}^H\Tilde{{\boldsymbol{\Phi}}}(-x)^{-1} \overline{\mathbf{S}}
\\
&{\Tilde{{\boldsymbol{\Phi}}}}(-x) = \mathbf{I}_N - \sum\limits_{k=1}^{K}{\tilde{\eta}}_{Q,k} (\mathcal{G}_k(-x))
\label{eq:cauchy_transform_tilde_phi_general_Q} \\
&{{\boldsymbol{\Phi}}}(-x)= {\rm{diag}}\left(\mathbf{I}_{M_1} \!-\!  \eta_{Q,1} (\mathcal{G}_{\boldsymbol{\mathcal{B}}_N}^{\mathcal{M}_N}(-x\mathbf{I}_N)),\right.
\nonumber \\
&~~~~~~~~~~~~~~~~~~~~~~~\mathbf{I}_{M_2} - \eta_{Q,2} (\mathcal{G}_{\boldsymbol{\mathcal{B}}_N}^{\mathcal{M}_N}(-x\mathbf{I}_N)),   \cdots,
 \nonumber\\
&~~~~~~~~~~~~~~~~~~~~~~~~~~\left.\mathbf{I}_{M_K} \!-\!  \eta_{Q,K} (\mathcal{G}_{\boldsymbol{\mathcal{B}}_N}^{\mathcal{M}_N}(-x\mathbf{I}_N))\right)
\label{eq:cauchy_transform_phi_general_Q} \\
&\mathcal{G}_k(-x) = \left(\!\left({-x}{{\boldsymbol{\Phi}}}(-x)
- \mathbf{Q}^{\frac{1}{2}}\overline{\mathbf{S}}{}^H{\Tilde{{\boldsymbol{\Phi}}}}(-x)^{-1} \overline{\mathbf{S}}\mathbf{Q}^{\frac{1}{2}}\right)^{-1}\!\right)_{\!\!k}
\label{eq:cauchy_transform_of_mathcal_HkHkH_general_Q} \\
&\mathcal{G}_{\boldsymbol{\mathcal{B}}_N}^{\mathcal{M}_N}(-x\mathbf{I}_N) = \left({-x}{\Tilde{{\boldsymbol{\Phi}}}}(-x)
-  \overline{\mathbf{S}}\mathbf{Q}^{\frac{1}{2}}{{\boldsymbol{\Phi}}}(-x)^{-1}\mathbf{Q}^{\frac{1}{2}}\overline{\mathbf{S}}{}^H\right)^{-1}\!\!.
\nonumber \\
\label{eq:cauchy_transform_of_mathcal_BN_general_Q}
\end{IEEEeqnarray}
By using a procedure similar to that in  \cite{zhang2013capacity}, \cite{couillet2011deterministic}, \cite{wen2013deterministic} and \cite{dumont2010capacity},
we obtain the following theorem.

\begin{theorem}
\label{th:capaicty_achieving matrix_theorem}
The optimal input covariance matrices
\begin{equation}
(\mathbf{Q}_{1}^{\star},\mathbf{Q}_{2}^{\star},\cdots,\mathbf{Q}_{K}^{\star}) \nonumber
\end{equation}
are the
solutions of the standard waterfilling maximization problem:
    \begin{IEEEeqnarray}{Rl}
        &\max\limits_{\mathbf{Q}_k}   \log\det(\mathbf{I}_{M_k}+\mathbf{\Gamma}_k\mathbf{Q}_k)
        \nonumber \\
        &{\rm s.t.} ~{\rm{tr}}(\mathbf{Q}_k)\leq M_k,\mathbf{Q}_k\succeq 0
    \end{IEEEeqnarray}
where
\begin{eqnarray}
& &\!\!\!\!\!\!\!\!\!\!\!\!\!\!\!\!\mathbf{\Gamma}_k= \langle(\mathbf{I}_{M}+\mathbf{\Gamma}\mathbf{Q}_{\backslash k})^{-1}\mathbf{\Gamma}\rangle_k \\
& &\!\!\!\!\!\!\!\!\!\!\!\!\!\!\!\!\mathbf{Q}_{\backslash k}={\rm{diag}}(\mathbf{Q}_1,\cdots,\mathbf{Q}_{k-1},\mathbf{0}_{M_k},\mathbf{Q}_{k+1},\cdots,\mathbf{Q}_{K}).
\end{eqnarray}
\end{theorem}

\begin{IEEEproof}
The proof is given in Appendix {\ref{sec:proof_of_capaicty_achieving matrix_theorem}}.
\end{IEEEproof}

\begin{remark}
When $L=1$, we have $\mathbf{H}_k=\overline{\mathbf{H}}_{1k}+\mathbf{U}_{1k}(\mathbf{M}_{1k}\odot{\mathbf{W}}_{1k})\mathbf{V}_{1k}^H$ \cite{wen2011on}.
Let $\mathbf{G}_k$ denote $\mathbf{M}_{1k}\odot\mathbf{M}_{1k}$.
We define $\boldsymbol{\psi}_k$ by
$[\boldsymbol{\psi}_k]_j=[\frac{P_k}{M_k}\mathbf{V}_{1k}^H\mathbf{Q}_k^{\frac{1}{2}}\mathcal{G}_k(-x)\mathbf{Q}_k^{\frac{1}{2}}\mathbf{V}_{1k}]_{jj}$. Then, we have that
\begin{equation}
{\tilde{\eta}}_{Q,k}(\mathcal{G}_k(-x))= \mathbf{U}_{1k}{\rm{diag}}({\mathbf{G}}_k\boldsymbol{\psi}_k)\mathbf{U}_{1k}^H. \nonumber
\end{equation}
Similarly, defining $\boldsymbol{\gamma}_{k}$ by
\begin{equation}
[\boldsymbol{\gamma}_{k}]_j=[\frac{P_k}{M_k}\mathbf{U}_{1k}^H\mathcal{G}_{\boldsymbol{\mathcal{B}}_N}^{\mathcal{M}_N}(-x\mathbf{I}_N)\mathbf{U}_{1k}]_{jj} \nonumber
\end{equation}
we have that
\begin{equation}
\eta_k(\mathcal{G}_{{\boldsymbol{\mathcal{B}}_N}}^{\mathcal{M}_N}(-x\mathbf{I}_N)) = \mathbf{V}_{1k}{\rm{diag}}({\mathbf{G}}_k^T\boldsymbol{\gamma}_{k})\mathbf{V}_{1k}^H. \nonumber
\end{equation}
Let $\mathbf{R}_k=-{\tilde{\eta}}_{Q,k}(\mathcal{G}_k(-x))$ and $\mathbf{T}_k=-\eta_k(\mathcal{G}_{{\boldsymbol{\mathcal{B}}_N}}^{\mathcal{M}_N}(-x\mathbf{I}_N))$.
With
\begin{equation}
{\rm{tr}}\left(x\sum_{k=1}^{K}{{\eta}}_{k,Q} (\mathcal{G}_{\boldsymbol{\mathcal{B}}_N}^{\mathcal{M}_N}(-x\mathbf{I}_N))\mathcal{G}_k(-x) \right)=\boldsymbol{\gamma}_k^T{\mathbf{G}}_k\boldsymbol{\psi}_k \nonumber
\end{equation}
and the previous results, it is easy to show that $\mathcal{V}_{\boldsymbol{\mathcal{B}}_N}(\sigma_z^2)$ provided by  \eqref{eq:shannon_transform_general_Q} reduces to that in Proposition 1 of \cite{wen2011on}, and the capacity achieving input covariance matrices provided by Theorem \ref{th:capaicty_achieving matrix_theorem} reduce to that in Proposition 2 of \cite{wen2011on}.

When $\widetilde{\mathbf{H}}_{lk}$ reduces to $\mathbf{R}_{lk}^{\frac{1}{2}}\mathbf{W}_{lk}\mathbf{T}_{lk}^{\frac{1}{2}}$, we have shown in the previous section that $\mathcal{V}_{\boldsymbol{\mathcal{B}}_N}(\sigma_z^2)$ provided by  \eqref{eq:theorem_2_present_1} reduces to that provided by Theorem 2 of \cite{zhang2013capacity}. It follows naturally that the capacity achieving input covariance  matrices presented in Theorem \ref{th:capaicty_achieving matrix_theorem} also reduce to that provided by Proposition 2 of \cite{zhang2013capacity}.
\end{remark}

To obtain $(\mathbf{Q}_{1}^{\star}, \mathbf{Q}_{2}^{\star}, \cdots, \mathbf{Q}_{K}^{\star})$, we need to iteratively compute $\mathbf{\Gamma}$ via (97)-(101). When each $N_l$ goes to a very large value, we can make these equations simpler with the assumption that $\mathbf{U}_{l1}=\mathbf{U}_{l2}=\cdots=\mathbf{U}_{lK}$ under some scenarios. Consider $\overline{\mathbf{H}} = \mathbf{0}$. The diagonal entries of the $N_l \times N_l$ diagonal matrix valued function
${\widetilde{\boldsymbol{\Lambda}}}_l(z)$ in \eqref{eq:reduce_formula_lambda_tilde_diagonal} become
\begin{IEEEeqnarray}{Rl}
&\!\!\!\!\!\!\left[{\widetilde{\boldsymbol{\Lambda}}}_l(z)\right]_{ii}
\nonumber \\
&\!\!= \left(1 -  \sum\limits_{k=1}^{K}\frac{P_k}{M_k}\left[\widetilde{\mathbf{\Pi}}_{lk}(\mathbf{Q}_k^{\frac{1}{2}}\mathcal{G}_k(z)\mathbf{Q}_k^{\frac{1}{2}})\right]_{ii}\right)^{-1}\!\!.
\end{IEEEeqnarray}
Then, equations (98)-(101) reduce to
\begin{IEEEeqnarray}{Rl}
&\mathcal{G}_{\boldsymbol{\mathcal{B}}_N}^{\mathcal{M}_N}(z\mathbf{I}_N) = z^{-1}\mathbf{U}_{R}{\widetilde{\boldsymbol{\Lambda}}}(z)\mathbf{U}_{R}^H
\label{eq:cauchy_transform_of_mathcal_BN_large_mimo_Q} \\
&{\widetilde{\boldsymbol{\Lambda}}}(z)={\rm{diag}}\left({\widetilde{\boldsymbol{\Lambda}}}_1(z), {\widetilde{\boldsymbol{\Lambda}}}_2(z), \cdots, {\widetilde{\boldsymbol{\Lambda}}}_L(z)\right) \\
&\mathcal{G}_k(z)= z^{-1}\left(\mathbf{I}_{M_k} - \eta_{Q,k}(\mathcal{G}_{\boldsymbol{\mathcal{B}}_N}^{\mathcal{M}_N}(z\mathbf{I}_N))\right)^{-1}
\end{IEEEeqnarray}
where $\mathbf{U}_R$ denotes ${\rm{diag}}(\mathbf{U}_1, \mathbf{U}_2, \cdots, \mathbf{U}_L)$.
In the above equations, we have avoided the matrix inversion in \eqref{eq:cauchy_transform_of_mathcal_BN_general_Q}.
However, due to the existence of $\mathbf{Q}$, (98)-(101)
can not be further reduced when $L=1$.
Let $\tilde{\lambda}_{li}(x)$ denote
\begin{equation}
1-\sum_{k=1}^{K}\frac{P_k}{M_k}[\widetilde{\mathbf{\Pi}}_{lk}(\mathbf{Q}_k^{\frac{1}{2}}\mathcal{G}_k(-x)\mathbf{Q}_k^{\frac{1}{2}})]_{ii}.
\nonumber
\end{equation}
We can rewrite $\mathcal{V}_{\boldsymbol{\mathcal{B}}_N}(x)$ for general $\mathbf{Q}$ as
\begin{IEEEeqnarray}{Rl}
  \!\!\!\!\!\!\!\mathcal{V}_{\boldsymbol{\mathcal{B}}_N}(x)
  =& \sum\limits_{i=1}^{K}\log\det\left(\mathbf{I}_{M_k} - \eta_{Q,k} (\mathcal{G}_{\boldsymbol{\mathcal{B}}_N}^{\mathcal{M}_N}(-x\mathbf{I}_N))\right)
  \nonumber \\
  &\!\!\!+\sum\limits_{l=1}^{L}\sum\limits_{i=1}^{N_l}\log(\tilde{\lambda}_{li}(x))
  + \sum\limits_{l=1}^{L}\sum\limits_{i=1}^{N_l}\frac{1-\tilde{\lambda}_{li}(x)}{\tilde{\lambda}_{li}(x)}.
\end{IEEEeqnarray}
In the case of $L=1$ and $\overline{\mathbf{H}}_{1k}=\mathbf{U}_{1}\mathbf{\Sigma}_{1k}\mathbf{V}_{1k}^H$, similar results can be obtained and are omitted here for brevity.
As shown in \cite{wen2011on}, it is easy to prove that the eigenvectors of the optimal input covariance matrix of the $k$-th user are aligned with ${\mathbf{V}_{1k}}$ when $L=1$ and $\overline{\mathbf{H}}_{1k}=\mathbf{0}$.
However, for $L\neq 1$, unless $\mathbf{V}_{lk}$ for different AS are the same, the eigenvectors of the optimal input covariance matrix of the $k$-th user are not aligned with ${\mathbf{V}_{lk}}$ even when $\overline{\mathbf{H}}_k=\mathbf{0}$.

\section{Simulation Results}

In this section, we provide simulation results to show the performance of the proposed free deterministic equivalent approach. Two simulation models are used. One consists of randomly generated jointly correlated channels.
The other is the WINNER II model \cite{meinila2009winner}. The WINNER II channel model is a geometry-based stochastic channel model (GSCM), where the channel parameters are determined stochastically based on statistical distributions extracted from channel measurements. Since the jointly correlated channel is a good approximation of the measured channel \cite{bonek2005experimental,weichselberger2006stochastic}, we assume that it can well approximate the WINNER II channel model. In all simulations, we set ${P_k}={M_k}$, $K=3$ and $L=2$ for simplicity. The signal-to-noise ratio (SNR) is given by SNR$=\frac{1}{M\sigma_z^2}$.

\vspace{0.1em}
For the first simulation model, $\mathbf{M}_{lk}$, ${\mathbf{U}}_{lk}$ and $\mathbf{V}_{lk}$ are all randomly generated. The matrices ${\mathbf{U}}_{lk}$ and ${\mathbf{V}}_{lk}$ are extracted from randomly generated Gaussian matrices with i.i.d. entries via singular value decomposition (SVD), and the entries $[\mathbf{M}_{lk}]_{ij}$ are first generated as uniform random variables with range $[0 ~ 1]$ and then normalized according to \eqref{eq:constraint_of_channel_matrix}. Each deterministic channel matrix $\overline{\mathbf{H}}_{lk}$ is set to a zero matrix for simplicity.

For the WINNER II model,  we use the cluster delay line (CDL) model of the Matlab implementation in \cite{hentila2007matlab} directly. The Fourier transform is used to convert the time-delay channel to a time-frequency channel. The Simulation scenario is set to B1 (typical urban microcell) with line of sight (LOS). The carrier frequency is 5.25GHz. The antenna arrays of both the BS and the users are uniform linear arrays (ULAs) with 1-cm spacing. For other detailed parameters, see \cite{meinila2009winner}. When the simulation model under consideration becomes the WINNER II model, we extract $\overline{\mathbf{H}}_{lk}$, $\mathbf{M}_{lk}$, ${\mathbf{U}}_{lk}$ and $\mathbf{V}_{lk}$ first.

\subsection{Extraction of $\overline{\mathbf{H}}_{lk}$, $\mathbf{M}_{lk}$, ${\mathbf{U}}_{lk}$ and $\mathbf{V}_{lk}$ from WINNER II Model}
We denote  by $S$ the number of samples, and by ${\mathbf{H}}_{lk}(s)$ the $s$-th sample of ${\mathbf{H}}_{lk}$. Then, each deterministic channel matrix
$\overline{\mathbf{H}}_{lk}$ is obtained from
\begin{equation}
\overline{\mathbf{H}}_{lk}=\frac{1}{S}\sum\limits_{s=1}^S{\mathbf{H}}_{lk}(s)
\end{equation}
and each random channel matrix $\widetilde{\mathbf{H}}_{lk}$ is given by
\begin{equation}
\widetilde{\mathbf{H}}_{lk}(s)={\mathbf{H}}_{lk}(s) - \overline{\mathbf{H}}_{lk}.
\end{equation}
Then, we normalize the channel matrices ${\mathbf{H}}_{lk}(s)$ according to \eqref{eq:constraint_of_channel_matrix}.
Furthermore, from the correlation matrices
\begin{IEEEeqnarray}{Rl}
{\mathbf{R}}_{r,{lk}}=&\frac{1}{S}\sum\limits_{s=1}^S\widetilde{\mathbf{H}}_{lk}(s)\widetilde{\mathbf{H}}_{lk}^H(s) \\
{\mathbf{R}}_{t,{lk}}=&\frac{1}{S}\sum\limits_{s=1}^S\widetilde{\mathbf{H}}_{lk}^H(s)\widetilde{\mathbf{H}}_{lk}(s)
\end{IEEEeqnarray}
and their eigenvalue decompositions
\begin{IEEEeqnarray}{Rl}
{\mathbf{R}}_{r,{lk}}=&{\mathbf{U}}_{lk}{\mathbf{\Sigma}}_{r,{lk}}{\mathbf{U}}_{lk}^H \\
{\mathbf{R}}_{t,{lk}}=&{\mathbf{V}}_{lk}{\mathbf{\Sigma}}_{t,{lk}}{\mathbf{V}}_{lk}^H
\end{IEEEeqnarray}
the eigenvector matrices ${\mathbf{U}}_{lk}$ and ${\mathbf{V}}_{lk}$ are obtained. Then,
 the coupling matrices $\mathbf{G}_{lk}=\mathbf{M}_{lk} \odot \mathbf{M}_{lk} $ are computed as \cite{weichselberger2006stochastic}
\begin{equation}
\mathbf{G}_{lk}=\frac{1}{S}\sum\limits_{s=1}^S\left({\mathbf{U}}_{lk}^H{\mathbf{H}}_{lk}(s){\mathbf{V}}_{lk}\right)\odot\left({\mathbf{U}}_{lk}^T{\mathbf{H}}_{lk}^*(s){\mathbf{V}}_{lk}^*\right).
\end{equation}

\subsection{Simulation Results}
\begin{figure}
\centering
\includegraphics[scale=0.55]{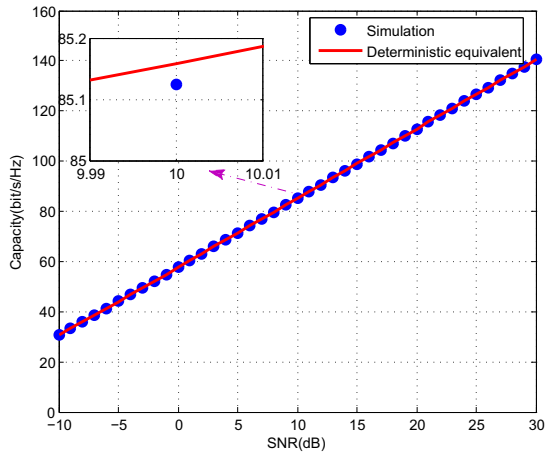}
\caption{Ergodic input-output mutual information versus SNRs of the randomly generated jointly correlated channels with
$N_1=N_2=64, M_1=M_2=M_3=4$. The line plots the deterministic equivalent
results, while the circle markers denote  the simulation results.}
\label{fig:capacity_simulation_analytic_one}
\end{figure}
We first consider the randomly generated jointly correlated channels with $N_1=N_2=64$, $M_1=M_2=M_3=4$ and $\mathbf{Q}_1=\mathbf{Q}_2=\mathbf{Q}_3=\mathbf{I}_{4}$. The results of the simulated ergodic mutual information   $N\mathcal{V}_{\mathbf{B}_N}(\sigma_z^2)$ and their deterministic equivalents $N\mathcal{V}_{{\boldsymbol{\mathcal{B}}}_N}(\sigma_z^2)$ are depicted in Fig.~\ref{fig:capacity_simulation_analytic_one}. The ergodic mutual information $N\mathcal{V}_{\mathbf{B}_N}(\sigma_z^2)$ in Fig.~\ref{fig:capacity_simulation_analytic_one} and the following figures is evaluated by Monte-Carlo simulations, where $10^4$ channel realizations are used for averaging.
As depicted in Fig.~\ref{fig:capacity_simulation_analytic_one}, the deterministic equivalent results are virtually the same as the simulation results.

\begin{figure}
\centering
\subfigure[]{
\includegraphics[scale=0.55]{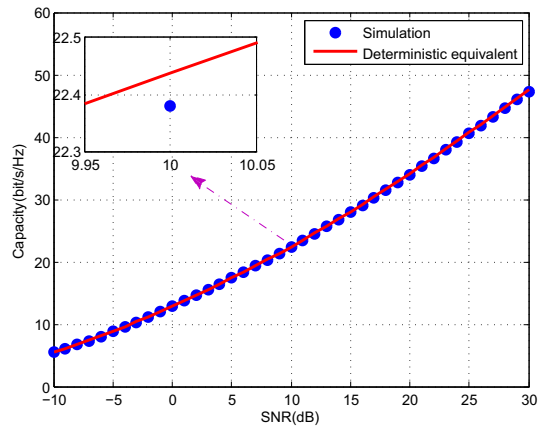}}
\subfigure[]{
\includegraphics[scale=0.55]{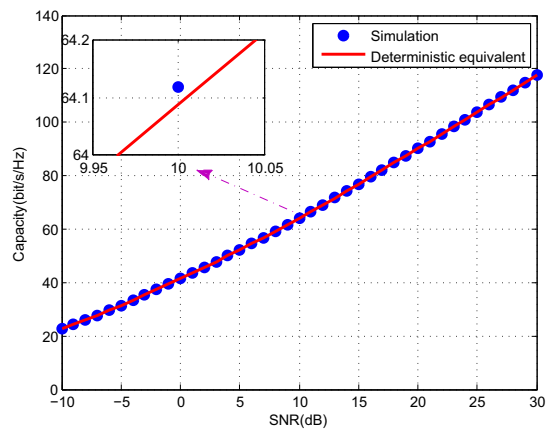}}
\caption{Ergodic input-output mutual information versus SNRs of the WINNER II channel with
(a) $ N_1=N_2=4, M_1=M_2=M_3=4$ and (b) $ N_1=N_2=64, M_1=M_2=M_3=4$. The lines plot the deterministic equivalent
results, while the circle markers denote  the simulation results.}
\label{fig:capacity_simulation_analytic_two}
\end{figure}
We then consider the WINNER II model for the case with $N_1=N_2=4, M_1=M_2=M_3=4$ and the case with $N_1=N_2=64, M_1=M_2=M_3=4$, respectively. For simplicity, we also set $\mathbf{Q}_1=\mathbf{Q}_2=\mathbf{Q}_3=\mathbf{I}_{4}$.
In Fig.~\ref{fig:capacity_simulation_analytic_two}, the ergodic mutual information $N\mathcal{V}_{\mathbf{B}_N}(\sigma_z^2)$ and their deterministic equivalents $N\mathcal{V}_{{\boldsymbol{\mathcal{B}}}_N}(\sigma_z^2)$ are represented. As shown in both  Fig.~\ref{fig:capacity_simulation_analytic_two}(a) and  Fig.~\ref{fig:capacity_simulation_analytic_two}(b), the differences between the deterministic equivalent results and the simulation results are negligible.

\begin{table}
\renewcommand{\arraystretch}{1.3}
\caption{Average execution time in seconds}
\label{table_example}
\centering
\begin{tabular}{|c|c|c|c|}
  \hline
                         & $N_1$=$N_2$=4 & $N_1$=$N_2$=64 & $N_1$=$N_2$=64  \\
                         & \!$M_1$=$M_2$=$M_3$=4\!& \!$M_1$=$M_2$=$M_3$=4\! &\!$M_1$=$M_2$=$M_3$=8\! \\
  \hline
  Monte-Carlo   & 9.74 & 12.9014 & 24.6753\\
  \hline
  DE  & 0.0269 & 0.3671 & 0.4655 \\
  \hline
\end{tabular}
\label{lb:average_execution_time_1}
\end{table}
To show the computational efficiency of the proposed deterministic equivalent $N\mathcal{V}_{{\boldsymbol{\mathcal{B}}}_N}(\sigma_z^2)$, we provide in Table~\ref{lb:average_execution_time_1} the average execution time for both the Monte-Carlo simulation and the proposed
algorithm, on a 1.8 GHz Intel quad core i5 processor with 4 GB of RAM, under different system sizes.
As shown in Table~\ref{lb:average_execution_time_1}, the proposed deterministic equivalent results are
much more efficient.
Moreover, the comparison indicates that the proposed deterministic equivalent provides a promising
foundation to derive efficient algorithms for system optimization.

\begin{figure*}
\centering
\subfigure[]{
\includegraphics[scale=0.55]{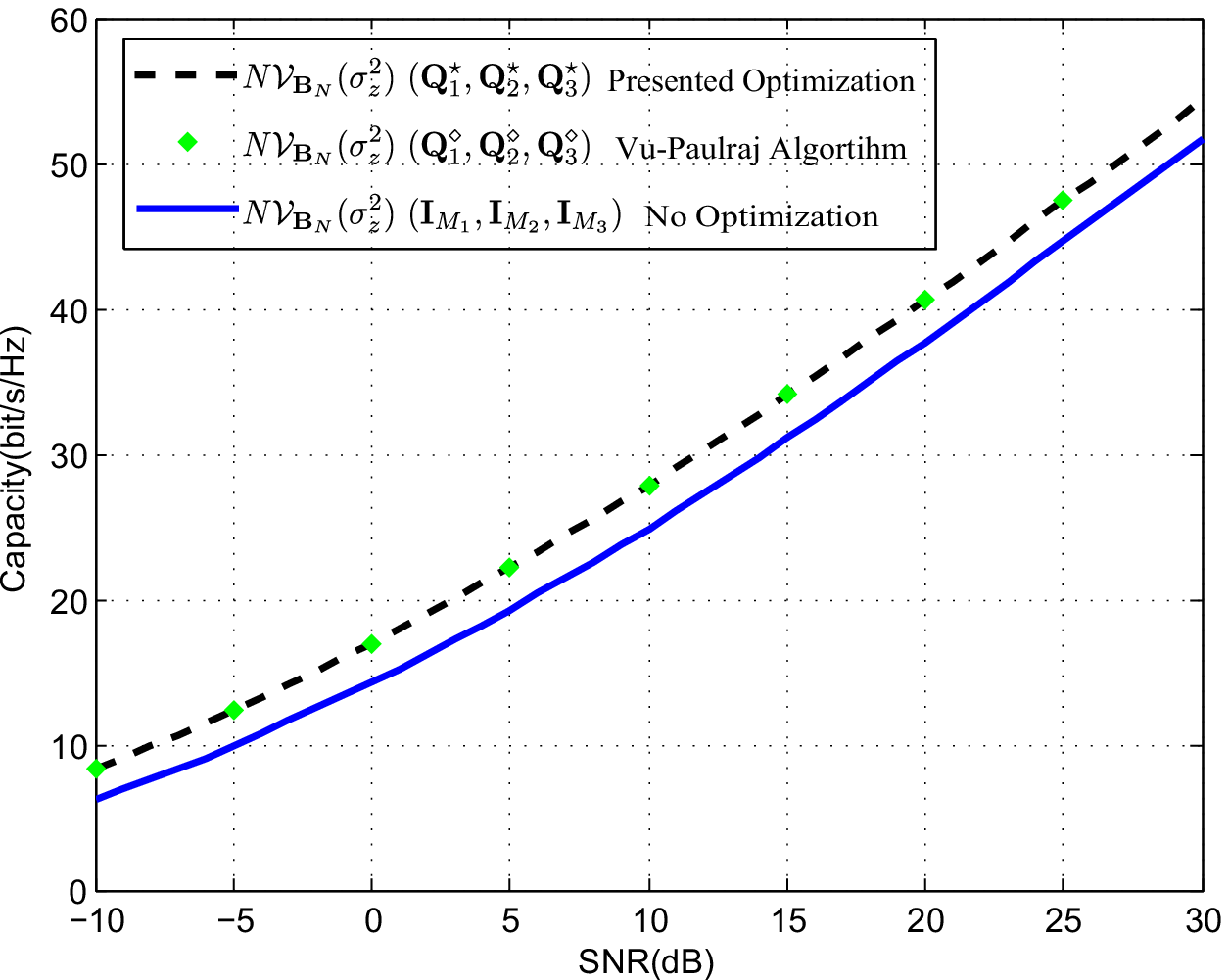}
}
\subfigure[]{
\includegraphics[scale=0.55]{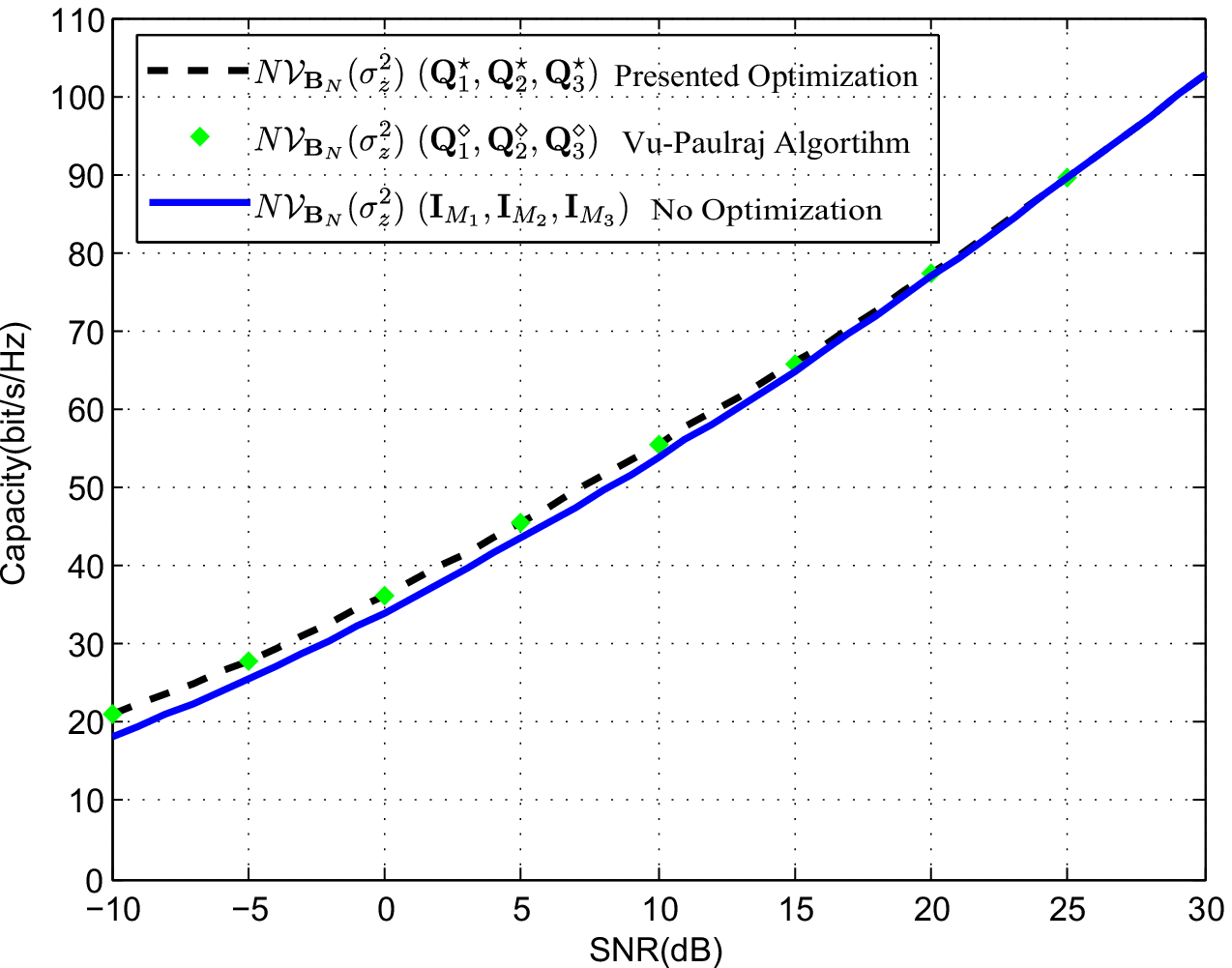}
}
\subfigure[]{
\includegraphics[scale=0.55]{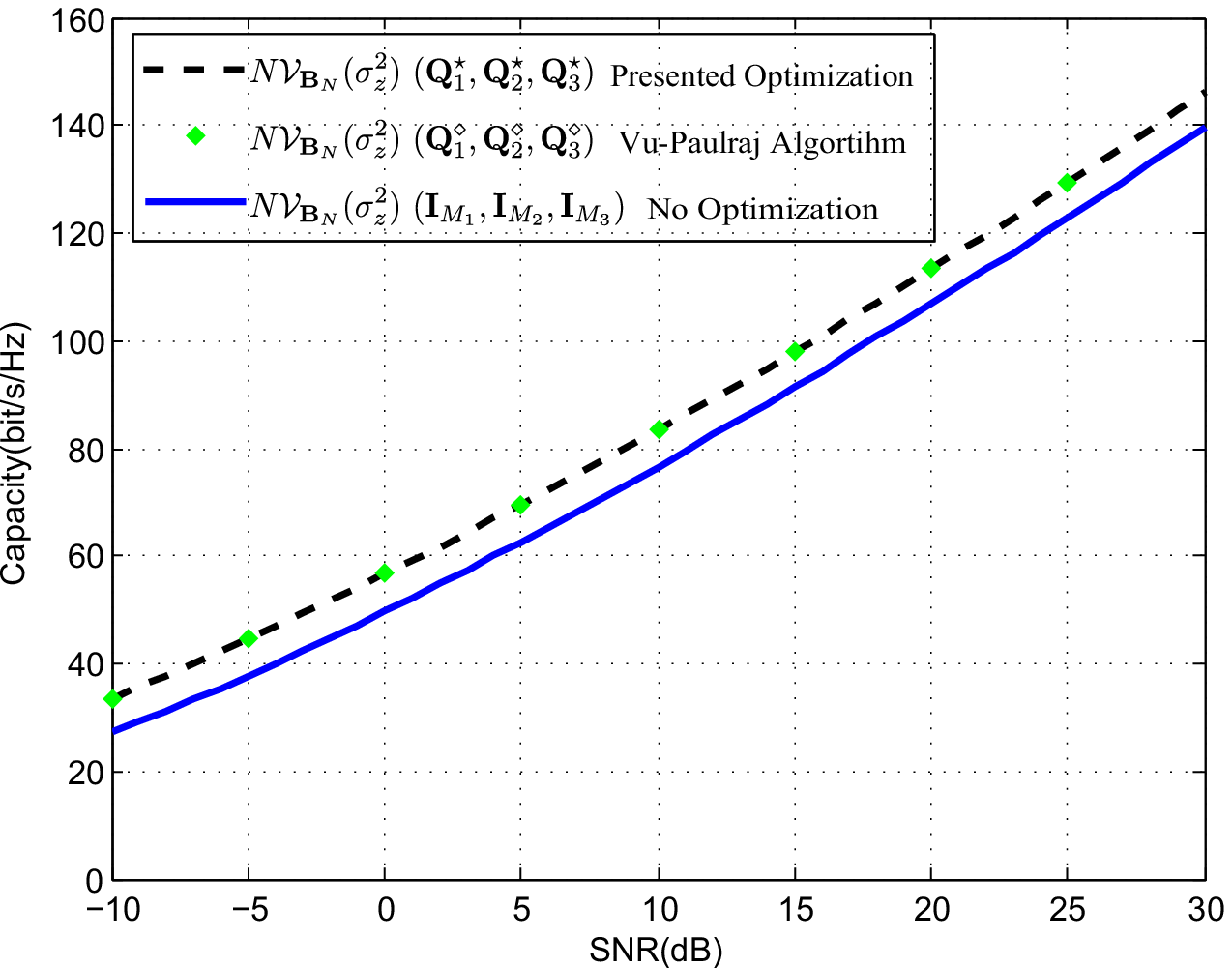}
}
\subfigure[]{
\includegraphics[scale=0.55]{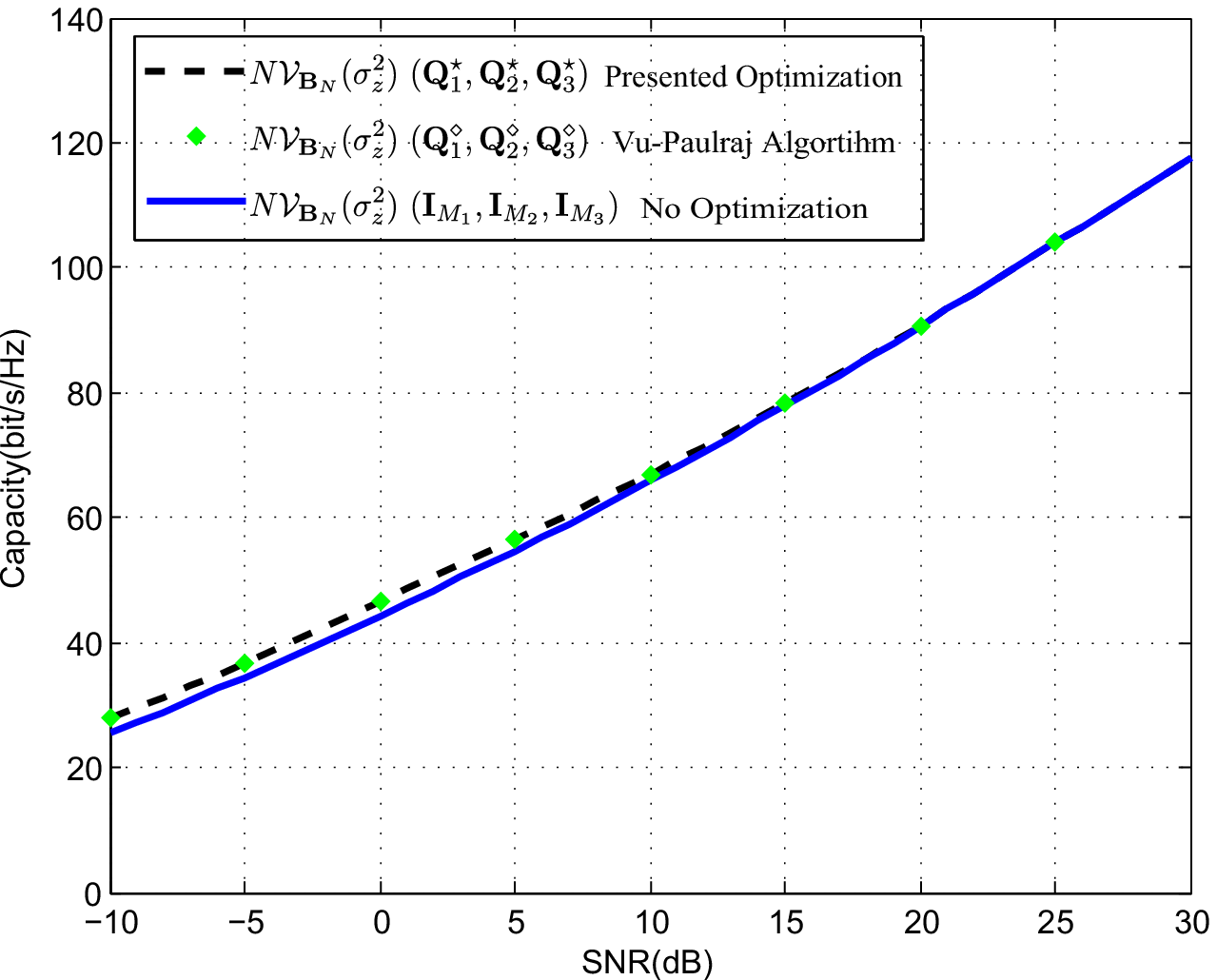}
}
\caption{Ergodic input-output mutual information versus SNRs of the WINNER II channel with
(a) $N_1=N_2=4, M_1=M_2=M_3=4$, (b) $N_1=N_2=32, M_1=M_2=M_3=4$,
(c) $N_1=N_2=32, M_1=M_2=M_3=8$ and (d) $N_1=N_2=64, M_1=M_2=M_3=4$. The solid lines plot the simulation results without optimization.
The dashed lines denote the simulation results of the proposed algorithm, while the diamond markers denote the simulation results of the Vu-Paulraj algorithm.}
\label{fig:precoding_simulation_analytic}
\end{figure*}

Simulations are also carried out to evaluate the performance of the capacity achieving input covariance matrices    $(\mathbf{Q}_{1}^{\star},\mathbf{Q}_{2}^{\star},\mathbf{Q}_{3}^{\star})$. Fig.~\ref{fig:precoding_simulation_analytic} depicts the results of the WINNER II channel models with various system sizes.
In Fig.~1 and Fig.~2, we have shown that the deterministic equivalent $N\mathcal{V}_{{\boldsymbol{\mathcal{B}}}_N}(\sigma_z^2)$ and the simulated ergodic mutual information
$N\mathcal{V}_{{\mathbf{B}}_N}(\sigma_z^2)$ are nearly the same. Since
the latter represents the actual performance of the input covariance matrices,
we use it for producing the numerical results in Fig.~3.
In all four subfigures of Fig.~\ref{fig:precoding_simulation_analytic}, both the ergodic mutual information $N\mathcal{V}_{\mathbf{B}_N}(\sigma_z^2)$ for $(\mathbf{Q}_{1}^{\star},\mathbf{Q}_{2}^{\star},\mathbf{Q}_{3}^{\star})$  and the ergodic mutual information $N\mathcal{V}_{\mathbf{B}_N}(\sigma_z^2)$ without optimization (\textit{i.e.}, for $(\mathbf{I}_{M_1},\mathbf{I}_{M_2},\mathbf{I}_{M_3})$) are shown. Let $(\mathbf{Q}_{1}^{\diamond},\mathbf{Q}_{2}^{\diamond},\mathbf{Q}_{3}^{\diamond})$ denote the solution of the Vu-Paulraj algorithm. The ergodic mutual information  $N\mathcal{V}_{\mathbf{B}_N}(\sigma_z^2)$ for $(\mathbf{Q}_{1}^{\diamond}, \mathbf{Q}_{2}^{\diamond}, \mathbf{Q}_{3}^{\diamond})$ are also given for comparison. We note that the ergodic mutual information $N\mathcal{V}_{\mathbf{B}_N}(\sigma_z^2)$ for $(\mathbf{Q}_{1}^{\star}, \mathbf{Q}_{2}^{\star}, \mathbf{Q}_{3}^{\star})$ and that for $(\mathbf{Q}_{1}^{\diamond}, \mathbf{Q}_{2}^{\diamond}, \mathbf{Q}_{3}^{\diamond})$ are indistinguishable. We also observe that increasing the number of receive antennas decreases the optimization gain when the number of transmit antennas is fixed, whereas increasing the number of transmit antennas provides a larger gain when the number of receive antennas is fixed. The main reason behind this phenomenon is as the following: If the number of transmit antennas is fixed, then more receive antennas means lower correlations between the received channel vectors from each transmit antenna (columns of the channel matrices), and thus the performance gain provided by the optimization algorithm becomes smaller. On the other hand, if the number of receive antennas is fixed, then the received channel vectors from each transmit antenna become more correlated as the number of transmit antennas increases, and thus a larger optimization gain can be observed.

\section{Conclusion}
In this paper, we proposed a free deterministic equivalent for the capacity analysis of a MIMO MAC with a more general channel model compared to previous works. The analysis is based on operator-valued free probability theory.
We explained why the free deterministic equivalent method for the considered channel model is reasonable, and also showed how to obtain the free deterministic equivalent of the channel Gram matrix.
The obtained free deterministic equivalent is an operator-valued random variable.
Then, we derived the Cauchy transform of the free deterministic equivalent, the approximate Shannon transform and hence the approximate ergodic mutual information.
Furthermore, we maximized the approximate ergodic mutual information to obtain the sum-rate capacity achieving input covariance matrices. Simulation results showed that the approximations are not only numerically accurate but also computationally efficient. The results of this paper can be used to design optimal precoders and evaluate the capacity or ergodic mutual information for massive MIMO uplinks with multiple antenna users.

\appendices

\section{Prerequisites and Free Deterministic Equivalents}
Free probability theory was introduced by Voiculescu as a non-commutative probability theory equipped with a notion of freeness. Voiculescu pointed out that freeness should be seen as an analogue to independence in classical probability theory \cite{speicher2014free}. Operator-valued free probability theory was also presented by Voiculescu from the very beginning in \cite{voiculescu1985symmetries}. In this appendix, we briefly review definitions and results of free probability theory and operator-valued free probability theory, and introduce the free deterministic equivalents used in this paper with a rigorous mathematical justification.

\subsection{Free Probability and Operator-valued Free Probability}
\label{sec:Free Probability and Operator-valued Free Probability}
In this subsection, we briefly review definitions and results of free probability theory \cite{nica2006lectures, speicher2014free} and operator-valued free probability theory \cite{shlyakhtenko1998gaussian, speicher2014free, shlyakhtenko1996random, belinschi2013analytic, roland1998combinatorial}.

Let $\mathcal{A}$ be a unital algebra. A non-commutative probability space $(\mathcal{A},\phi)$ consists of $\mathcal{A}$ and a linear functional $\phi: \mathcal{A} \rightarrow \mathbb{C}$. The elements of a non-commutative probability space are called non-commutative random variables. If $\mathcal{A}$ is also a $C^*$-algebra and $\phi(a^*a) \geq 0 $ for all $a \in \mathcal{A}$, then $(\mathcal{A},\phi)$ is a $C^*$-probability space. An element $a$ of $\mathcal{A}$ is called a selfadjoint random variable  if $a = a^*$; an element $u$ of $\mathcal{A}$ is called a unitary random variable  if $uu^* = u^*u=1$; an element $a$ of $\mathcal{A}$ is called a normal random variable if $aa^* = a^*a$.

Let $(\mathcal{A}, \phi)$ be a $C^*$-probability space and $a \in \mathcal{A}$ be a normal random variable. If there exists a compactly supported probability measure $\mu_a$ on $\mathbb{C}$ such that
    \begin{equation}
        \int z^k({z}^*)^l d \mu_a(z)=\phi(a^k(a^*)^l), k, l \in \mathbb{N}
    \end{equation}
then $\mu_a$ is uniquely determined and called the $*$-distribution of $a$. If $a$ is selfadjoint, then $\mu_a$ is simply called the distribution of $a$.

Let $\mathcal{A}_1, \mathcal{A}_2, \cdots, \mathcal{A}_n$ be a family of unital subalgebras of $\mathcal{A}$ and $k$ be a positive integer. The subalgebras $\mathcal{A}_i$ are called free or freely independent,  if $\phi(x_1x_2\cdots x_k) = 0$  for any $k$, whenever $\phi(x_j) = 0$ and $x_j \in \mathcal{A}_{i(j)}$ for all $j$, and $i(j) \neq i(j + 1)$ for $j = 1, \cdots , k-1$. Let $y_1,y_2,\cdots,y_n \in \mathcal{A}$. The non-commutative random variables $y_i$ are called free, if the unital subalgebras ${\rm alg}(1,y_i)$ are free, where ${\rm alg}(1,y_i)$ denotes the unital algebra generated by the random variable $y_i$.

Let $(\mathcal{A}, \phi)$ be a $C^*$-probability space, $s  \in \mathcal{A}$ be a selfadjoint element and $r$ be a positive real number. If the distribution of $s$ is determined by \cite{nica1996multiplication}
    \begin{equation}
        \phi(s^n) = \frac{2}{\pi r^2}\int_{-r}^{r}t^n\sqrt{r^2 - t^2}dt
    \end{equation}
then $s$ is a semicircular element of radius $r$.
An element $c$ with the definition $c=\frac{1}{\sqrt{2}}(s_1+is_2)$ is called a circular element, if $s_1$ and $s_2$ are two freely independent semicircular elements with the same variance.

Let $\mathcal{B} \subset \mathcal{A}$ be a unital subalgebra. A linear map $F: \mathcal{A} \rightarrow \mathcal{B}$ is a conditional expectation, if $F[b] = b$ for all $b \in \mathcal{B}$ and $F[b_1\boldsymbol{\mathcal{X}}b_2] = b_1F[\boldsymbol{\mathcal{X}}]b_2$ for all
$\boldsymbol{\mathcal{X}} \in \mathcal{A}$ and $b_1, b_2 \in \mathcal{B}$.
An operator-valued probability space $(\mathcal{A},F)$, also called $\mathcal{B}$-valued probability space, consists of $\mathcal{B} \subset \mathcal{A}$ and a conditional expectation $F: \mathcal{A} \rightarrow \mathcal{B}$. The elements of a $\mathcal{B}$-valued probability space are called $\mathcal{B}$-valued random variables. If in addition $\mathcal{A}$ is a $C^*$-algebra, $\mathcal{B}$ is a $C^*$-subalgebra and $F$ is completely positive, then $(\mathcal{A},F)$ is a $\mathcal{B}$-valued $C^*$-probability space. Let $\boldsymbol{\mathcal{X}}$ be a $\mathcal{B}$-valued random variable of $(\mathcal{A},F)$. The $\mathcal{B}$-valued distribution of $\boldsymbol{\mathcal{X}}$ is given by all $\mathcal{B}$-valued moments $F[\boldsymbol{\mathcal{X}}b_1\boldsymbol{\mathcal{X}}b_2 \cdots \boldsymbol{\mathcal{X}}b_{n-1}\boldsymbol{\mathcal{X}}]$, where $b_1,b_2,\cdots,b_{n-1} \in \mathcal{B}$.

We denote by $\mathbf{M}_n(\mathcal{P})$ the algebra of $n \times n$ complex random matrices.
The mathematical expectation operator $\mathbb{E}$ over $\mathbf{M}_n(\mathcal{P})$ is a conditional expectation from $\mathbf{M}_n(\mathcal{P})$ to $\mathcal{M}_n$. Thus, $(\mathbf{M}_n(\mathcal{P}), \mathbb{E})$ is an $\mathcal{M}_n$-valued $C^*$-probability space. Furthermore, $(\mathbf{M}_n(\mathcal{P}), \mathbb{E}_{\mathcal{D}_n})$ is a $\mathcal{D}_n$-valued probability space, and $(\mathbf{M}_n(\mathcal{P}),  \frac{1}{n}{\rm{tr}} \circ \mathbb{E})$ or $(\mathbf{M}_n(\mathcal{P}),  \frac{1}{n}{\rm{tr}} \circ \mathbb{E}_{\mathcal{D}_n})$ is a $C^*$-probability space. Let $\mathbf{X} \in \mathbf{M}_n(\mathcal{P})$ be a random Hermitian matrix. Then, $\mathbf{X}$ is at the same time an $\mathcal{M}_n$-valued, a $\mathcal{D}_n$-valued and a scalar valued $C^*$-random variable. The $\mathcal{M}_n$-valued distribution of $\mathbf{X}$ determines the $\mathcal{D}_n$-valued distribution of $\mathbf{X}$, which determines also the expected eigenvalue distribution of $\mathbf{X}$.

Let $\boldsymbol{\mathcal{X}}_1,\boldsymbol{\mathcal{X}}_2,\cdots,\boldsymbol{\mathcal{X}}_k \in (\mathcal{A},F)$ denote a family of $\mathcal{B}$-valued random variables and $n$ be a positive integer. Let $A_i$ denote the polynomials in some $\boldsymbol{\mathcal{X}}_{j(i)}$ with coefficients from $\mathcal{B}$, \textit{i.e.}, $A_i \in \mathcal{B}\langle \boldsymbol{\mathcal{X}}_{j(i)} \rangle$ for $i=1,2,\cdots,n$. The $\mathcal{B}$-valued random variables $\boldsymbol{\mathcal{X}}_i $ are free with amalgamation over $\mathcal{B}$, if $F(A_1A_2\cdots A_n) = 0$ for any $n$,
whenever $F(A_i) = 0$ for all $i$, and $j(i) \neq j(i + 1)$ for $i = 1, \cdots , n-1$.

Let $S(n)$ be the finite totally ordered set $\{1,2,\cdots,n\}$ and $V_i(1 \leq i \leq r)$ be pairwise disjoint subsets of $S(n)$. A set $\pi = \{V_1,V_2,\cdots,V_r\}$ is called a partition if $V_1 \cup V_2 \cdots \cup V_r = S(n)$. The subsets $V_1,V_2,\cdots,V_r$ are called blocks of $\pi$. The set of non-crossing partitions of $S(n)$ is denoted by $NC(n)$.

The $\mathcal{B}$-valued multiplicative maps $\{f_{\pi}^{\mathcal{B}}\}_{\pi \in NC(n)}:\mathcal{A}^n \rightarrow \mathcal{B}$ are defined recursively as
    \begin{IEEEeqnarray}{Rl}
        &\!\!\!\!f_{\pi_1\sqcup\pi_2}^{\mathcal{B}}(\boldsymbol{\mathcal{X}}_1, \boldsymbol{\mathcal{X}}_2, \cdots, \boldsymbol{\mathcal{X}}_n)
        \nonumber \\
        &=
        f_{\pi_1}^{\mathcal{B}}(\boldsymbol{\mathcal{X}}_1, \boldsymbol{\mathcal{X}}_2, \cdots, \boldsymbol{\mathcal{X}}_p) f_{\pi_2}^{\mathcal{B}}(\boldsymbol{\mathcal{X}}_{p+1}, \boldsymbol{\mathcal{X}}_{p+2}, \cdots, \boldsymbol{\mathcal{X}}_n)
        \\
        &\!\!\!\!f_{{\rm ins}(p,\pi_2\rightarrow\pi_1)}^{\mathcal{B}}(\boldsymbol{\mathcal{X}}_1, \boldsymbol{\mathcal{X}}_2,
        \cdots, \boldsymbol{\mathcal{X}}_n)
         \nonumber \\
        &=
        f_{\pi_1}^{\mathcal{B}}(\boldsymbol{\mathcal{X}}_1, \boldsymbol{\mathcal{X}}_2, \cdots,
        \boldsymbol{\mathcal{X}}_p f_{\pi_2}^{\mathcal{B}}(\boldsymbol{\mathcal{X}}_{p+1},\boldsymbol{\mathcal{X}}_{p+2},
        \cdots, \boldsymbol{\mathcal{X}}_{p+q}), \nonumber \\
     &~~~~~~~~~~~~~~~~~\boldsymbol{\mathcal{X}}_{p+q+1},\boldsymbol{\mathcal{X}}_{p+q+2},\cdots,\boldsymbol{\mathcal{X}}_n)
    \end{IEEEeqnarray}
where $\pi_1$ and $\pi_2$ are two non-crossing partitions, $\pi_1\sqcup\pi_2$ denotes the disjoint union with $\pi_2$ after $\pi_1$, and ${\rm ins}(p,\pi_2\rightarrow\pi_1)$ denotes the partition obtained from $\pi_1$ by inserting the partition $\pi_2$ after the $p$-th element of the set on which $\pi_1$ determines a partition. Let $\mathbf{1}_n$ denote $\{\{1,2,\cdots,n\}\}$, $\mathbf{0}_n$ denote $\{\{1\},\{2\},\cdots,\{n\}\}$ and $f_{n}^{\mathcal{B}}(\boldsymbol{\mathcal{X}}_1,\boldsymbol{\mathcal{X}}_2,\cdots,\boldsymbol{\mathcal{X}}_n)$ denote $
f_{\mathbf{1}_n}^{\mathcal{B}}(\boldsymbol{\mathcal{X}}_1,\boldsymbol{\mathcal{X}}_2,\cdots,\boldsymbol{\mathcal{X}}_n)$.

Let $\nu_{\pi}^{\mathcal{B}}:\mathcal{A}^n \rightarrow \mathcal{B}$ be defined by $\nu_{n}^{\mathcal{B}}(\boldsymbol{\mathcal{X}}_1,\boldsymbol{\mathcal{X}}_2,\cdots,\boldsymbol{\mathcal{X}}_n)
=F(\boldsymbol{\mathcal{X}}_1\boldsymbol{\mathcal{X}}_2\cdots\boldsymbol{\mathcal{X}}_n)$. The $\mathcal{B}$-valued cumulants $\kappa_{\pi}^{\mathcal{B}}:\mathcal{A}^n \rightarrow \mathcal{B}$, also $\mathcal{B}$-valued multiplicative maps, are indirectly and inductively defined by
    \begin{equation}
        F(\boldsymbol{\mathcal{X}}_1\boldsymbol{\mathcal{X}}_2\cdots\boldsymbol{\mathcal{X}}_n) = \sum\limits_{\pi \in NC(n)} \kappa_{\pi}^{\mathcal{B}}(\boldsymbol{\mathcal{X}}_1, \boldsymbol{\mathcal{X}}_2, \cdots, \boldsymbol{\mathcal{X}}_{n}).
        \label{eq:operator_valued_moments_from_cumulants}
    \end{equation}
Furthermore, the $\mathcal{B}$-valued cumulants can be obtained from the $\mathcal{B}$-valued moments by
    \begin{IEEEeqnarray}{Rl}
        &\!\!\!\!\kappa_{\pi}^{\mathcal{B}}(\boldsymbol{\mathcal{X}}_1, \boldsymbol{\mathcal{X}}_2, \cdots, \boldsymbol{\mathcal{X}}_{n})
        \nonumber \\
        &= \sum\limits_{\sigma \leq \pi, \sigma \in NC(n)} \nu_{\sigma}^{\mathcal{B}}(\boldsymbol{\mathcal{X}}_1, \boldsymbol{\mathcal{X}}_2, \cdots, \boldsymbol{\mathcal{X}}_{n})\mu(\sigma,\pi)
        \label{eq:computation_of_opeartor-valued_cumulant_for_any_partition_from_moment}
    \end{IEEEeqnarray}
where $\sigma \leq \pi$ denotes that each block of $\sigma$ is completely contained in one of the blocks of
 $\pi$, and $\mu(\sigma,\pi)$ is the M\"{o}bius function over the non-crossing partition set $NC(n)$.

Freeness over $\mathcal{B}$ can also be defined by using the $\mathcal{B}$-valued cumulants. Let $S_1, S_2$ be two subsets of $\mathcal{A}$ and ${\mathcal{A}}_i$ be the algebra generated by $S_i$ and $\mathcal{B}$ for $i = 1, 2$. Then ${\mathcal{A}}_1$ and ${\mathcal{A}}_2$ are free with amalgamation over $\mathcal{B}$ if and only if whenever
$\boldsymbol{\mathcal{X}}_1, \cdots ,\boldsymbol{\mathcal{X}}_n \in S_1 \bigcup S_2$,
    \begin{equation}
        \kappa_n^{\mathcal{B}}(\boldsymbol{\mathcal{X}}_1, \cdots ,\boldsymbol{\mathcal{X}}_n) = 0
    \end{equation}
unless either all $\boldsymbol{\mathcal{X}}_1, \cdots ,\boldsymbol{\mathcal{X}}_n \in S_1 $ or all $\boldsymbol{\mathcal{X}}_1, \cdots ,\boldsymbol{\mathcal{X}}_n \in S_2$.

Let $(\mathcal{A}, \phi)$ be a non-commutative probability space and $d$ be a positive integer. A matrix $\mathbf{A}  \in \mathbf{M}_d(\mathcal{A})$ is said to be R-cyclic if the following condition holds,
$\kappa_n^{\mathbb{C}}([\mathbf{A}]_{i_1j_1}, \cdots, [\mathbf{A}]_{i_nj_n})=0$,
for every $n \geq 1$ and every $1 \leq i_1 , j_1 , \cdots, i_n , j_n \leq d$ for which it is not true that
$j_1=i_2, \cdots, j_{n-1}=i_n, j_n=i_1$ \cite{nica2002r}.

Let the operator upper half plane $\mathbb{H}_{+}(\mathcal{B})$ be defined by $\mathbb{H}_{+}(\mathcal{B}) = \{b \in \mathcal{B}: \Im(b) \succ 0\}$. For a selfadjoint random variable $\boldsymbol{\mathcal{X}} \in \mathcal{A}$ and $b \in \mathbb{H}_{+}(\mathcal{B})$, the $\mathcal{B}$-valued Cauchy transform $\mathcal{G}_{\boldsymbol{\mathcal{X}}}^{\mathcal{B}}(b)$ is defined by
    \begin{eqnarray}
        \mathcal{G}_{\boldsymbol{\mathcal{X}}}^\mathcal{B}(b) \!\!\!\!&=&\!\!\!\! F\{(b-\boldsymbol{\mathcal{X}})^{-1}\}
         \nonumber \\
         \!\!\!\!&=&\!\!\!\! \sum\limits_{n\geq0}F\{b^{-1}(\boldsymbol{\mathcal{X}}b^{-1})^n\},\|b^{-1}\| \leq \|\boldsymbol{\mathcal{X}}\|^{-1}.
    \end{eqnarray}
Let the operator lower half plane $\mathbb{H}_{-}(\mathcal{B})$ be defined by $\mathbb{H}_{-}(\mathcal{B}) = \{b \in \mathcal{B}: \Im(b) \prec 0\}$. We have that $\mathcal{G}_{\boldsymbol{\mathcal{X}}}^\mathcal{B}(b) \in \mathbb{H}_{-}(\mathcal{B})$.
The $\mathcal{B}$-valued R-transform of $\boldsymbol{\mathcal{X}}$ is defined by
    \begin{equation}
        \mathcal{R}_{\boldsymbol{\mathcal{X}}}^\mathcal{B}(b) = \sum\limits_{n\geq0}\kappa_{n+1}^{\mathcal{B}}(\boldsymbol{\mathcal{X}}b, \cdots, \boldsymbol{\mathcal{X}}b, \boldsymbol{\mathcal{X}}b,\boldsymbol{\mathcal{X}})
        \label{eq:definition_of_R_transform}
    \end{equation}
where $b \in \mathbb{H}_{-}(\mathcal{B})$.

Let $\boldsymbol{\mathcal{X}}$ and $\boldsymbol{\mathcal{Y}}$ be two $\mathcal{B}$-valued random variables.
The $\mathcal{B}$-valued freeness relation between $\boldsymbol{\mathcal{X}}$ and $\boldsymbol{\mathcal{Y}}$ is actually a rule for calculating the mixed $\mathcal{B}$-valued moments in $\boldsymbol{\mathcal{X}}$ and $\boldsymbol{\mathcal{Y}}$ from the $\mathcal{B}$-valued moments of $\boldsymbol{\mathcal{X}}$ and the $\mathcal{B}$-valued moments of $\boldsymbol{\mathcal{Y}}$. Furthermore, if $\boldsymbol{\mathcal{X}}$ and $\boldsymbol{\mathcal{Y}}$ are free over $\mathcal{B}$, then their mixed $\mathcal{B}$-valued cumulants in $\boldsymbol{\mathcal{X}}$ and $\boldsymbol{\mathcal{Y}}$ vanish. This further implies
    \begin{equation}
        \mathcal{R}_{\boldsymbol{\mathcal{X}}+\boldsymbol{\mathcal{Y}}}^\mathcal{B}(b) = \mathcal{R}_{\boldsymbol{\mathcal{X}}}^\mathcal{B}(b) + \mathcal{R}_{\boldsymbol{\mathcal{Y}}}^\mathcal{B}(b).
        \label{eq:operator_r_transform_of_sum_of_free_varaible}
    \end{equation}
The relation between the $\mathcal{B}$-valued Cauchy transform and R-transform is given by
    \begin{equation}
        \mathcal{R}_{\boldsymbol{\mathcal{X}}}^\mathcal{B}(b) = {\mathcal{G}^\mathcal{B}_{\boldsymbol{\mathcal{X}}}}^{\langle-1\rangle}(b) - b^{-1}
        \label{eq:operator_r_transform_cauchy_transform_relation}
    \end{equation}
where ${\mathcal{G}^\mathcal{B}_{\boldsymbol{\mathcal{X}}}}^{\langle-1\rangle}: \mathbb{H}_{-}(\mathcal{B}) \rightarrow \mathbb{H}_{+}(\mathcal{B}) $ is the inverse function of $\mathcal{G}_{\boldsymbol{\mathcal{X}}}^\mathcal{B}$.
According to \eqref{eq:operator_r_transform_cauchy_transform_relation}, \eqref{eq:operator_r_transform_of_sum_of_free_varaible} becomes
    \begin{equation}
        {\mathcal{G}_{\boldsymbol{\mathcal{X}}+\boldsymbol{\mathcal{Y}}}^\mathcal{B}}^{\langle-1\rangle}(b) - b^{-1} = {\mathcal{G}_{\boldsymbol{\mathcal{X}}}^\mathcal{B}}^{\langle-1\rangle}(b) - b^{-1} + \mathcal{R}_{\boldsymbol{\mathcal{Y}}}^\mathcal{B}(b).
        \label{eq:Cauchy_transform_of_sum_of_two_varaiable_temp1}
    \end{equation}
By substituting $\mathcal{G}_{\boldsymbol{\mathcal{X}}+\boldsymbol{\mathcal{Y}}}^\mathcal{B}(b)$ for each $b$, \eqref{eq:Cauchy_transform_of_sum_of_two_varaiable_temp1} becomes
    \begin{equation}
        b = {\mathcal{G}_{\boldsymbol{\mathcal{X}}}^\mathcal{B}}^{\langle-1\rangle}
        \left(\mathcal{G}_{\boldsymbol{\mathcal{X}}+\boldsymbol{\mathcal{Y}}}^\mathcal{B}(b)\right) + \mathcal{R}_{\boldsymbol{\mathcal{Y}}}^\mathcal{B}\left(\mathcal{G}_{\boldsymbol{\mathcal{X}} +
        \boldsymbol{\mathcal{Y}}}^\mathcal{B}(b)\right)
    \end{equation}
which further leads to
\begin{equation}
        \mathcal{G}_{\boldsymbol{\mathcal{X}}+\boldsymbol{\mathcal{Y}}}^\mathcal{B}(b) = \mathcal{G}_{\boldsymbol{\mathcal{X}}}^\mathcal{B}\left(b - \mathcal{R}_{\boldsymbol{\mathcal{Y}}}^\mathcal{B}\left(\mathcal{G}_{\boldsymbol{\mathcal{X}}
        +\boldsymbol{\mathcal{Y}}}^\mathcal{B}(b)\right)\right).
        \label{eq:operator_cauchy_transform_of_sum_of_free_varaible}
    \end{equation}

A $\mathcal{B}$-valued random variable $\boldsymbol{\mathcal{X}} \in \mathcal{A}$ is called a $\mathcal{B}$-valued semicircular variable if its $\mathcal{B}$-valued R-transform is given by
    \begin{equation}
        \mathcal{R}_{\boldsymbol{\mathcal{X}}}^\mathcal{B}(b) = \kappa_{2}^\mathcal{B}(\boldsymbol{\mathcal{X}}b,\boldsymbol{\mathcal{X}}).
    \end{equation}
According to \eqref{eq:operator_valued_moments_from_cumulants} and \eqref{eq:definition_of_R_transform}, the higher order $\mathcal{B}$-valued moments of $\boldsymbol{\mathcal{X}}$ are given in terms of the second order moments by summing over the non-crossing pair partitions.

Let $\boldsymbol{\mathcal{X}}_1,\boldsymbol{\mathcal{X}}_2,\cdots, \boldsymbol{\mathcal{X}}_n$ be a family of $\mathcal{B}$-valued random variables, the maps
\begin{equation}
\eta_{ij}:\mathbf{C} \rightarrow F\{\boldsymbol{\mathcal{X}}_i\mathbf{C}\boldsymbol{\mathcal{X}}_j\} \nonumber
\end{equation}
are called the covariances of the family, where $\mathbf{C} \in \mathcal{B}$.

\subsection{Free Deterministic Equivalents}
\label{sec:Free Deterministic Equivalent}
In this subsection, we introduce the free deterministic equivalents for the case where all the matrices are square and have the same size, and the random matrices are Hermitian and composed of independent Gaussian entries with different variances.

Let $\mathbf{Y}_1,\mathbf{Y}_2,\cdots,\mathbf{Y}_t$ be a $t$-tuple of $n \times n$ Hermitian random matrices. The entries $[\mathbf{Y}_k]_{ij}$ are Gaussian random variables. For fixed $k$, the entries $[\mathbf{Y}_k]_{ij}$ on and above the diagonal are independent, and $[\mathbf{Y}_k]_{ij}=[\mathbf{Y}_k]_{ji}^*$. Moreover, the entries from different matrices are also independent. Let $\frac{1}{n}\sigma_{ij,k}^2(n)$ denote the variance of $[\mathbf{Y}_k]_{ij}$. Then, we have $\sigma_{ij,k}(n)=\sigma_{ji,k}(n)$ and
    \begin{equation}
        \mathbb{E}\{[\mathbf{Y}_k]_{ij}[\mathbf{Y}_l]_{rs}\} = \frac{1}{n}\sigma_{ij,k}(n)\sigma_{rs,l}(n)\delta_{jr}\delta_{is}\delta_{kl}
        \label{eq:variance_of_matrices_entries_for_shlyakhtenko1996random}
    \end{equation}
where $1 \leq k,l \leq t$ and $1 \leq i,j,r,s \leq n $.
Let $\mathbf{A}_1,\mathbf{A}_2,\cdots,\mathbf{A}_s$ be a family of $n \times n$ deterministic matrices
and
\begin{equation} P_c :=  P(\mathbf{A}_1,\mathbf{A}_2,\cdots,\mathbf{A}_s,{\boldsymbol{\mathbf{Y}}}_{1},{\boldsymbol{\mathbf{Y}}}_{2},\cdots,{\boldsymbol{\mathbf{Y}}}_t) \nonumber
\end{equation}
be a selfadjoint polynomial. In the following, we will give the definition of
the free deterministic equivalent of $P_c$.


Let $\mathcal{A}$ be a unital algebra, $(\mathcal{A},\phi)$ be a scalar-valued probability space and $\boldsymbol{\mathcal{Y}}_1, \boldsymbol{\mathcal{Y}}_2,\cdots, \boldsymbol{\mathcal{Y}}_t \in \mathbf{M}_n(\mathcal{A})$ be a family of selfadjoint matrices with non-commutative random variables.
The entries $[\boldsymbol{\mathcal{Y}}_{k}]_{ii}$ are centered semicircular elements, and the entries $[\boldsymbol{\mathcal{Y}}_{k}]_{ij}, i \neq j$, are centered circular elements. The variance of the entry $[\boldsymbol{\mathcal{Y}}_{k}]_{ij}$ is given by
\begin{equation}
\phi([\boldsymbol{\mathcal{Y}}_{k}]_{ij}[\boldsymbol{\mathcal{Y}}_{k}]_{ij}^*) = \mathbb{E}\{[\mathbf{Y}_{k}]_{ij}[\mathbf{Y}_{k}]_{ij}^*\}. \nonumber
\end{equation}
Moreover, the entries on and above the diagonal
of $\boldsymbol{\mathcal{Y}}_{k}$ are free, and the entries from different $\boldsymbol{\mathcal{Y}}_{k}$ are also
free. Thus, we have
\begin{equation}
\phi([\boldsymbol{\mathcal{Y}}_k]_{ij}[\boldsymbol{\mathcal{Y}}_l]_{rs}) = \mathbb{E}\{[\mathbf{Y}_k]_{ij}[\mathbf{Y}_l]_{rs}\} \nonumber
\end{equation}
where $k \ne l$, $1 \leq k,l \leq t$ and $1 \leq i,j,r,s \leq n $.

According to Definition $2.9$ of \cite{nica2002r}, $\boldsymbol{\mathcal{Y}}_1, \boldsymbol{\mathcal{Y}}_2, \cdots, \boldsymbol{\mathcal{Y}}_t$ form an R-cyclic family of  matrices.
Then, from Theorem $8.2$ of \cite{nica2002r} it follows that $\mathcal{M}_n,  \boldsymbol{\mathcal{Y}}_1, \boldsymbol{\mathcal{Y}}_2, \cdots, \boldsymbol{\mathcal{Y}}_t$ are free over $\mathcal{D}_n$. According to Theorem 7.2 of \cite{nica2002r}, we have that
\begin{IEEEeqnarray}{Rl}
&\!\!\!\!\!\!\!\!\kappa_t^{\mathcal{D}_n}(\boldsymbol{\mathcal{Y}}_{k}\mathbf{C}_1,
\cdots,\boldsymbol{\mathcal{Y}}_{k}\mathbf{C}_{t-1},\boldsymbol{\mathcal{Y}}_{k})
\nonumber \\
&=
\sum\limits_{i_1,\cdots,i_t=1}^{n}[\mathbf{C}_1]_{i_2i_2}\cdots[\mathbf{C}_{t-1}]_{i_{t}i_{t}}
\nonumber \\
&
~~~~~~~~~~~~\kappa_t^{\mathbb{C}}([\boldsymbol{\mathcal{Y}}_{k}]_{i_1i_2},[\boldsymbol{\mathcal{Y}}_{k}]_{i_2i_3},\cdots, [\boldsymbol{\mathcal{Y}}_{k}]_{i_{t}i_1})\mathbf{P}_{i_1}
\end{IEEEeqnarray}
where $\mathbf{C}_1, \cdots, \mathbf{C}_{t-1} \in \mathcal{D}_n$ and $\mathbf{P}_{i_1}$ denotes the $n \times n$ matrix containing zeros in all entries except for the $i_1$-th diagonal entry, which is $1$. Since the entries on and above the diagonal of $\boldsymbol{\mathcal{Y}}_{lk}$ are a family of free (semi)circular elements and $[\boldsymbol{\mathcal{Y}}_{k}]_{ij}=[\boldsymbol{\mathcal{Y}}_{k}]_{ji}^*$, we have
\begin{equation}
\kappa_t^{\mathbb{C}}([\boldsymbol{\mathcal{Y}}_{k}]_{i_1i_2},[\boldsymbol{\mathcal{Y}}_{k}]_{i_2i_3},\cdots , [\boldsymbol{\mathcal{Y}}_{k}]_{i_ti_1})=0 \nonumber
\end{equation}
unless $t=2$. Then, we obtain
\begin{equation}
\kappa_t^{\mathcal{D}_n}(\boldsymbol{\mathcal{Y}}_{k}\mathbf{C}_1,
\cdots,\boldsymbol{\mathcal{Y}}_{k}\mathbf{C}_{t-1},\boldsymbol{\mathcal{Y}}_{k})=\mathbf{0}_n \nonumber
\end{equation}
unless $t=2$.
Thus, $\boldsymbol{\mathcal{Y}}_{1}, \cdots, \boldsymbol{\mathcal{Y}}_{t}$ are $\mathcal{D}_n$-valued semicircular elements.

In \cite{shlyakhtenko1996random}, Shlyakhtenko has proved that $\mathbf{Y}_1, \mathbf{Y}_2, \cdots, \mathbf{Y}_t$ are asymptotically free over $L^{\infty}[0, 1]$, and the asymptotic $L^{\infty}[0, 1]$-valued joint distribution of $\mathbf{Y}_1, \mathbf{Y}_2, \cdots, \mathbf{Y}_t$ and that of $\boldsymbol{\mathcal{Y}}_1, \boldsymbol{\mathcal{Y}}_2,\cdots, \boldsymbol{\mathcal{Y}}_t$ are the same. However, the proof of \cite{shlyakhtenko1996random} is based on operator algebra and might be hard to understand. Thus, we present Theorem \ref{th:diagonal_valued_free_results} in the following and prove it ourselves.
\begin{assumption}
\label{assump:variance_bounded}
The variances $\sigma_{ij,k}(n)$ are uniformly bounded in $n$.
\end{assumption}
Let $\psi_{k}[n]:\mathcal{D}_n \rightarrow \mathcal{D}_n$ be defined by
$\psi_{k}[n](\mathbf{\Delta}_n)=\mathbb{E}_{\mathcal{D}_n}\{\mathbf{Y}_k\mathbf{\Delta}_n\mathbf{Y}_k\}$,
where $\mathbf{\Delta}_n  \in \mathcal{D}_n$.
\begin{assumption}
There exist maps $\psi_{k}: L^{\infty}[0, 1] \rightarrow L^{\infty}[0, 1]$ such that whenever $i_n(\mathbf{\Delta}_n) \rightarrow d \in L^{\infty}[0, 1]$ in norm, then also $\lim_{n\rightarrow \infty}\psi_{k}[n](\mathbf{\Delta}_n) = \psi_{k}(d)$.
\label{assump:variance_operator_valued_limit}
\end{assumption}
\begin{theorem}
Let $m$ be a positive integer. Assume that Assumption \ref{assump:variance_bounded} holds. Then we have that
    \begin{IEEEeqnarray}{Rl}
        &\!\!\!\!\!\!\!\!\lim\limits_{n \rightarrow \infty} i_n  (\mathbb{E}_{\mathcal{D}_n}\{\mathbf{Y}_{p_1}\mathbf{C}_{1}\cdots\mathbf{Y}_{p_{m-1}} \mathbf{C}_{m-1}\mathbf{Y}_{p_m}\}
        \nonumber \\
        &- E_{\mathcal{D}_n}\{\boldsymbol{\mathcal{Y}}_{p_1}\mathbf{C}_{1}\cdots\boldsymbol{\mathcal{Y}}_{p_{m-1}} \mathbf{C}_{m-1}\boldsymbol{\mathcal{Y}}_{p_m}\})=0_{L^{\infty}[0, 1]}
    \end{IEEEeqnarray}
where $1 \leq p_1,\cdots, p_m \leq t$
and $\mathbf{C}_1,\cdots,\mathbf{C}_{m-1}$ is a family of $n \times n$ deterministic diagonal matrices with uniformly bounded entries.
Furthermore, if Assumption \ref{assump:variance_operator_valued_limit} holds, then
$\mathbf{Y}_1, \mathbf{Y}_2, \cdots, \mathbf{Y}_t$ are asymptotically free over $L^{\infty}[0, 1]$.
\label{th:diagonal_valued_free_results}
\end{theorem}

\begin{IEEEproof}
In \cite{nica2006lectures}, a proof of asymptotic freeness between Gaussian random
matrices is presented. Extending the proof therein, we obtain the following results.

We first prove the special case when $p_1=p_2=\cdots=p_m=k$, \textit{i.e.},
    \begin{IEEEeqnarray}{Rl}
        &\!\!\!\!\!\!\!\!\!\!\!\!\lim\limits_{n \rightarrow \infty} i_n( \mathbb{E}_{\mathcal{D}_n}\{ \mathbf{Y}_k\mathbf{C}_{1}\cdots\mathbf{Y}_k\mathbf{C}_{m-1}\mathbf{Y}_k\}
        \nonumber \\
        &-  E_{\mathcal{D}_n}\{ \boldsymbol{\mathcal{Y}}_k\mathbf{C}_{1}\cdots\boldsymbol{\mathcal{Y}}_k\mathbf{C}_{m-1} \boldsymbol{\mathcal{Y}}_k\})= 0_{L^{\infty}[0, 1]}.
        \label{eq:limit_moments_of_random_matrix_and_deterministic_diagonal_matrix_equal_to_free_deterministic_equivalent}
    \end{IEEEeqnarray}
The ${\mathcal{D}_n}$-valued moment $\mathbb{E}_{\mathcal{D}_n}\{\mathbf{Y}_k\mathbf{C}_{1}\cdots\mathbf{Y}_k \mathbf{C}_{m-1}\mathbf{Y}_k\}$ is given by
    \begin{IEEEeqnarray}{Rl}
        &\!\!\!\!\!\!\!\!\!\mathbb{E}_{\mathcal{D}_n}\{\mathbf{Y}_k\mathbf{C}_{1}\cdots\mathbf{Y}_k \mathbf{C}_{m-1}\mathbf{Y}_k\} \nonumber \\
        &=\sum\limits_{i_1,\cdots,i_m=1}^n \mathbb{E}\{[\mathbf{Y}_k]_{i_1i_2}[\mathbf{C}_{1}]_{i_2i_2}\cdots[\mathbf{Y}_k]_{i_{m-1}i_{m}}
        \nonumber \\
        &~~~~~~~~~~~~~~~~[\mathbf{C}_{m-1}]_{i_mi_m}[\mathbf{Y}_k]_{i_mi_1}\} \mathbf{P}_{i_1}  \nonumber \\
        &=\sum\limits_{i_1,\cdots,i_m=1}^n \mathbb{E}\{[\mathbf{Y}_k]_{i_1i_2}\cdots [\mathbf{Y}_k]_{i_{m-1}i_{m}}[\mathbf{Y}_k]_{i_mi_1}\}\nonumber \\
        &~~~~~~~~~~~~~~~~[\mathbf{C}_{1}]_{i_2i_2}\cdots [\mathbf{C}_{m-1}]_{i_mi_m}  \mathbf{P}_{i_1}.
        \label{eq:moments_of_matrix_yk}
    \end{IEEEeqnarray}
According to the Wick formula (Theorem 22.3 of \cite{nica2006lectures}), we have that
    \begin{IEEEeqnarray}{Rl}
        &\!\!\!\!\!\!\!\!\mathbb{E}\{[\mathbf{Y}_k]_{i_1i_2}\cdots [\mathbf{Y}_k]_{i_{m-1}i_m}[\mathbf{Y}_k]_{i_mi_1}\}
        \nonumber \\
        &=\sum\limits_{\pi \in \mathcal{P}_2(m)}
        \prod \limits_{(r,s) \in \pi}\mathbb{E}\{[\mathbf{Y}_k]_{i_ri_{\gamma(r)}}[\mathbf{Y}_k]_{i_si_{\gamma(s)}}\}
    \end{IEEEeqnarray}
where $\mathcal{P}_2(m)$ denotes the set of pair partitions of $S(m)$, and $\gamma$ is the cyclic permutation of $S(m)$ defined by $\gamma(i)=i+1 \mod m$.
Then, \eqref{eq:moments_of_matrix_yk} can be rewritten as
    \begin{IEEEeqnarray}{Rl}
        &\!\!\mathbb{E}_{\mathcal{D}_n}\{\mathbf{Y}_k\mathbf{C}_{1}\cdots\mathbf{Y}_k \mathbf{C}_{m-1}\mathbf{Y}_k\} \nonumber \\
        &=\!\!\sum\limits_{i_1,\cdots,i_m=1}^n \sum\limits_{\pi \in \mathcal{P}_2(m)}
        \!\!\left(\prod \limits_{(r,s) \in \pi}\!\!\mathbb{E}\{[\mathbf{Y}_k]_{i_ri_{\gamma(r)}}[\mathbf{Y}_k]_{i_si_{\gamma(s)}}\}\right)
        \nonumber \\
        &~~~~~~~~~~~~~~~~~~~~~~~~[\mathbf{C}_{1}]_{i_2i_2} \cdots [\mathbf{C}_{m-1}]_{i_m i_m}  \mathbf{P}_{i_1} \nonumber \\
        &=\!\!\sum\limits_{\pi \in NC_2(m)} \sum\limits_{i_1,\cdots,i_m=1}^n
        \!\!\left(\prod \limits_{(r,s) \in \pi}\!\!\mathbb{E}\{[\mathbf{Y}_k]_{i_ri_{\gamma(r)}}[\mathbf{Y}_k]_{i_si_{\gamma(s)}}\}\right)
        \nonumber \\
        &~~~~~~~~~~~~~~~~~~~~~~~~[\mathbf{C}_{1}]_{i_2i_2} \cdots [\mathbf{C}_{m-1}]_{i_m i_m}  \mathbf{P}_{i_1} \nonumber \\
        &~~~~+\!\!\!\!\sum\limits_{\substack{\pi \in \mathcal{P}_2(m)\\ \pi \notin NC_2(m)}}\!\! \sum\limits_{i_1,\cdots,i_m=1}^n
        \!\!\left(\prod \limits_{(r,s) \in \pi}\!\!\mathbb{E}\{[\mathbf{Y}_k]_{i_ri_{\gamma(r)}}[\mathbf{Y}_k]_{i_si_{\gamma(s)}}\}\right)
        \nonumber \\
        &~~~~~~~~~~~~~~~~~~~~~~~~~~~~[\mathbf{C}_{1}]_{i_2i_2} \cdots [\mathbf{C}_{m-1}]_{i_m i_m}  \mathbf{P}_{i_1}
        \label{eq:mix_moments_of_matrix_w_and_deterministic_diagonal_matrix}
    \end{IEEEeqnarray}
where $NC_2(m) \subset \mathcal{P}_2(m)$ denotes the set of non-crossing pair partitions of $S(m)$.
Meanwhile, the ${\mathcal{D}_n}$-valued moment $E_{\mathcal{D}_n}\{\boldsymbol{\mathcal{Y}}_k\mathbf{C}_{1}\cdots \boldsymbol{\mathcal{Y}}_k\mathbf{C}_{m-1}\boldsymbol{\mathcal{Y}}_k\}$ is given by
     \begin{IEEEeqnarray}{Rl}
        & \!\!\!\!\!\!\!\!\!\!\!\!E_{\mathcal{D}_n}\{\boldsymbol{\mathcal{Y}}_k\mathbf{C}_{1}\cdots \boldsymbol{\mathcal{Y}}_k\mathbf{C}_{m-1}\boldsymbol{\mathcal{Y}}_k\} \nonumber \\
        &=\sum\limits_{i_1,\cdots,i_m=1}^n \phi([\boldsymbol{\mathcal{Y}}_k]_{i_1i_2}[\mathbf{C}_{1}]_{i_2i_2}\cdots [\boldsymbol{\mathcal{Y}}_k]_{i_{m-1}i_{m}}
        \nonumber \\
        &~~~~~~~~~~~~~~~~[\mathbf{C}_{m-1}]_{i_mi_m}[\boldsymbol{\mathcal{Y}}_k]_{i_mi_1}) \mathbf{P}_{i_1}
        \nonumber \\
        &=\sum\limits_{i_1,\cdots,i_m=1}^n \phi([\boldsymbol{\mathcal{Y}}_k]_{i_1i_2}\cdots [\boldsymbol{\mathcal{Y}}_k]_{i_{m-1}i_{m}}[\boldsymbol{\mathcal{Y}}_k]_{i_mi_1})
        \nonumber \\
        &~~~~~~~~~~~~~~~~[\mathbf{C}_{1}]_{i_2i_2}\cdots[\mathbf{C}_{m-1}]_{i_mi_m} \mathbf{P}_{i_1}.
        \label{eq:moments_of_free_matrix_yk}
    \end{IEEEeqnarray}
The entries of $\boldsymbol{\mathcal{Y}}_k$ are a family of semicircular and circular elements.
From (8.8) and (8.9) in \cite{nica2006lectures}, we obtain
    \begin{IEEEeqnarray}{Rl}
        &\!\!\!\!\!\!\!\!\phi([\boldsymbol{\mathcal{Y}}_k]_{i_1i_2}\cdots [\boldsymbol{\mathcal{Y}}_k]_{i_{m-1}i_{m}}[\boldsymbol{\mathcal{Y}}_k]_{i_mi_1})
        \nonumber \\
        &= \sum\limits_{\pi \in NC_2(m)}\kappa_{\pi}^{\mathbb{C}}([\boldsymbol{\mathcal{Y}}_k]_{i_1i_2},\cdots, [\boldsymbol{\mathcal{Y}}_k]_{i_{m-1}i_{m}},[\boldsymbol{\mathcal{Y}}_k]_{i_mi_1})
        \nonumber \\
        &=
        \sum\limits_{\pi \in NC_2(m)}
        \prod \limits_{(r,s) \in \pi}\phi([\boldsymbol{\mathcal{Y}}_k]_{i_ri_{\gamma(r)}}[\boldsymbol{\mathcal{Y}}_k]_{i_si_{\gamma(s)}}).
    \end{IEEEeqnarray}
Then, $\eqref{eq:moments_of_free_matrix_yk}$ can be rewritten as
    \begin{eqnarray}
        & &\!\!\!\!\!\!\!\!\!\!\!\!\!\!\!\!E_{\mathcal{D}_n}\{\boldsymbol{\mathcal{Y}}_k\mathbf{C}_{1} \cdots\boldsymbol{\mathcal{Y}}_k\mathbf{C}_{m-1}\boldsymbol{\mathcal{Y}}_k\} \nonumber \\
        &=&\!\!\!\!\!\!\!\!\sum\limits_{\pi \in NC_2(m)} \sum\limits_{i_1,\cdots,i_m=1}^n
        \left(\prod \limits_{(r,s) \in \pi}
        \phi([\boldsymbol{\mathcal{Y}}_k]_{i_ri_{\gamma(r)}}[\boldsymbol{\mathcal{Y}}_k]_{i_si_{\gamma(s)}})\right)
        \nonumber \\
        &&~~~~~~~~~~~~~~~~~~~~[\mathbf{C}_{1}]_{i_2i_2}\cdots [\mathbf{C}_{m-1}]_{i_mi_m} \mathbf{P}_{i_1}.
        \label{eq:mix_moments_of_free_matrix_w_and_deterministic_diagonal_matrix}
    \end{eqnarray}
If $m$ is odd, then $\mathcal{P}_2(m)$ and $NC_2(m)$ are empty sets. Thus, we obtain that
both $\mathbb{E}_{\mathcal{D}_n}\{\mathbf{Y}_k\mathbf{C}_{1}\cdots\mathbf{Y}_k\mathbf{C}_{m-1}\mathbf{Y}_k\}$ and $E_{\mathcal{D}_n}\{ \boldsymbol{\mathcal{Y}}_k \mathbf{C}_{1} \cdots \boldsymbol{\mathcal{Y}}_k \mathbf{C}_{m-1} \boldsymbol{\mathcal{Y}}_k\}$
are equal to zero matrices for odd $m$. Thus, we assume that $m$ is even for the remainder of the proof.

According to $\phi([\boldsymbol{\mathcal{Y}}_k]_{i_rj_r}[\boldsymbol{\mathcal{Y}}_k]_{i_sj_s}) = \mathbb{E}\{[\mathbf{Y}_k]_{i_rj_r}[\mathbf{Y}_k]_{i_sj_s}\}$, \eqref{eq:mix_moments_of_matrix_w_and_deterministic_diagonal_matrix} and
\eqref{eq:mix_moments_of_free_matrix_w_and_deterministic_diagonal_matrix},
\eqref{eq:limit_moments_of_random_matrix_and_deterministic_diagonal_matrix_equal_to_free_deterministic_equivalent} is equivalent to that
    \begin{IEEEeqnarray}{Rl}
        &i_n (\sum\limits_{\substack{\pi \in \mathcal{P}_2(m)\\ \pi \notin NC_2(m)}} \sum\limits_{i_1,\cdots,i_m=1}^n
        (\prod \limits_{(r,s) \in \pi}\mathbb{E}\{[\mathbf{Y}_k]_{i_ri_{\gamma(r)}}[\mathbf{Y}_k]_{i_si_{\gamma(s)}}\})
        \nonumber \\
        &~~~~~~~~~~~~~~~~~~~~~~~~~~~~~~~~~~[\mathbf{C}_{1}]_{i_2i_2}\cdots [\mathbf{C}_{m-1}]_{i_m i_m}  \mathbf{P}_{i_1})  \nonumber
    \end{IEEEeqnarray}
vanishes as $n \rightarrow \infty$.
It is convenient to identify a pair partition $\pi$ with a special permutation
by declaring the blocks of $\pi$ to be cycles \cite{nica2006lectures}. Then,
$(r, s) \in \pi$ means $\pi(r) = s$ and $\pi(s) = r$.
Applying \eqref{eq:variance_of_matrices_entries_for_shlyakhtenko1996random}, we obtain equation \eqref{eq:crossing_partition_of_moments_for_matrices_free_over_diagonal} at the top
of the following page,
\begin{figure*}[!t]
\normalsize
\setcounter{tempequationcounter}{\value{equation}}
\begin{IEEEeqnarray}{Rl}
        &\!\!\!\!\!\!\!\!\sum\limits_{\substack{\pi \in \mathcal{P}_2(m)\\ \pi \notin NC_2(m)}} \!\!\sum\limits_{i_1,\cdots,i_m=1}^n\!\!
        \left(\prod \limits_{(r,s) \in \pi}\mathbb{E}\{[\mathbf{Y}_k]_{i_ri_{\gamma(r)}}[\mathbf{Y}_k]_{i_si_{\gamma(s)}}\}\right)
        [\mathbf{C}_{1}]_{i_2i_2}\cdots [\mathbf{C}_{m-1}]_{i_m i_m}  \mathbf{P}_{i_1}
        \nonumber \\
        &= n^{-\frac{m}{2}}\!\!\!\! \sum\limits_{\substack{\pi \in \mathcal{P}_2(m)\\ \pi \notin NC_2(m)}} \!\!\sum\limits_{i_1,\cdots,i_m=1}^n\!\!
        \left(\prod \limits_{(r,s) \in \pi}\sigma_{i_ri_{\gamma(r)},k}(n)\sigma_{i_{s}i_{\gamma(s)},k}(n)
        \delta_{i_ri_{\gamma(s)}}\delta_{i_{s} i_{\gamma(r)}}\right)
        [\mathbf{C}_{1}]_{i_2i_2}\cdots [\mathbf{C}_{m-1}]_{i_m i_m}  \mathbf{P}_{i_1}
        \nonumber \\
        &= n^{-\frac{m}{2}}\!\!\!\! \sum\limits_{\substack{\pi \in \mathcal{P}_2(m)\\ \pi \notin NC_2(m)}} \!\!\sum\limits_{i_1,\cdots,i_m=1}^n\!\!
        \left(\prod \limits_{r=1}^m \sigma_{i_ri_{\gamma(r)},k}(n)\delta_{i_ri_{\gamma\pi(r)}}\right)
        [\mathbf{C}_{1}]_{i_2i_2}\cdots [\mathbf{C}_{m-1}]_{i_m i_m}  \mathbf{P}_{i_1}
        \nonumber \\
        &= n^{-\frac{m}{2}}\!\!\!\! \sum\limits_{\substack{\pi \in \mathcal{P}_2(m)\\ \pi \notin NC_2(m)}} \!\!\sum\limits_{i_1,\cdots,i_m=1}^n \!\!\left(\prod \limits_{r=1}^m\delta_{i_ri_{\gamma\pi(r)}}\right)
        \left(\prod \limits_{r=1}^m \sigma_{i_ri_{\gamma(r)},k}(n)\right)[\mathbf{C}_{1}]_{i_2i_2}\cdots [\mathbf{C}_{m-1}]_{i_m i_m}  \mathbf{P}_{i_1}
        \label{eq:crossing_partition_of_moments_for_matrices_free_over_diagonal}
    \end{IEEEeqnarray}
\addtocounter{tempequationcounter}{1}
\setcounter{equation}{\value{tempequationcounter}}
\hrulefill
\end{figure*}
where $\gamma\pi$ denotes the product of the two permutations $\gamma$ and $\pi$, and is defined as their composition as functions, \textit{i.e.}, $\gamma\pi(r)$ denotes $\gamma(\pi(r))$.
Applying the triangle inequality, we then obtain
    \begin{eqnarray}
        &&\!\!\!\!\!\!\!\!\left| n^{-\frac{m}{2}} \!\!\!\! \sum\limits_{i_2,\cdots,i_m=1}^n \left(\prod \limits_{r=1}^m\delta_{i_ri_{\gamma\pi(r)}}\right)
        \left(\prod \limits_{r=1}^m \sigma_{i_ri_{\gamma(r)},k}(n)\right) \right.
        \nonumber \\
        &&~~~~~~~~~~~~~~~~~\left.
        \vphantom{\left| n^{-\frac{m}{2}}\sum\limits_{i_2,\cdots,i_m=1}^n \left(\prod \limits_{r=1}^m\delta_{i_ri_{\gamma\pi(r)}}\right)
        \left(\prod \limits_{r=1}^m \sigma_{i_ri_{\gamma(r)},k}(n)\right) \right.}
        [\mathbf{C}_{1}]_{i_2i_2}\cdots [\mathbf{C}_{m-1}]_{i_m i_m}\right|
        \nonumber \\
        && \leq  n^{-\frac{m}{2}}\sum\limits_{i_2,\cdots,i_m=1}^n \left(\prod \limits_{r=1}^m\delta_{i_ri_{\gamma\pi(r)}}\right)
        \left(\prod \limits_{r=1}^m \sigma_{i_ri_{\gamma(r)},k}(n)\right)
        \nonumber \\
        &&~~~~~~~~~~~~~~~~\left|[\mathbf{C}_{1}]_{i_2i_2}\cdots [\mathbf{C}_{m-1}]_{i_m i_m}\right|
   \end{eqnarray}
where $i_1$ is fixed.
Since the entries of $\mathbf{C}_1,\cdots,\mathbf{C}_{m-1}$ and $\sigma_{ij,k}(n)$ are uniformly bounded in $n$,
there must exists a positive real number $c_0$ such that
    \begin{eqnarray}
        &&\!\!\!\!\!\!\!\!\!\!\!\!\left| n^{-\frac{m}{2}}\sum\limits_{i_2,\cdots,i_m=1}^n \left(\prod \limits_{r=1}^m\delta_{i_ri_{\gamma\pi(r)}}\right)
        \left(\prod \limits_{r=1}^m \sigma_{i_ri_{\gamma(r)},k}(n)\right)\right.
        \nonumber \\
        &&~~~~~~~~~~~~~~~~\left.[\mathbf{C}_{1}]_{i_2i_2}\cdots [\mathbf{C}_{m-1}]_{i_m i_m}\right|
        \nonumber \\
        &&\leq c_0  n^{-\frac{m}{2}} \sum\limits_{i_2,\cdots,i_m=1}^n \left(\prod \limits_{r=1}^m\delta_{i_ri_{\gamma\pi(r)}}\right).
                \label{eq:diagonal_matrix_valued_free_inequality}
    \end{eqnarray}
In \cite{nica2006lectures} (p.365), it is shown that
\begin{eqnarray}
\sum\limits_{i_1,i_2,\cdots,i_m=1}^n \left(\prod \limits_{r=1}^m\delta_{i_ri_{\gamma\pi(r)}}\right)
=n^{\#(\gamma\pi)}
\label{eq:combinatorical_results_of_lectures_of_free_probability}
\end{eqnarray}
where $\#(\gamma\pi)$ is the number of cycles in the permutation $\gamma\pi$.
The interpretation of \eqref{eq:combinatorical_results_of_lectures_of_free_probability} is as follows: For each cycle of $\gamma\pi$, one can choose one of
the numbers $1, \cdots ,n$ for the constant value of $i_r$ on this orbit, and all
these choices are independent from each other.
Following the same interpretation, we have that
\begin{eqnarray}
 \sum\limits_{i_2,\cdots,i_m=1}^n \left(\prod \limits_{r=1}^m\delta_{i_ri_{\gamma\pi(r)}}\right)
=n^{\#(\gamma\pi)-1}
\label{eq:combinatorical_results_of_lectures_of_free_probability_variation}
\end{eqnarray}
when $i_r$ on the orbit of one cycle of $\gamma\pi$ is fixed on $i_1$.
If $\pi \in \mathcal{P}_2(m)$, we have $\#(\gamma\pi) - 1 - \frac{m}{2} = -2g$ as stated below Theorem 22.12
of \cite{nica2006lectures}, where $g \geq 0$ is called genus in the geometric language of genus expansion. The result comes from Proposition 4.2 of \cite{zvonkin1997matrix}. If $\pi \in NC_2(m)$, then $g=0$ as stated in Exercise 22.14 of \cite{nica2006lectures}. Furthermore,  for $\pi \in \mathcal{P}_2(m)$ and $\pi \notin NC_2(m)$, we have $\#(\gamma\pi) - 1 - \frac{m}{2} \leq -2$. Thus, the RHS of the inequality in \eqref{eq:diagonal_matrix_valued_free_inequality} is of order $n^{-2}$, and the left-hand side (LHS) of the inequality in \eqref{eq:diagonal_matrix_valued_free_inequality} vanishes as $n \rightarrow \infty$. Furthermore, \eqref{eq:crossing_partition_of_moments_for_matrices_free_over_diagonal} also vanishes and we have proven \eqref{eq:limit_moments_of_random_matrix_and_deterministic_diagonal_matrix_equal_to_free_deterministic_equivalent}.

Then, we prove the general case that
\begin{IEEEeqnarray}{Rl}
&\!\!\!\!\!\!\!\!\!\!\lim\limits_{n \rightarrow \infty} i_n (\mathbb{E}_{\mathcal{D}_n}\{\mathbf{Y}_{p_1}\mathbf{C}_{1}\cdots\mathbf{Y}_{p_{m-1}}\mathbf{C}_{m-1}\mathbf{Y}_{p_m}\}
\nonumber \\
&\!\!
- E_{\mathcal{D}_n}\{\boldsymbol{\mathcal{Y}}_{p_1}\mathbf{C}_{1}\cdots\boldsymbol{\mathcal{Y}}_{p_{m-1}} \mathbf{C}_{m-1}\boldsymbol{\mathcal{Y}}_{p_m}\})=0_{L^{\infty}[0, 1]}.
\label{eq:limit_moments_of_random_matrix_and_deterministic_diagonal_matrix_equal_to_free_deterministic_equivalent_general}
\end{IEEEeqnarray}
The $\mathcal{D}_n$-valued moment $\mathbb{E}_{\mathcal{D}_n}\{\mathbf{Y}_{p_1}\mathbf{C}_{1}\cdots\mathbf{Y}_{p_{m-1}}\mathbf{C}_{m-1}\mathbf{Y}_{p_m}\}$ is given by
\begin{IEEEeqnarray}{Rl}
& \mathbb{E}_{\mathcal{D}_n}\{\mathbf{Y}_{p_1}\mathbf{C}_{1}\cdots\mathbf{Y}_{p_{m-1}}\mathbf{C}_{m-1}\mathbf{Y}_{p_m}\}
\nonumber \\
&=\sum\limits_{\pi \in NC_2(m)} \sum\limits_{i_1,\cdots,i_m=1}^n
\prod \limits_{(r,s) \in \pi}\!\!\!\!\mathbb{E}\{[\mathbf{Y}_{p_r}]_{i_ri_{\gamma(r)}}[\mathbf{Y}_{p_s}]_{i_si_{\gamma(s)}}\}
\nonumber \\
&~~~~~~~~~~~~~~~~~~~~~~~~~~~~[\mathbf{C}_{1}]_{i_2i_2}\cdots [\mathbf{C}_{m-1}]_{i_m i_m}  \mathbf{P}_{i_1} \nonumber \\
&~~~~+\sum\limits_{\substack{\pi \in \mathcal{P}_2(m)\\ \pi \notin NC_2(m)}}\!\! \sum\limits_{i_1,\cdots,i_m=1}^n
\prod \limits_{(r,s) \in \pi}\!\!\!\!\mathbb{E}\{[\mathbf{Y}_{p_r}]_{i_ri_{\gamma(r)}}[\mathbf{Y}_{p_s}]_{i_si_{\gamma(s)}}\}
\nonumber \\
&~~~~~~~~~~~~~~~~~~~~~~~~~~~~[\mathbf{C}_{1}]_{i_2i_2}\cdots [\mathbf{C}_{m-1}]_{i_m i_m}  \mathbf{P}_{i_1}.
\label{eq:mix_moments_of_matrix_w_and_deterministic_diagonal_matrix_general}
\end{IEEEeqnarray}
 To prove \eqref{eq:limit_moments_of_random_matrix_and_deterministic_diagonal_matrix_equal_to_free_deterministic_equivalent_general} is equivalent to prove that the second term on the RHS of \eqref{eq:mix_moments_of_matrix_w_and_deterministic_diagonal_matrix_general} vanishes as $n \rightarrow \infty$. Then, according to \eqref{eq:variance_of_matrices_entries_for_shlyakhtenko1996random}, we have that
\begin{IEEEeqnarray}{Rl}
&\!\!\!\!\sum\limits_{\substack{\pi \in \mathcal{P}_2(m)\\ \pi \notin NC_2(m)}} \!\! \sum\limits_{i_1,\cdots,i_m=1}^n \!\!
\left(\prod \limits_{(r,s) \in \pi} \!\!\!\! \mathbb{E}\{[\mathbf{Y}_{p_r}]_{i_ri_{\gamma(r)}} [\mathbf{Y}_{p_s}]_{i_si_{\gamma(s)}}\}\right)
\nonumber \\
&~~~~~~~~~~~~~~~~~~~~~~~~~[\mathbf{C}_{1}]_{i_2i_2} \cdots [\mathbf{C}_{m-1}]_{i_m i_m}  \mathbf{P}_{i_1}
\nonumber \\
&= n^{-\frac{m}{2}} \!\!\!\!\!\!\sum\limits_{\substack{\pi \in \mathcal{P}_2(m)\\ \pi \notin NC_2(m)}} \!\!\sum\limits_{i_1,\cdots,i_m=1}^n\!\!
\left(\prod \limits_{(r,s) \in \pi} \!\!\!\! \sigma_{i_ri_{\gamma(r)},p_r}(n)\sigma_{i_{s}i_{\gamma(s)},p_s}(n) \right.
\nonumber \\
&~~~~~~\left.\vphantom{\left(\prod \limits_{(r,s) \in \pi}\sigma_{i_ri_{\gamma(r)},p_r}(n)\sigma_{i_{s}i_{\gamma(s)},p_s}(n) \right.}\delta_{i_ri_{\gamma(s)}}\delta_{i_{s} i_{\gamma(r)}}\delta_{p_{r}p_{s}}\!\!\right)\!\!
[\mathbf{C}_{1}]_{i_2i_2}\cdots [\mathbf{C}_{m-1}]_{i_m i_m}  \mathbf{P}_{i_1}.
\label{eq:crossing_partition_of_moments_for_matrices_free_over_diagonal_general}
\end{IEEEeqnarray}
The above equation is similar to \eqref{eq:crossing_partition_of_moments_for_matrices_free_over_diagonal}, the only difference is the extra factor $\delta_{p_rp_s}$, which just indicates that we have an extra condition
on the partitions $\pi$. A similar situation has been given in the proof of Proposition $22.22$ of \cite{nica2006lectures}.
Let $\mathcal{P}_2^{(p)}(m)$  and $NC_2^{(p)}(m)$ be defined by
\begin{equation}
\mathcal{P}_2^{(p)}(m) = \{\pi \in \mathcal{P}_2(m):p_r = p_{{\pi(r)}} ~\forall r = 1, \cdots ,m\} \nonumber
\end{equation}
and
\begin{equation}
NC_2^{(p)}(m) = \{\pi \in NC_2(m):p_r = p_{\pi(r)} ~\forall r = 1, \cdots ,m\}. \nonumber
\end{equation}
Then,
\eqref{eq:crossing_partition_of_moments_for_matrices_free_over_diagonal_general} becomes
\begin{IEEEeqnarray}{Rl}
&\!\!\!\!\sum\limits_{\substack{\pi \in \mathcal{P}_2(m)\\ \pi \notin NC_2(m)}} \sum\limits_{i_1,\cdots,i_m=1}^n
\left(\prod \limits_{(r,s) \in \pi}\mathbb{E}\{[\mathbf{Y}_{p_r}]_{i_ri_{\gamma(r)}}[\mathbf{Y}_{p_s}]_{i_si_{\gamma(s)}}\}\right)
\nonumber \\
&~~~~~~~~~~~~~~~~~~~~~~~~~~~~[\mathbf{C}_{1}]_{i_2i_2}\cdots [\mathbf{C}_{m-1}]_{i_m i_m}  \mathbf{P}_{i_1}
\nonumber \\
&= n^{-\frac{m}{2}} \sum\limits_{\substack{\pi \in \mathcal{P}_2^{(p)}(m)\\ \pi \notin NC_2^{(p)}(m)}} \sum\limits_{i_1,\cdots,i_m=1}^n
\left(\prod \limits_{r=1}^m \sigma_{i_ri_{\gamma(r)},p_r}(n)\delta_{i_ri_{\gamma\pi(r)}}\right)
\nonumber \\
&~~~~~~~~~~~~~~~~~~~~~~~~~~~~[\mathbf{C}_{1}]_{i_2i_2}\cdots [\mathbf{C}_{m-1}]_{i_m i_m}  \mathbf{P}_{i_1}
\label{eq:crossing_partition_of_moments_for_matrices_free_over_diagonal_general_2}.
\end{IEEEeqnarray}
For all partitions $\pi \in \mathcal{P}_2^{(p)}(m) \backslash NC_2^{(p)}(m)$, we have that $\#(\gamma\pi) - 1 - \frac{m}{2} \leq -2$.
Comparing \eqref{eq:crossing_partition_of_moments_for_matrices_free_over_diagonal} with \eqref{eq:crossing_partition_of_moments_for_matrices_free_over_diagonal_general_2},
we obtain that \eqref{eq:crossing_partition_of_moments_for_matrices_free_over_diagonal_general_2} vanishes as $n \rightarrow \infty$ and furthermore
\eqref{eq:limit_moments_of_random_matrix_and_deterministic_diagonal_matrix_equal_to_free_deterministic_equivalent_general}
 holds.

 Since $\boldsymbol{\mathcal{Y}}_1, \boldsymbol{\mathcal{Y}}_2, \cdots, \boldsymbol{\mathcal{Y}}_t$ are $\mathcal{D}_n$-valued semicircular elements and also free over $\mathcal{D}_n$, their asymptotic $L^{\infty}[0, 1]$-valued joint distribution is only determined by $\psi_{k}, 1 \leq k \leq t$. Thus, the asymptotic $L^{\infty}[0, 1]$-valued joint distribution of $\boldsymbol{\mathcal{Y}}_1, \boldsymbol{\mathcal{Y}}_2, \cdots, \boldsymbol{\mathcal{Y}}_t$ exists.
Furthermore, the asymptotic $L^{\infty}[0, 1]$-valued joint moments
\begin{equation}
\lim\limits_{n \rightarrow \infty} i_n ( \mathbb{E}_{\mathcal{D}_n}\{\mathbf{Y}_{p_1}\mathbf{C}_{1}\cdots\mathbf{Y}_{p_{m-1}}\mathbf{C}_{m-1}\mathbf{Y}_{p_m}\}) \nonumber
\end{equation}
include all the information about the asymptotic $L^{\infty}[0, 1]$-valued joint distribution of $\mathbf{Y}_1, \mathbf{Y}_2, \cdots, \mathbf{Y}_t$. Thus, we obtain from \eqref{eq:limit_moments_of_random_matrix_and_deterministic_diagonal_matrix_equal_to_free_deterministic_equivalent_general} that the asymptotic $L^{\infty}[0, 1]$-valued joint distributions of $\mathbf{Y}_1, \mathbf{Y}_2, \cdots, \mathbf{Y}_t$ and $\boldsymbol{\mathcal{Y}}_1, \boldsymbol{\mathcal{Y}}_2,\cdots, \boldsymbol{\mathcal{Y}}_t$ are the same. Finally, we have that $\mathbf{Y}_1, \mathbf{Y}_2, \cdots, \mathbf{Y}_t$ are asymptotically free over $L^{\infty}[0, 1]$.
\end{IEEEproof}

The asymptotic $L^{\infty}[0,1]$-valued distribution of the polynomial $P({\boldsymbol{\mathcal{Y}}}_{1},{\boldsymbol{\mathcal{Y}}}_{2},\cdots,{\boldsymbol{\mathcal{Y}}}_{t})$ is the same as the expected asymptotic $L^{\infty}[0,1]$-valued distribution of $P(\mathbf{Y}_1, \mathbf{Y}_2, \cdots, \mathbf{Y}_t)$ in the sense that
    \begin{IEEEeqnarray}{Rl}
        &\!\!\!\!\lim_{n \rightarrow \infty} i_n(\mathbb{E}_{\mathcal{D}_n}\{(P(\mathbf{Y}_1, \mathbf{Y}_2, \cdots, \mathbf{Y}_t))^k\}
        \nonumber \\
        &~~~~- E_{\mathcal{D}_n}\{(P({\boldsymbol{\mathcal{Y}}}_{1}, {\boldsymbol{\mathcal{Y}}}_{2}, \cdots, {\boldsymbol{\mathcal{Y}}}_{t}))^k\}) = 0_{L^{\infty}[0, 1]}.
    \end{IEEEeqnarray}

When the $n \times n$ deterministic matrices $\mathbf{A}_1,\mathbf{A}_2,\cdots,\mathbf{A}_s$ are also considered, we will present Theorem \ref{th:determinstic_matrices_gaussian_matrices_asymptotic_operator_valued_free} in the following subsection to show the
asymptotic $L^{\infty}[0,1]$-valued freeness of
\begin{equation}
\{\mathbf{A}_1,\mathbf{A}_2,\cdots,\mathbf{A}_s\}, \mathbf{Y}_1,\mathbf{Y}_2,\cdots,\mathbf{Y}_t.  \nonumber
\end{equation}
Furthermore, Theorem \ref{th:determinstic_matrices_gaussian_matrices_asymptotic_operator_valued_free} implies that
the asymptotic $L^{\infty}[0,1]$-valued distribution of
\begin{eqnarray}
P_f := P(\mathbf{A}_1,\mathbf{A}_2,\cdots,\mathbf{A}_s,{\boldsymbol{\mathcal{Y}}}_{1},{\boldsymbol{\mathcal{Y}}}_{2},\cdots,{\boldsymbol{\mathcal{Y}}}_t) \nonumber
\end{eqnarray}
and the expected asymptotic $L^{\infty}[0,1]$-valued distribution of $P_c$
are the same.
The polynomial $P_f$ is called the free deterministic equivalent of $P_c$.

For finite dimensional random matrices, the difference between the $\mathcal{D}_n$-valued distribution of $P_f$ and $P_c$ is given by the deviation from $\mathcal{D}_n$-valued freeness of
\begin{equation}
\{\mathbf{A}_1,\mathbf{A}_2,\cdots,\mathbf{A}_s\}, \mathbf{Y}_1,\mathbf{Y}_2,\cdots,\mathbf{Y}_t  \nonumber
\end{equation}
and the deviation of the expected $\mathcal{D}_n$-valued distribution of $\mathbf{Y}_1,\mathbf{Y}_2,\cdots,\mathbf{Y}_t$ from being the same as the $\mathcal{D}_n$-valued distribution of ${\boldsymbol{\mathcal{Y}}}_{1}, {\boldsymbol{\mathcal{Y}}}_{2}, \cdots, {\boldsymbol{\mathcal{Y}}}_t$. For large dimensional matrices, these deviations become smaller and the $\mathcal{D}_n$-valued distribution of $P_f$ provides a better approximation for the expected $\mathcal{D}_n$-valued distribution of $P_c$.

\subsection{New Asymptotic $L^{\infty}[0,1]$-valued Freeness Results }
\label{New_asymptotic_freeness_results}
Reference \cite{nica2006lectures} presents a proof of asymptotic free independence between Gaussian random
matrices and deterministic matrices. We extend the proof therein and obtain the following theorem.
\begin{assumption}
The spectral norms of the deterministic matrices $\mathbf{A}_1,\mathbf{A}_2,\cdots,\mathbf{A}_s$ are uniformly bounded.
\label{assump:bounded_spectral norms}
\end{assumption}
\begin{theorem}
\label{th:determinstic_matrices_gaussian_matrices_asymptotic_operator_valued_free}
Let $\mathcal{E}_n$ denote the algebra of $n \times n$ diagonal matrices with uniformly bounded entries
and $\mathcal{F}_n$ denote the algebra generated by $\mathbf{A}_{1}, \mathbf{A}_{2}, \cdots,\mathbf{A}_{s}$ and $\mathcal{E}_n$. Let $m$ be a positive integer and $\mathbf{C}_0, \mathbf{C}_1,  \cdots, \mathbf{C}_m \in \mathcal{F}_n$ be a family of $n \times n$ deterministic matrices. Assume that Assumptions \ref{assump:variance_bounded} and \ref{assump:bounded_spectral norms} hold. Then,
\begin{IEEEeqnarray}{Rl}
&\!\!\!\!\!\!\!\!\!\!\lim\limits_{n \rightarrow \infty}i_n (\mathbb{E}_{\mathcal{D}_n}\{\mathbf{C}_{0}\mathbf{Y}_{p_1}\mathbf{C}_{1}\mathbf{Y}_{p_2}\mathbf{C}_2\cdots \mathbf{Y}_{p_m}\mathbf{C}_m\}
\nonumber \\
&- E_{\mathcal{D}_n}\{\mathbf{C}_{0}\boldsymbol{\mathcal{Y}}_{p_1}\mathbf{C}_{1}\boldsymbol{\mathcal{Y}}_{p_2}\mathbf{C}_2 \cdots\boldsymbol{\mathcal{Y}}_{p_m}\mathbf{C}_m\})=0_{L^{\infty}[0, 1]}
\end{IEEEeqnarray}
where $1 \leq p_1,\cdots, p_m \leq t$. Furthermore, if Assumption \ref{assump:variance_operator_valued_limit} also holds, then $\mathbf{Y}_1, \mathbf{Y}_2, \cdots, \mathbf{Y}_t$, $\mathcal{F}_n$ are asymptotically free over $L^{\infty}[0, 1]$.
\end{theorem}
\begin{IEEEproof}
We first prove the special case when $p_1=p_2=\cdots=p_m=k$, \textit{i.e.},
\begin{IEEEeqnarray}{Rl}
&\!\!\!\!\!\!\!\!\lim\limits_{n \rightarrow \infty} i_n( \mathbb{E}_{\mathcal{D}_n}\!\{ \mathbf{C}_{0}\mathbf{Y}_k\mathbf{C}_{1}\mathbf{Y}_k\mathbf{C}_2\cdots\mathbf{Y}_k\mathbf{C}_m\}
\nonumber \\
&~- E_{\mathcal{D}_n}\{ \mathbf{C}_{0}\boldsymbol{\mathcal{Y}}_k\mathbf{C}_{1}\boldsymbol{\mathcal{Y}}_k\mathbf{C}_2\cdots \boldsymbol{\mathcal{Y}}_k\mathbf{C}_m\})= 0_{L^{\infty}[0, 1]}.
\label{eq:limit_moments_of_random_matrix_and_deterministic_matrix_equal_to_free_deterministic_equivalent}
\end{IEEEeqnarray}
Using steps similar to those used to derive \eqref{eq:mix_moments_of_matrix_w_and_deterministic_diagonal_matrix} and \eqref{eq:mix_moments_of_free_matrix_w_and_deterministic_diagonal_matrix} in the proof of Theorem \ref{th:diagonal_valued_free_results}, we obtain
\begin{IEEEeqnarray}{Rl}
&\!\!\!\mathbb{E}_{\mathcal{D}_n}\{\mathbf{C}_{0}\mathbf{Y}_k\mathbf{C}_{1}\mathbf{Y}_k\mathbf{C}_2\cdots\mathbf{Y}_k\mathbf{C}_m\} \nonumber \\
&= \sum\limits_{\pi \in NC_2(m)} \sum\limits_{\substack{i_1,\cdots,i_m\\j_0,j_1,\cdots,j_m=1}}^n
\left(\prod \limits_{(r,s) \in \pi}\mathbb{E}\{[\mathbf{Y}_k]_{i_rj_r}[\mathbf{Y}_k]_{i_sj_s}\}\right)
\nonumber \\
&~~~~~~~~~~~~~~~~~[\mathbf{C}_{0}]_{j_0i_1}\cdots [\mathbf{C}_{m-1}]_{j_{m-1}i_m}[\mathbf{C}_m]_{j_{m}j_0} \mathbf{P}_{j_0} \nonumber \\
&~ +\sum\limits_{\substack{\pi \in \mathcal{P}_2(m)\\ \pi \notin NC_2(m)}} \sum\limits_{\substack{i_1,\cdots,i_m\\j_0,j_1,\cdots,j_m=1}}^n
\left(\prod \limits_{(r,s) \in \pi}\mathbb{E}\{[\mathbf{Y}_k]_{i_rj_r}[\mathbf{Y}_k]_{i_sj_s}\}\right)
\nonumber \\
&~~~~~~~~~~~~~~~~~[\mathbf{C}_{0}]_{j_0i_1}\cdots [\mathbf{C}_{m-1}]_{j_{m-1}i_m}[\mathbf{C}_m]_{j_{m}j_0} \mathbf{P}_{j_0}
\label{eq:mix_moments_of_matrix_w_and_deterministic_matrix_d}
\end{IEEEeqnarray}
and
\begin{IEEEeqnarray}{Rl}
&\!\!\!\!E_{\mathcal{D}_n}\{\mathbf{C}_{0}\boldsymbol{\mathcal{Y}}_k\mathbf{C}_{1}\boldsymbol{\mathcal{Y}}_k\mathbf{C}_2\cdots\boldsymbol{\mathcal{Y}}_k\mathbf{C}_m\} \nonumber \\
&= \sum\limits_{\pi \in NC_2(m)} \sum\limits_{\substack{i_1,\cdots,i_m\\j_0,j_1,\cdots,j_m=1}}^n
\left(\prod \limits_{(r,s) \in \pi}\phi([\boldsymbol{\mathcal{Y}}_k]_{i_rj_r}[\boldsymbol{\mathcal{Y}}_k]_{i_sj_s})\right)
\nonumber \\
&~~~~~~~~~~~~~~[\mathbf{C}_{0}]_{j_0i_1}\cdots [\mathbf{C}_{m-1}]_{j_{m-1}i_m}[\mathbf{C}_m]_{j_{m}j_0} \mathbf{P}_{j_0}
\label{eq:mix_moments_of_free_matrix_w_and_deterministic_matrix_d}
\end{IEEEeqnarray}
respectively. Furthermore, both
\begin{equation}
\mathbb{E}_{\mathcal{D}_n}\{\mathbf{C}_{0}\mathbf{Y}_k\mathbf{C}_{1}\mathbf{Y}_k\mathbf{C}_2\cdots\mathbf{Y}_k\mathbf{C}_m\}
\nonumber
\end{equation}
and
\begin{equation}
E_{\mathcal{D}_n}\{\mathbf{C}_{0}\boldsymbol{\mathcal{Y}}_k\mathbf{C}_{1}\boldsymbol{\mathcal{Y}}_k\mathbf{C}_2\cdots\boldsymbol{\mathcal{Y}}_k\mathbf{C}_m\}
\nonumber
\end{equation}
are equal to zero matrices for odd $m$. Thus, we also assume that $m$ is even for the remainder of the proof.

According to $\phi([\boldsymbol{\mathcal{Y}}_k]_{i_rj_r}[\boldsymbol{\mathcal{Y}}_k]_{i_sj_s}) = \mathbb{E}\{[\mathbf{Y}_k]_{i_rj_r}[\mathbf{Y}_k]_{i_sj_s}\}$, \eqref{eq:mix_moments_of_matrix_w_and_deterministic_matrix_d} and
\eqref{eq:mix_moments_of_free_matrix_w_and_deterministic_matrix_d},
\eqref{eq:limit_moments_of_random_matrix_and_deterministic_matrix_equal_to_free_deterministic_equivalent} is equivalent to that
\begin{IEEEeqnarray}{Rl}
&\!\!\!\!\!i_n (\sum\limits_{\substack{\pi \in \mathcal{P}_2(m)\\ \pi \notin NC_2(m)}} \sum\limits_{\substack{i_1,\cdots,i_m\\j_0,j_1,\cdots,j_m=1}}^n
(\prod \limits_{(r,s) \in \pi}\mathbb{E}\{[\mathbf{Y}_k]_{i_rj_r}[\mathbf{Y}_k]_{i_sj_s}\})
\nonumber \\
&~~~~~~~~~~~~~~~~~[\mathbf{C}_{0}]_{j_0i_1}\cdots [\mathbf{C}_{m-1}]_{j_{m-1}i_m}[\mathbf{C}_m]_{j_{m}j_0} \mathbf{P}_{j_0})  \nonumber
\end{IEEEeqnarray}
vanishes as $n \rightarrow \infty$.  From \eqref{eq:variance_of_matrices_entries_for_shlyakhtenko1996random}, we then obtain equation \eqref{eq:inequality_formula_of_crossing_partition_of_moments} at the top of the following page.
\begin{figure*}[!t]
\normalsize
\setcounter{tempequationcounter}{\value{equation}}
\begin{IEEEeqnarray}{Rl}
&\!\!\!\!\!\!\!\!\sum\limits_{\substack{\pi \in \mathcal{P}_2(m)\\ \pi \notin NC_2(m)}} \sum\limits_{\substack{i_1,\cdots,i_m\\j_0,j_1,\cdots,j_m=1}}^n
\left(\prod \limits_{(r,s) \in \pi}\mathbb{E}\{[\mathbf{Y}_k]_{i_rj_r}[\mathbf{Y}_k]_{i_sj_s}\}\right)
[\mathbf{C}_{0}]_{j_0i_1}\cdots [\mathbf{C}_{m-1}]_{j_{m-1}i_m}[\mathbf{C}_m]_{j_{m}j_0} \mathbf{P}_{j_0}
\nonumber \\
&= n^{-\frac{m}{2}} \sum\limits_{\substack{\pi \in \mathcal{P}_2(m)\\ \pi \notin NC_2(m)}} \sum\limits_{\substack{i_1,\cdots,i_m\\j_0,j_1,\cdots,j_m=1}}^n
\left(\prod \limits_{(r,s) \in \pi}\sigma_{i_rj_r,k}(n)\sigma_{i_sj_s,k}(n)\delta_{i_rj_s}\delta_{i_sj_r}\right)[\mathbf{C}_{0}]_{j_0i_1}\cdots[\mathbf{C}_{m-1}]_{j_{m-1}i_m}  [\mathbf{C}_m]_{j_{m}j_0}  \mathbf{P}_{j_0} \nonumber \\
&= n^{-\frac{m}{2}} \sum\limits_{\substack{\pi \in \mathcal{P}_2(m)\\ \pi \notin NC_2(m)}} \sum\limits_{\substack{i_1,\cdots,i_m\\j_0,j_1,\cdots,j_m=1}}^n
\left(\prod \limits_{r=1}^m
\sigma_{i_rj_r,k}(n)\delta_{i_rj_{{\pi(r)}}}\right)
[\mathbf{C}_{0}]_{j_0i_1}\cdots [\mathbf{C}_{m-1}]_{j_{m-1}i_m}[\mathbf{C}_m]_{j_{m}j_0}  \mathbf{P}_{j_0} \nonumber \\
&= n^{-\frac{m}{2}} \!\!\!\!\!\! \sum\limits_{\substack{\pi \in \mathcal{P}_2(m)\\ \pi \notin NC_2(m)}} \sum\limits_{j_0,j_1,\cdots,j_m=1}^n \left(\prod \limits_{r=1}^m \sigma_{j_{\pi(r)}j_r,k}(n)\right)
[\mathbf{C}_{0}]_{j_0j_{\pi\gamma(m)}}[\mathbf{C}_{1}]_{j_1j_{\pi\gamma(1)}}\cdots [\mathbf{C}_{m-1}]_{j_{m-1}j_{\pi\gamma(m-1)}}
[\mathbf{C}_m]_{j_{m}j_0}  \mathbf{P}_{j_0}
\label{eq:inequality_formula_of_crossing_partition_of_moments}
\end{IEEEeqnarray}
\addtocounter{tempequationcounter}{1}
\setcounter{equation}{\value{tempequationcounter}}
\hrulefill
\end{figure*}
Since $\mathbf{C}_0,\mathbf{C}_1,\cdots,\mathbf{C}_m$ are not diagonal matrices,  \eqref{eq:inequality_formula_of_crossing_partition_of_moments} is different from \eqref{eq:crossing_partition_of_moments_for_matrices_free_over_diagonal} in the proof of Theorem \ref{th:diagonal_valued_free_results}. Thus, the method used to prove the LHS of \eqref{eq:crossing_partition_of_moments_for_matrices_free_over_diagonal} vanishes is no longer suitable here.
In the following, we use a different method  to prove the LHS of \eqref{eq:inequality_formula_of_crossing_partition_of_moments} vanishes as $n \rightarrow \infty$.

If all $\sigma_{i_rj_r,k}(n)=1$, then \eqref{eq:inequality_formula_of_crossing_partition_of_moments} becomes
\begin{IEEEeqnarray}{Rl}
&\!\!\!\!\sum\limits_{\substack{\pi \in \mathcal{P}_2(m)\\ \pi \notin NC_2(m)}} \sum\limits_{\substack{i_1,\cdots,i_m\\j_0,j_1,\cdots,j_m=1}}^n
\left(\prod \limits_{(r,s) \in \pi}\mathbb{E}\{[\mathbf{Y}_k]_{i_rj_r}[\mathbf{Y}_k]_{i_sj_s}\}\right)
\nonumber \\
&~~~~~~~~~~~~~~~~~~~~[\mathbf{C}_{0}]_{j_0i_1}\cdots [\mathbf{C}_{m-1}]_{j_{m-1}i_m}[\mathbf{C}_m]_{j_{m}j_0} \mathbf{P}_{j_0} \nonumber \\
&~~= n^{-\frac{m}{2}} \!\!\!\!\!\! \sum\limits_{\substack{\pi \in \mathcal{P}_2(m)\\ \pi \notin NC_2(m)}} \sum\limits_{j_0,j_1,\cdots,j_m=1}^n
[\mathbf{C}_{0}]_{j_0j_{\pi\gamma(m)}}[\mathbf{C}_{1}]_{j_1j_{\pi\gamma(1)}}\cdots
\nonumber \\
&~~~~~~~~~~~~~~~~~~~~ [\mathbf{C}_{m-1}]_{j_{m-1}j_{\pi\gamma(m-1)}}
[\mathbf{C}_m]_{j_{m}j_0}  \mathbf{P}_{j_0}.
\label{eq:equality_formula_of_crossing_partition_of_moments_variances_one}
\end{IEEEeqnarray}
Let $\rho_1,\rho_2,\cdots,\rho_u$ be cycles of $\pi\gamma$ and ${\rm{tr}}_{\pi\gamma}(\mathbf{C}_{1},\cdots,\mathbf{C}_m)$ be defined by
\begin{eqnarray}
&&\!\!\!\!\!\!\!\!{\rm{tr}}_{\pi\gamma}(\mathbf{C}_{1},\cdots,\mathbf{C}_m)={\rm{tr}}_{\rho_1}(\mathbf{C}_{1},\cdots,\mathbf{C}_m){\rm{tr}}_{\rho_2}(\mathbf{C}_{1},\cdots,\mathbf{C}_m)
\nonumber \\
&&~~~~~~~~~~~~~~~~~~~~~~~~\cdots{\rm{tr}}_{\rho_u}(\mathbf{C}_{1},\cdots\mathbf{C}_m)
\label{eq:definition_of_joint_moments_of_deterministic_matrices_2}
\end{eqnarray}
where
\begin{equation}
{\rm{tr}}_{\rho_i}(\mathbf{C}_{1},\cdots,\mathbf{C}_m)=\frac{1}{n}{\rm{tr}}(\mathbf{C}_{v_1}\mathbf{C}_{v_2}\cdots\mathbf{C}_{v_a})\nonumber
\end{equation}
if $\rho_i = (v_1,v_2,\cdots,v_a)$.
Lemma 22.31 of \cite{nica2006lectures} shows that
\begin{eqnarray}
&&\!\!\!\!\!\!\!\!\sum\limits_{j_1,\cdots,j_m=1}^n [\mathbf{C}_{1}]_{j_1j_{\pi\gamma(1)}}\cdots [\mathbf{C}_{m-1}]_{j_{m-1}j_{\pi\gamma(m-1)}}[\mathbf{C}_{m}]_{j_mj_{\pi\gamma(m)}}
\nonumber \\
&&~~~~~~~~~~~~~~~~~~~~= n^{ \# (\pi\gamma)}{\rm{tr}}_{\pi\gamma}(\mathbf{C}_{1},\cdots,\mathbf{C}_m).
\label{eq:lemma_22_31_of_nica2006lectures}
\end{eqnarray}
For example, let $m=8$ and $\pi=(1,4)(3,6)(2,7)(5,8)$.
Then, we have
\begin{IEEEeqnarray}{Rl}
&\pi\gamma(1)=\pi(\gamma(1))=\pi(2)=7 \nonumber \\
&\pi\gamma(2)=\pi(\gamma(2))=\pi(3)=6 \nonumber \\
&~~~~~~~~~~~~~~\cdots \nonumber  \\
&\pi\gamma(8)=\pi(\gamma(8))=\pi(1)=4. \nonumber
\end{IEEEeqnarray}
Then, we obtain $\pi\gamma=(4, 8)(1,7,5,3)(2, 6)$, $\#(\pi\gamma)=3$ and
\begin{IEEEeqnarray}{Rl}
&\!\!\!\!\!\!\!\!\sum\limits_{j_1,\cdots,j_8=1}^n [\mathbf{C}_{1}]_{j_1j_7}[\mathbf{C}_{2}]_{j_2j_6}[\mathbf{C}_{3}]_{j_3j_1}
[\mathbf{C}_{4}]_{j_4j_8}
\IEEEnonumber \\
&~~~~~~~~~~~~~~~~~~~~~~~~[\mathbf{C}_{5}]_{j_5j_3}[\mathbf{C}_{6}]_{j_6j_2}[\mathbf{C}_{7}]_{j_7j_5}
[\mathbf{C}_{8}]_{j_8j_4}    \nonumber
\\
&=\sum\limits_{j_1,j_3,j_5,j_7=1}^n [\mathbf{C}_{1}]_{j_1j_7}[\mathbf{C}_{7}]_{j_7j_5}[\mathbf{C}_{5}]_{j_5j_3}[\mathbf{C}_{3}]_{j_3j_1}
\IEEEnonumber \\
&~~~~~~~~\sum\limits_{j_2,j_6=1}^n[\mathbf{C}_{2}]_{j_2j_6}[\mathbf{C}_{6}]_{j_6j_2}
\sum\limits_{j_4,j_8=1}^n [\mathbf{C}_{4}]_{j_4j_8}[\mathbf{C}_{8}]_{j_8j_4}
\IEEEnonumber
\\
&
=n^3\frac{1}{n}{\rm{tr}}(\mathbf{C}_{4}\mathbf{C}_8)\frac{1}{n}{\rm{tr}}(\mathbf{C}_{1}\mathbf{C}_{7}\mathbf{C}_{5}\mathbf{C}_{3})\frac{1}{n}{\rm{tr}}(\mathbf{C}_{2}\mathbf{C}_6)
\IEEEnonumber \\
&
=n^{\#(\pi\gamma)}{\rm{tr}}_{\pi\gamma}(\mathbf{C}_{1},\cdots,\mathbf{C}_8).
\label{eq:example}
\end{IEEEeqnarray}
From Remarks 23.8 and Proposition 23.11 of \cite{nica2006lectures}, we have that $\#(\pi\gamma)=\#(\gamma\pi)$.
Without loss of generality, let $\rho_1=(w_1,w_2,\cdots,w_b)$ be the cycle of $\pi\gamma$ containing $m$ and $w_b=m$. We denote by $\alpha$ the permutation $\rho_2 \cup \cdots \cup \rho_u$.
Then, we obtain a result similar  to \eqref{eq:lemma_22_31_of_nica2006lectures} that
\begin{eqnarray}
&&\!\!\!\!\!\!\!\!n^{-\frac{m}{2}}\sum\limits_{j_1,\cdots,j_m=1}^n [\mathbf{C}_{0}]_{j_0j_{\pi\gamma(m)}}[\mathbf{C}_{1}]_{j_1j_{\pi\gamma(1)}}\cdots
\nonumber \\
&&~~~~~~~~~~~~~~~~~~~~~~~~[\mathbf{C}_{m-1}]_{j_{m-1}j_{\pi\gamma(m-1)}}[\mathbf{C}_m]_{j_{m}j_0}
\nonumber \\
& &=n^{ \# (\gamma\pi-\frac{m}{2}-1)}{\rm{tr}}_{\alpha}(\mathbf{C}_{1},\cdots,\mathbf{C}_m)[\mathbf{C}_{0}\mathbf{C}_{w_1} \cdots\mathbf{C}_{w_b}]_{j_0j_0}.
\nonumber \\
\label{eq:combinatoric_results_of_sum_of_deterministic_matrices}
\end{eqnarray}
Under the assumptions on $\mathbf{C}_{0},\mathbf{C}_{1},\cdots,\mathbf{C}_m$, the limits of all
\begin{equation}
{\rm{tr}}_{\alpha}(\mathbf{C}_{1},\cdots,\mathbf{C}_m)[\mathbf{C}_{0}\mathbf{C}_{w_1} \cdots\mathbf{C}_{w_b}]_{j_0j_0}
\nonumber
\end{equation}
exist. For each crossing pair partition $\pi$, we have that $\#(\gamma\pi) - 1 - \frac{m}{2} \leq -2$. Thus, the RHS of
\eqref{eq:equality_formula_of_crossing_partition_of_moments_variances_one} is of order $n^{-2}$, and the LHS of  \eqref{eq:equality_formula_of_crossing_partition_of_moments_variances_one} vanishes as $n \rightarrow \infty$.

For general $\sigma_{i_rj_r,k}(n)$, the formula
\begin{eqnarray}
&&\!\!\!\!\!\!\!\!\!\!\!\!\!\!\!\!n^{-\frac{m}{2}}\sum\limits_{j_1,\cdots,j_m=1}^n \left(\prod \limits_{r=1}^m \sigma_{j_{\pi(r)}j_r,k}(n)\right)
[\mathbf{C}_{0}]_{j_0j_{\pi\gamma(m)}}
\nonumber \\
&&[\mathbf{C}_{1}]_{j_1j_{\pi\gamma(1)}}\cdots [\mathbf{C}_{m-1}]_{j_{m-1}j_{\pi\gamma(m-1)}}
[\mathbf{C}_m]_{j_{m}j_0}
\label{eq:general_variance_combin}
\end{eqnarray}
is still a product of elements similar to \eqref{eq:combinatoric_results_of_sum_of_deterministic_matrices} along the cycles of $\pi\gamma$.
For example, let $\pi=(1, 4)(2, 6)(3, 7)(5, 8)$, $m=8$ and $\pi\gamma=(4, 8)(1, 6, 3)(2, 7, 5)$.
Then, we obtain equation \eqref{eq:general_sigama_rules} at the top of the following page,
\begin{figure*}[!t]
\normalsize
\setcounter{tempequationcounter}{\value{equation}}
\begin{IEEEeqnarray}{Rl}
&\!\!\!\!n^{-4}\sum\limits_{j_1,\cdots,j_8=1}^n \left(\prod \limits_{r=1}^8 \sigma_{j_{\pi(r)}j_r,k}(n)\right)
[\mathbf{C}_{0}]_{j_0j_{\pi\gamma(8)}}[\mathbf{C}_{1}]_{j_1j_{\pi\gamma(1)}}\cdots [\mathbf{C}_{m-1}]_{j_{7}j_{\pi\gamma(7)}}
[\mathbf{C}_m]_{j_{8}j_0}
\nonumber \\
&=n^{-4}\sum\limits_{j_1,\cdots,j_8=1}^n\left([\mathbf{C}_{3}]_{j_3j_1
}[\mathbf{\Lambda}_{j_4}]_{j_1j_1}[\mathbf{C}_{1}]_{j_1j_6}[\mathbf{\Lambda}_{j_2}]_{j_6j_6} [\mathbf{C}_{6}]_{j_6j_3}\right)
\left([\mathbf{C}_2]_{j_2j_7}[\mathbf{\Lambda}_{j_3}]_{j_7j_7} [\mathbf{C}_{7}]_{j_7j_5}[\mathbf{\Lambda}_{j_8}]_{j_5j_5}[\mathbf{C}_{5}]_{j_5j_2}\right)
\nonumber \\
& ~~~~~~~~~~~~~~~~~~~~~~~~~~~~~~~~~
\left([\mathbf{C}_{0}]_{j_0j_4}[\mathbf{C}_{4}]_{j_4j_8}[\mathbf{C}_{8}]_{j_8j_0}\right)
\nonumber \\
&=n^{-4}\sum\limits_{j_0,j_2,j_3,j_4,j_8=1}^n\left([\mathbf{C}_{3}
\mathbf{\Lambda}_{j_4}\mathbf{C}_{1}\mathbf{\Lambda}_{j_2}\mathbf{C}_{6}]_{j_3j_3}\right)
\left([\mathbf{C}_2\mathbf{\Lambda}_{j_3}\mathbf{C}_{7}\mathbf{\Lambda}_{j_8}\mathbf{C}_{5}]_{j_2j_2}\right)
\left([\mathbf{C}_{0}]_{j_0j_4}[\mathbf{C}_{4}]_{j_4j_8}[\mathbf{C}_{8}]_{j_8j_0}\right)
\nonumber \\
&= n^{-2}\sum\limits_{j_2,j_3=1}^n \frac{1}{n^2}
[\mathbf{C}_{0}\mathbf{\Xi}_{j_2j_3}\mathbf{C}_{4}\mathbf{\Sigma}_{j_2j_3}\mathbf{C}_{8}]_{j_0j_0}
\label{eq:general_sigama_rules}
\end{IEEEeqnarray}
\addtocounter{tempequationcounter}{1}
\setcounter{equation}{\value{tempequationcounter}}
\hrulefill
\vspace*{4pt}
\end{figure*}
where
\begin{IEEEeqnarray} {Rl}
&\mathbf{\Lambda}_{j_r}={\rm{diag}}(\sigma_{1j_r,k}^2(n), \sigma_{2j_r,k}^2(n), \cdots, \sigma_{nj_r,k}^2(n))  \IEEEnonumber \\
&\mathbf{\Xi}_{j_2j_3}={\rm{diag}}(
[\mathbf{C}_{3}\mathbf{\Lambda}_{1}\mathbf{C}_{1}\mathbf{\Lambda}_{j_2}\mathbf{C}_{6}]_{j_3j_3},
[\mathbf{C}_{3}\mathbf{\Lambda}_{2}\mathbf{C}_{1}\mathbf{\Lambda}_{j_2}\mathbf{C}_{6}]_{j_3j_3}
\IEEEnonumber \\
&~~~~~~~~~~~~~~~~~~~~~~~~~~~~~~~~\cdots, [\mathbf{C}_{3}
\mathbf{\Lambda}_{n}\mathbf{C}_{1}\mathbf{\Lambda}_{j_2}\mathbf{C}_{6}]_{j_3j_3}) \IEEEnonumber \\
&\mathbf{\Sigma}_{j_2j_3}={\rm{diag}}(
[\mathbf{C}_2\mathbf{\Lambda}_{j_3}\mathbf{C}_{7}\mathbf{\Lambda}_{1}\mathbf{C}_{5}]_{j_2j_2},
[\mathbf{C}_2\mathbf{\Lambda}_{j_3}\mathbf{C}_{7}\mathbf{\Lambda}_{2}\mathbf{C}_{5}]_{j_2j_2}
\IEEEnonumber \\
&~~~~~~~~~~~~~~~~~~~~~~~~~~~~~~~~\cdots, [\mathbf{C}_2\mathbf{\Lambda}_{j_3}\mathbf{C}_{7}\mathbf{\Lambda}_{n}\mathbf{C}_{5}]_{j_2j_2}). \IEEEnonumber
\end{IEEEeqnarray}
Thus, \eqref{eq:general_variance_combin} is still of order $n^{\#(\gamma\pi) - 1 - \frac{m}{2}}$, and
the LHS of \eqref{eq:equality_formula_of_crossing_partition_of_moments_variances_one} is of order $n^{-2}$.
Furthermore, we have proven that
\eqref{eq:limit_moments_of_random_matrix_and_deterministic_matrix_equal_to_free_deterministic_equivalent} holds.


Then, we continue to prove the situation with more than one random matrix that
    \begin{IEEEeqnarray}{Rl}
       &\lim\limits_{n \rightarrow \infty} i_n (\mathbb{E}_{\mathcal{D}_n}\{\mathbf{C}_{0}\mathbf{Y}_{p_1}\mathbf{C}_{1}\mathbf{Y}_{p_2} \mathbf{C}_2\cdots \mathbf{Y}_{p_m}\mathbf{C}_m\}
         \IEEEnonumber \\
       &~~~~~~~~-\mathbb{E}_{\mathcal{D}_n}\{\mathbf{C}_{0}\boldsymbol{\mathcal{Y}}_{p_1}\mathbf{C}_{1} \boldsymbol{\mathcal{Y}}_{p_2} \mathbf{C}_2\cdots\boldsymbol{\mathcal{Y}}_{p_m}\mathbf{C}_m\})= 0_{L^{\infty}[0, 1]}. \IEEEnonumber \\
        \label{eq:limit_moments_of_random_matrix_and_deterministic_matrix_equal_to_free_deterministic_equivalent_general}
    \end{IEEEeqnarray}
The proof of \eqref{eq:limit_moments_of_random_matrix_and_deterministic_matrix_equal_to_free_deterministic_equivalent_general} is similar to that of \eqref{eq:limit_moments_of_random_matrix_and_deterministic_diagonal_matrix_equal_to_free_deterministic_equivalent_general}
in the proof of Theorem \ref{th:diagonal_valued_free_results} and omitted here for brevity.

Since $\mathcal{M}_n, \boldsymbol{\mathcal{Y}}_1, \boldsymbol{\mathcal{Y}}_2, \cdots, \boldsymbol{\mathcal{Y}}_t$ are free over $\mathcal{D}_n$ and $\mathcal{F}_n \subset \mathcal{M}_n$, we obtain that $\mathcal{F}_n, \boldsymbol{\mathcal{Y}}_1, \boldsymbol{\mathcal{Y}}_2, \cdots, \boldsymbol{\mathcal{Y}}_t$ are free over $\mathcal{D}_n$. Then,
since $\boldsymbol{\mathcal{Y}}_1, \boldsymbol{\mathcal{Y}}_2, \cdots, \boldsymbol{\mathcal{Y}}_t$ are $\mathcal{D}_n$-valued semicircular elements, we have that the asymptotic $L^{\infty}[0, 1]$-valued joint distribution of
 $\mathcal{F}_n, \boldsymbol{\mathcal{Y}}_1, \boldsymbol{\mathcal{Y}}_2, \cdots, \boldsymbol{\mathcal{Y}}_t$
 is only determined by $\psi_{k}$ and the asymptotic $L^{\infty}[0, 1]$-valued joint distribution of elements from $\mathcal{F}_n$.
Furthermore, the elements of $\mathcal{F}_n$ have uniformly bounded spectral norm. Thus,
the asymptotic $L^{\infty}[0, 1]$-valued joint distribution of
 $\mathcal{F}_n, \boldsymbol{\mathcal{Y}}_1, \boldsymbol{\mathcal{Y}}_2, \cdots, \boldsymbol{\mathcal{Y}}_t$ exists.
Then, since the asymptotic $L^{\infty}[0, 1]$-valued joint moments
\begin{equation}
\lim\limits_{n \rightarrow \infty} i_n( \mathbb{E}_{\mathcal{D}_n}\{\mathbf{C}_{0}\mathbf{Y}_{p_1}\mathbf{C}_{1}\cdots\mathbf{Y}_{p_{m-1}}\mathbf{C}_{m-1} \mathbf{Y}_{p_m}\mathbf{C}_{m}\}) \nonumber
\end{equation}
include all the information about the asymptotic $L^{\infty}[0, 1]$-valued joint distribution of $\mathcal{F}_n, \mathbf{Y}_1, \mathbf{Y}_2, \cdots, \mathbf{Y}_t$, we obtain from \eqref{eq:limit_moments_of_random_matrix_and_deterministic_matrix_equal_to_free_deterministic_equivalent_general} that the asymptotic $L^{\infty}[0, 1]$-valued joint distributions of $\mathcal{F}_n, \mathbf{Y}_1, \mathbf{Y}_2, \cdots, \mathbf{Y}_t$ and $\mathcal{F}_n, \boldsymbol{\mathcal{Y}}_1, \boldsymbol{\mathcal{Y}}_2, \cdots, \boldsymbol{\mathcal{Y}}_t$ are the same.  Thus,  we have that $\mathcal{F}_n, \mathbf{Y}_1, \mathbf{Y}_2, \cdots, \mathbf{Y}_t$ are asymptotically free over $L^{\infty}[0, 1]$.
\end{IEEEproof}

\section{Proof of Lemma \ref{lm:semicircular_lemma}}
\label{sec:proof_of_semicircular_lemma}

From Definition $2.9$ of \cite{nica2002r}, we have that $\boldsymbol{\mathcal{Y}}_{11}, \cdots, \boldsymbol{\mathcal{Y}}_{LK}$  form an R-cyclic family of  matrices.
Applying Theorem $8.2$ of \cite{nica2002r}, we then obtain $\mathcal{M}_n, \boldsymbol{\mathcal{Y}}_{11}, \cdots, \boldsymbol{\mathcal{Y}}_{LK}$ are free over $\mathcal{D}_n$.
The joint $\mathcal{M}_n$-valued cumulants of $\widehat{\boldsymbol{\mathcal{X}}}_{11}, \cdots, \widehat{\boldsymbol{\mathcal{X}}}_{LK}$ are given by
\begin{IEEEeqnarray}{Rl}
&\!\!\!\!\kappa_t^{\mathcal{M}_n}(\widehat{\boldsymbol{\mathcal{X}}}_{i_1j_1}\mathbf{C}_1,\widehat{\boldsymbol{\mathcal{X}}}_{i_2j_2}\mathbf{C}_2,
\cdots,\widehat{\boldsymbol{\mathcal{X}}}_{i_tj_t})
\nonumber \\
&~=\kappa_t^{\mathcal{M}_n}(\mathbf{A}_{i_1j_1}\boldsymbol{\mathcal{Y}}_{i_1j_1}\mathbf{A}_{i_1j_1}^H\mathbf{C}_1,\mathbf{A}_{i_2j_2}\boldsymbol{\mathcal{Y}}_{i_2j_2}\mathbf{A}_{i_2j_2}^H\mathbf{C}_2,
\nonumber \\
&~~~~~~~~~~~~~~~~~~~~~~~~~~~~~~~~~~\cdots,\mathbf{A}_{i_tj_t}\boldsymbol{\mathcal{Y}}_{i_tj_t}\mathbf{A}_{i_tj_t}^H)
\nonumber \\
&~= \mathbf{A}_{i_1j_1}\kappa_t^{\mathcal{M}_n}(\boldsymbol{\mathcal{Y}}_{i_1j_1}\mathbf{A}_{i_1j_1}^H\mathbf{C}_1\mathbf{A}_{i_2j_2},\boldsymbol{\mathcal{Y}}_{i_2j_2}\mathbf{A}_{i_2j_2}^H\mathbf{C}_2\mathbf{A}_{i_3j_3},
\nonumber \\
&~~~~~~~~~~~~~~~~~~~~~~~~~~~~~~~~~~\cdots,\boldsymbol{\mathcal{Y}}_{i_tj_t})\mathbf{A}_{i_tj_t}^H
\nonumber \\
&~= \mathbf{A}_{i_1j_1}\kappa_t^{\mathcal{D}_n}(\boldsymbol{\mathcal{Y}}_{i_1j_1}E_{\mathcal{D}_n}\{\mathbf{A}_{i_1j_1}^H\mathbf{C}_1\mathbf{A}_{i_2j_2}\}, \boldsymbol{\mathcal{Y}}_{i_2j_2}\nonumber \\
&~~~~~~~~~~~~~~E_{\mathcal{D}_n}\{\mathbf{A}_{i_2j_2}^H\mathbf{C}_2\mathbf{A}_{i_3j_3}\},\cdots,\boldsymbol{\mathcal{Y}}_{i_tj_t})\mathbf{A}_{i_tj_t}^H
 \label{eq:relation_between_Mn_valued_cumulants_and_En_valued_cumulants}
\end{IEEEeqnarray}
where $1 \leq i_t \leq L$, $1 \leq j_t \leq K$, $\mathbf{C}_1,\mathbf{C}_2,\cdots, \mathbf{C}_t \in \mathcal{M}_n$,
and the last equality is obtained by applying Theorem $3.6$ of \cite{nica2002operator}, which requires that $\mathcal{M}_n$ and $\{\boldsymbol{\mathcal{Y}}_{11}, \cdots, \boldsymbol{\mathcal{Y}}_{LK}\}$ are free over $\mathcal{D}_n$.
Since $\kappa_t^{\mathcal{D}_n} \in {\mathcal{D}_n}$, we obtain
\begin{equation}
\kappa_t^{\mathcal{M}_n}(\widehat{\boldsymbol{\mathcal{X}}}_{i_1j_1}\mathbf{C_1},\widehat{\boldsymbol{\mathcal{X}}}_{i_2j_2}\mathbf{C_2},\cdots,\widehat{\boldsymbol{\mathcal{X}}}_{i_tj_t})
\in \mathcal{D}. \nonumber \\
\end{equation} This implies
 the $\mathcal{D}$-valued cumulants of $\widehat{\boldsymbol{\mathcal{X}}}_{11}, \cdots, \widehat{\boldsymbol{\mathcal{X}}}_{LK}$ are the restrictions
of their $\mathcal{M}_n$-valued cumulants over $\mathcal{D}$ by applying Theorem $3.1$ of \cite{nica2002operator}. Thus, we have that
\begin{IEEEeqnarray}{Rl}
&\!\!\!\!\kappa_t^{\mathcal{D}}(\widehat{\boldsymbol{\mathcal{X}}}_{i_1j_1}\mathbf{C}_1,\widehat{\boldsymbol{\mathcal{X}}}_{i_2j_2}\mathbf{C}_2,\cdots,\widehat{\boldsymbol{\mathcal{X}}}_{i_tj_t})
\nonumber \\
&=\kappa_t^{\mathcal{M}_n}(\widehat{\boldsymbol{\mathcal{X}}}_{i_1j_1}\mathbf{C}_1,\widehat{\boldsymbol{\mathcal{X}}}_{i_2j_2}\mathbf{C}_2,\cdots,\widehat{\boldsymbol{\mathcal{X}}}_{i_tj_t})
\nonumber \\
&=\mathbf{A}_{i_1j_1}\kappa_t^{\mathcal{D}_n}(\boldsymbol{\mathcal{Y}}_{i_1j_1}E_{\mathcal{D}_n}\{\mathbf{A}_{i_1j_1}^H\mathbf{C}_1\mathbf{A}_{i_2j_2}\},\boldsymbol{\mathcal{Y}}_{i_2j_2}
\nonumber \\
&~~~~~~~~~~
E_{\mathcal{D}_n}\{\mathbf{A}_{i_2j_2}^H\mathbf{C}_2\mathbf{A}_{i_3j_3}\},\cdots,\boldsymbol{\mathcal{Y}}_{i_tj_t})\mathbf{A}_{i_tj_t}^H
\end{IEEEeqnarray}
where $\mathbf{C}_1,\mathbf{C}_2,\cdots, \mathbf{C}_t \in \mathcal{D}$ and the last equality is obtained by applying
\eqref{eq:relation_between_Mn_valued_cumulants_and_En_valued_cumulants}.
Since $\boldsymbol{\mathcal{Y}}_{11}, \cdots, \boldsymbol{\mathcal{Y}}_{LK}$ are free over $\mathcal{D}_n$, we have that
\begin{IEEEeqnarray}{Rl}
&\!\!\!\!\kappa_t^{\mathcal{D}_n}(\boldsymbol{\mathcal{Y}}_{i_1j_1}E_{\mathcal{D}_n}\{\mathbf{A}_{i_1j_1}^H\mathbf{C}_1\mathbf{A}_{i_2j_2}\},
\boldsymbol{\mathcal{Y}}_{i_2j_2}
\nonumber \\
&~~~~~~~~E_{\mathcal{D}_n}\{\mathbf{A}_{i_2j_2}^H\mathbf{C}_2\mathbf{A}_{i_3j_3}\},\cdots,\boldsymbol{\mathcal{Y}}_{i_tj_t})=\mathbf{0}_n
\end{IEEEeqnarray}
unless $i_1=i_2=\cdots=i_t$ and $j_1=j_2=\cdots=j_t$. Hence, $\widehat{\boldsymbol{\mathcal{X}}}_{11}, \cdots, \widehat{\boldsymbol{\mathcal{X}}}_{LK}$ are free over $\mathcal{D}$.
Moreover, since each $\boldsymbol{\mathcal{Y}}_{ij}$ is semicircular over $\mathcal{D}_n$, we obtain
\begin{IEEEeqnarray}{Rl}
&\!\!\!\!\kappa_t^{\mathcal{D}_n}(\boldsymbol{\mathcal{Y}}_{ij}E_{\mathcal{D}_n}\{\mathbf{A}_{ij}^H\mathbf{C}_1\mathbf{A}_{ij}\},
\boldsymbol{\mathcal{Y}}_{ij}
\nonumber \\
&~~~~~~~~E_{\mathcal{D}_n}\{\mathbf{A}_{ij}^H\mathbf{C}_2\mathbf{A}_{ij}\},\cdots,\boldsymbol{\mathcal{Y}}_{ij})=\mathbf{0}_n
\end{IEEEeqnarray}
except for $t=2$. This implies each $\widehat{\boldsymbol{\mathcal{X}}}_{lk}$ is also semicircular over $\mathcal{D}$.
Furthermore, since $\widehat{\boldsymbol{\mathcal{X}}}_{11}, \cdots, \widehat{\boldsymbol{\mathcal{X}}}_{LK}$ are free over $\mathcal{D}$, we obtain $\widetilde{\boldsymbol{\mathcal{X}}}$ is semicircular over $\mathcal{D}$.

According to \eqref{eq:relation_between_Mn_valued_cumulants_and_En_valued_cumulants}, we obtain
\begin{IEEEeqnarray}{Rl}
&\!\!\kappa_t^{\mathcal{M}_n}(\widehat{\boldsymbol{\mathcal{X}}}_{i_1j_1}\mathbf{C}_1, \widehat{\boldsymbol{\mathcal{X}}}_{i_2j_2}\mathbf{C}_2,\cdots,\widehat{\boldsymbol{\mathcal{X}}}_{i_tj_t})
\nonumber \\
&=E_{\mathcal{D}}\{\kappa_t^{\mathcal{M}_n}(\widehat{\boldsymbol{\mathcal{X}}}_{i_1j_1}E_{\mathcal{D}}\{\mathbf{C}_1\}, \widehat{\boldsymbol{\mathcal{X}}}_{i_2j_2}
E_{\mathcal{D}}\{\mathbf{C}_2\},\cdots,\widehat{\boldsymbol{\mathcal{X}}}_{i_tj_t})\}. \nonumber \\
\end{IEEEeqnarray}
Thus, we have that $\widehat{\boldsymbol{\mathcal{X}}}_{11}, \cdots, \widehat{\boldsymbol{\mathcal{X}}}_{LK}$ and ${\mathcal{M}_n}$ are free over $\mathcal{D}$ by applying Theorem $3.5$ of \cite{nica2002operator}.
It follows that $\widetilde{\boldsymbol{\mathcal{X}}}$ and ${\mathcal{M}_n}$ are free over $\mathcal{D}$.

\section{Proof of Theorem \ref{th:cauchy_transform}}
\label{sec:proof_of_cauchy_theorem}
{Recall that $\boldsymbol{\mathcal{X}} = \overline{\mathbf{X}} +  \widetilde{\boldsymbol{\mathcal{X}}}$.}
Since $\widetilde{\boldsymbol{\mathcal{X}}}$ and $\overline{\boldsymbol{\mathbf{X}}}$ are free over $\mathcal{D}$ by Lemma \ref{lm:semicircular_lemma}, we can apply \eqref{eq:operator_cauchy_transform_of_sum_of_free_varaible} and thus obtain
\begin{IEEEeqnarray}{Rl}
\!\!\!\!\!\!\!\!\mathcal{G}_{\boldsymbol{\mathcal{X}}}^{\mathcal{D}}(z\mathbf{I}_n)
=&\mathcal{G}_{\overline{\boldsymbol{\mathbf{X}}}}^{\mathcal{D}}\left(z\mathbf{I}_n - \mathcal{R}_{\widetilde{\boldsymbol{\mathcal{X}}}}^{\mathcal{D}}\left(\mathcal{G}_{\boldsymbol{\mathcal{X}}}^{\mathcal{D}}(z\mathbf{I}_n)\right)\right)
\nonumber \\
=&E_{\mathcal{D}}\left\{\left(z\mathbf{I}_{n} - \mathcal{R}_{\widetilde{\boldsymbol{\mathcal{X}}}}^{\mathcal{D}}\left(\mathcal{G}_{\boldsymbol{\mathcal{X}}}^{\mathcal{D}}(z\mathbf{I}_n)\right)- \overline{\boldsymbol{\mathbf{X}}}\right)^{-1}\right\}.
 \label{eq:equation_of_operator_valued_cauchy_transform}
\end{IEEEeqnarray}
Since $\boldsymbol{\mathcal{X}} = \boldsymbol{\mathcal{X}}^H$ and
\begin{IEEEeqnarray}{Rl}
\mathcal{G}_{\boldsymbol{\mathcal{X}}}^{\mathcal{D}}(z\mathbf{I}_n) =
E_{\mathcal{D}}\{(z\mathbf{I}_{n} - \boldsymbol{\mathcal{X}})^{-1}\}
\end{IEEEeqnarray}
we have that
\begin{IEEEeqnarray}{Rl}
&\!\!\!\!
\Im(\mathcal{G}_{\boldsymbol{\mathcal{X}}}^{\mathcal{D}}(z\mathbf{I}_n))
\nonumber \\
&= \frac{1}{2i}\left(\mathcal{G}_{\boldsymbol{\mathcal{X}}}^{\mathcal{D}}(z\mathbf{I}_n) - \left(\mathcal{G}_{\boldsymbol{\mathcal{X}}}^{\mathcal{D}}(z\mathbf{I}_n)\right)^H\right)
\nonumber \\
&=
\frac{1}{2i}E_{\mathcal{D}}\left\{\left(z\mathbf{I}_{n} - \boldsymbol{\mathcal{X}}\right)^{-1} - \left({z}^*\mathbf{I}_{n} - \boldsymbol{\mathcal{X}}\right)^{-1}\right\} \nonumber \\
&=-\Im(z)E_{\mathcal{D}}\left\{\left(z\mathbf{I}_{n} - \boldsymbol{\mathcal{X}}\right)^{-1}\left({z}^*\mathbf{I}_{n} - \boldsymbol{\mathcal{X}}\right)^{-1}\right\}.
\end{IEEEeqnarray}
It is obvious that $E\{(z\mathbf{I}_{n} - \boldsymbol{\mathcal{X}})^{-1}({z}^*\mathbf{I}_{n} - \boldsymbol{\mathcal{X}})^{-1}\}$ is positive definite.  Each block matrix of $E_{\mathcal{D}}\{(z\mathbf{I}_{n} - \boldsymbol{\mathcal{X}})^{-1}({z}^*\mathbf{I}_{n} - \boldsymbol{\mathcal{X}})^{-1}\}$ is a principal submatrix of $E\{(z\mathbf{I}_{n} - \boldsymbol{\mathcal{X}})^{-1}({z}^*\mathbf{I}_{n} - \boldsymbol{\mathcal{X}})^{-1}\}$
and thus positive definite by Theorem $3.4$ of \cite{bapat2012linear}. Then  $E_{\mathcal{D}}\{(z\mathbf{I}_{n} - \boldsymbol{\mathcal{X}})^{-1}({z}^*\mathbf{I}_{n} - \boldsymbol{\mathcal{X}})^{-1}\}$ is also positive definite. Thus, we obtain $\Im(\mathcal{G}_{\boldsymbol{\mathcal{X}}}^{\mathcal{D}}(z\mathbf{I}_n)) \prec 0$  for $z \in \mathbb{C}^+$.

This implies that $\mathcal{G}_{\boldsymbol{\mathcal{X}}}^{\mathcal{D}}(z\mathbf{I}_n)$ should be a solution of \eqref{eq:equation_of_operator_valued_cauchy_transform} with the property that $\Im(\mathcal{G}_{\boldsymbol{\mathcal{X}}}^{\mathcal{D}}(z\mathbf{I}_n)) \prec 0$ for $z \in \mathbb{C}^+$.
In the following, we will prove that \eqref{eq:equation_of_operator_valued_cauchy_transform} has
exactly one solution with $\Im(\mathcal{G}_{\boldsymbol{\mathcal{X}}}^{\mathcal{D}}(z\mathbf{I}_n)) \prec 0$ for $z \in \mathbb{C}^+$.
Replace $\mathcal{G}_{\boldsymbol{\mathcal{X}}}^{\mathcal{D}}(z\mathbf{I}_n)$ with $-i\mathbf{W}$, we have that $\Re(\mathbf{W}) \succ 0$. Then, \eqref{eq:equation_of_operator_valued_cauchy_transform} becomes
\begin{IEEEeqnarray}{Rl}
 \mathbf{W} =& iE_{\mathcal{D}}\left\{\left(z\mathbf{I}_{n} - \mathcal{R}_{\widetilde{\boldsymbol{\mathcal{X}}}}^{\mathcal{D}}(-i\mathbf{W})- \overline{\boldsymbol{\mathbf{X}}}\right)^{-1}\right\} \nonumber \\
  =& E_{\mathcal{D}}\left\{\left(\mathbf{V} + \mathcal{R}_{\widetilde{\boldsymbol{\mathcal{X}}}}^{\mathcal{D}}(\mathbf{W})\right)^{-1}\right\} \nonumber \\
    =& E_{\mathcal{D}}\{\mathfrak{F}_\mathbf{V}(\mathbf{W})\}
\end{IEEEeqnarray}
where $\mathbf{V}=-iz\mathbf{I}_{n}+i\overline{\boldsymbol{\mathbf{X}}}$. Since $z \in \mathbb{C}^+$ and $\overline{\boldsymbol{\mathbf{X}}}$ is Hermitian, we have that $\Re(\mathbf{V}) \succeq \epsilon \mathbf{I}_{n}$ for some $\epsilon > 0$.

Let ${\mathcal{M}_n}_{+}$ denote  $\{\mathbf{W} \in \mathcal{M}_n:\Re(\mathbf{W}) \succeq \epsilon\mathbf{I}~{\rm for~some~}\epsilon > 0\}$. We
define $R_a = \{\mathbf{W} \in  {\mathcal{M}_n}_{+}:\|\mathbf{W}\| \leq a \}$ for $a > 0$.
According to Proposition $3.2$ of \cite{helton2007operator}, $\mathfrak{F}_\mathbf{V}$ is well defined, $\|\mathfrak{F}_\mathbf{V}(\mathbf{W})\| \leq \|{\Re(\mathbf{V})}^{-1}\| $, and $\mathfrak{F}_\mathbf{V}$ maps $R_a$ strictly to itself for $\mathbf{V} \in  {\mathcal{M}_n}_{+}$ and $\|{\Re(\mathbf{V})}^{-1}\| <  a $.
Furthermore, by applying the Earle-Hamilton fixed point theorem \cite{earle1970fixed},  the statement in Theorem $2.1$ of \cite{helton2007operator} that
 there exists exactly one solution $\mathbf{W} \in {\mathcal{M}_n}_{+}$  to the equation $\mathbf{W} = \mathfrak{F}_\mathbf{V}(\mathbf{W})$ and
the solution is the limit of iterates $\mathbf{W}_n = \mathfrak{F}_\mathbf{V}^n(\mathbf{W}_0)$ for every $\mathbf{W}_0 \in {\mathcal{M}_n}_{+}$ is proven.

We herein extend the proof of \cite{helton2007operator}. First, we define $R_b = \{\mathbf{W} \in  {\mathcal{M}_n}_{+} \cap \mathcal{D} : \|\mathbf{W}\| \leq b \}$ for $b > 0$. Using Proposition $3.2$ of \cite{helton2007operator}, we have that $\|\mathfrak{F}_\mathbf{V}(\mathbf{W})\| \leq \|{\Re(\mathbf{V})}^{-1}\| $ and
$\Re(\mathfrak{F}_\mathbf{V}(\mathbf{W})) \succeq \epsilon\mathbf{I}$ for some $\epsilon > 0$ and $\mathbf{W} \in R_b$. Since $\|E_{\mathcal{D}}\{\mathfrak{F}_\mathbf{V}(\mathbf{W})\}\| \leq \|\mathfrak{F}_\mathbf{V}(\mathbf{W})\| $, we obtain $\|E_{\mathcal{D}}\{\mathfrak{F}_\mathbf{V}(\mathbf{W})\}\| \leq \|{\Re(\mathbf{V})}^{-1}\| $. Furthermore, because each diagonal block of $E_{\mathcal{D}}\{\mathfrak{F}_\mathbf{V}(\mathbf{W})\}$ is a principal submatrix of $\mathfrak{F}_\mathbf{V}(\mathbf{W})$, we also have that
$\lambda_{min}(\mathfrak{F}_\mathbf{V}(\mathbf{W})) \leq \lambda_{min}(E_{\mathcal{D}}\{\mathfrak{F}_\mathbf{V}(\mathbf{W})\})$ by applying Theorem $1$ of \cite{thompson1972principal}. Hence, we
have that $\Re(E_{\mathcal{D}}\{\mathfrak{F}_\mathbf{V}(\mathbf{W})\}) \succeq \epsilon\mathbf{I}$ for some $\epsilon > 0$, and that $E_{\mathcal{D}} \circ \mathfrak{F}_\mathbf{V}$ maps $R_b$ strictly to itself for $\mathbf{V} \in  {\mathcal{M}_n}_{+} \cap \mathcal{D} $ and $\|{\Re(\mathbf{V})}^{-1}\| <  b $.
Thus, applying the Earle-Hamilton fixed point theorem,
we obtain there exists exactly one solution $\mathbf{W} \in {\mathcal{M}_n}_{+} \cap \mathcal{D}$  to the equation $\mathbf{W} = E_{\mathcal{D}}\{\mathfrak{F}_\mathbf{V}(\mathbf{W})\}$ and
the solution is the limit of iterates $\mathbf{W}_n = (E_{\mathcal{D}} \circ \mathfrak{F}_\mathbf{V})^n(\mathbf{W}_0)$ for every $\mathbf{W}_0 \in {\mathcal{M}_n}_{+} \cap \mathcal{D}$.

Following a derivation similar to that of \eqref{eq:realtion_of_Cauchy_transform_of X_and_X2}, we have that
    \begin{eqnarray}
        \mathcal{G}_{\boldsymbol{\mathcal{X}}}^{\mathcal{D}}(z\mathbf{I}_n)  = z\mathcal{G}_{\boldsymbol{\mathcal{X}}^{2}}^{\mathcal{D}}(z^2\mathbf{I}_n)
    \end{eqnarray}
where $z,z^2 \in \mathbb{C}^+$.
Then, we obtain
\begin{IEEEeqnarray}{Rl}
&\!\!\!\!\!\!\!z\mathcal{G}_{\boldsymbol{\mathcal{X}}^2}^{\mathcal{D}}(z^2\mathbf{I}_n)
\nonumber \\
&= E_{\mathcal{D}}\left\{\left(z\mathbf{I}_{n} - \mathcal{R}_{\widetilde{\boldsymbol{\mathcal{X}}}}^{\mathcal{D}}\left(z\mathcal{G}_{\boldsymbol{\mathcal{X}}^2}^{\mathcal{D}}(z^2\mathbf{I}_n)\right)- \overline{\boldsymbol{\mathbf{X}}}\right)^{-1}\right\} \label{eq:equation_of_operator_valued_cauchy_transform_X2}
\end{IEEEeqnarray}
by substituting   $z\mathcal{G}_{\boldsymbol{\mathcal{X}}}^{\mathcal{D}}(z^2\mathbf{I}_n)$ for $\mathcal{G}_{\boldsymbol{\mathcal{X}}}^{\mathcal{D}}(z\mathbf{I}_n)$ in
 \eqref{eq:equation_of_operator_valued_cauchy_transform}.
{Furthermore, we have that $\Im(z^{-1}\mathcal{G}_{\boldsymbol{\mathcal{X}}}^{\mathcal{D}}(z\mathbf{I}_n)) \prec 0$
for $z,z^2 \in \mathbb{C}^+$.
  Thus, $z\mathcal{G}_{\boldsymbol{\mathcal{X}}^2}^{\mathcal{D}}(z^2\mathbf{I}_n)$ with  $\Im(\mathcal{G}_{\boldsymbol{\mathcal{X}}^2}^{\mathcal{D}}(z^2\mathbf{I}_n)) \prec 0$ for $z,z^2 \in \mathbb{C}^+$ is uniquely determined by \eqref{eq:equation_of_operator_valued_cauchy_transform_X2}.}

Since $\widetilde{\boldsymbol{\mathcal{X}}}$ is semicircular over $\mathcal{D}$ as shown in Lemma \ref{lm:semicircular_lemma}, we have that
\begin{IEEEeqnarray}{Rl}
\!\!\!\!\!\!\mathcal{R}_{\widetilde{\boldsymbol{\mathcal{X}}}}^{\mathcal{D}}(\mathbf{C})
=&E_{\mathcal{D}}\{{\widetilde{\boldsymbol{\mathcal{X}}}}\mathbf{C}{\widetilde{\boldsymbol{\mathcal{X}}}}\}
=E_{\mathcal{D}}\{{\widetilde{\boldsymbol{\mathbf{X}}}}\mathbf{C}{\widetilde{\boldsymbol{\mathbf{X}}}}\}\nonumber \\
=& \left(
            \begin{array}{ccccc}
              \sum\limits_{k=1}^{K}{\tilde{\eta}}_k(\mathbf{C}_k)     & \mathbf{0} & \cdots & \mathbf{0} \\
              \mathbf{0}    & \eta_{1} (\widetilde{\mathbf{C}}) & \ldots & \mathbf{0} \\
              \vdots  & \vdots &  \ddots & \vdots \\
              \mathbf{0}    & \mathbf{0} & \ldots & \eta_{K} (\widetilde{\mathbf{C}}) \\
            \end{array}
\right) \label{eq:operator_valued_r_transform_of_widetilde_mathcal_H}
\end{IEEEeqnarray}
where $\mathbf{C}={\rm{diag}}(\widetilde{\mathbf{C}},\mathbf{C}_1,\cdots,\mathbf{C}_K)$, ${\widetilde{\mathbf{C}}}\in \mathcal{M}_N$ and $\mathbf{C}_k\in \mathcal{M}_{M_k}$.
Then according to \eqref{eq:Cauchy_transform_of_Hfreesquare_in_detail}  and \eqref{eq:operator_valued_r_transform_of_widetilde_mathcal_H}, \eqref{eq:equation_of_operator_valued_cauchy_transform_X2}
 becomes
\begin{IEEEeqnarray}{Rl}
&\!\!\!\!\!\!\!\!\left(
            \begin{array}{ccccc}
              z\mathcal{G}_{\boldsymbol{\mathcal{B}}_N}^{\mathcal{M}_N}(z^2\mathbf{I}_N)      & \mathbf{0} & \cdots & \mathbf{0} \\
              \mathbf{0}    & z\mathcal{G}_1(z^2)  & \ldots & \mathbf{0} \\
              \vdots  & \vdots &  \ddots & \vdots \\
              \mathbf{0}    & \mathbf{0} &  \ldots & z\mathcal{G}_K(z^2) \\
            \end{array}
\right)
\nonumber \\
&\!\!\!\!= E_{\mathcal{D}}\!\!\left\{\!\left(\!
            \begin{array}{cccc}
              z{\tilde{{\boldsymbol{\Phi}}}}(z^2) & -\overline{\mathbf{H}}_1   & \cdots & -\overline{\mathbf{H}}_K \\
              -\overline{\mathbf{H}}{}_1^H    & z{{\boldsymbol{\Phi}}}_1(z^2) & \ldots & \mathbf{0} \\
              \vdots  & \vdots & \ddots & \vdots \\
              -\overline{\mathbf{H}}{}_K^H    & \mathbf{0} & \ldots & z{{\boldsymbol{\Phi}}}_K(z^2) \\
            \end{array}
\!\!\right)^{-1}\!\!\right\}  \label{eq:equation_of_operator_value_cauchy_transform_1}
\end{IEEEeqnarray}
where
\begin{IEEEeqnarray}{Rl}
&\tilde{\boldsymbol{\Phi}}(z^2)= \mathbf{I}_N - \sum\limits_{k=1}^{K}{\tilde{\eta}}_{k}(\mathcal{G}_k(z^2)) \\
&{\boldsymbol{\Phi}}_k(z^2) = \mathbf{I}_{M_k} - \eta_{k} (\mathcal{G}_{\boldsymbol{\mathcal{B}}_N}^{\mathcal{M}_N}(z^2\mathbf{I}_N)).
\end{IEEEeqnarray}
According to the block matrix inversion formula \cite{petersen2008matrix}
\begin{IEEEeqnarray}{Rl}
&\!\!\!\!\!\!\!\!\left(\begin{array}{cc}
  \mathbf{A}_{11} & \mathbf{A}_{12} \\
  \mathbf{A}_{21} & \mathbf{A}_{22} \\
\end{array}\right)^{-1}
\nonumber \\
&=
\left(\begin{array}{cc}
  \mathbf{C}_{1}^{-1} & -\mathbf{A}_{11}^{-1}\mathbf{A}_{12}\mathbf{C}_{2}^{-1} \\
  -\mathbf{C}_{2}^{-1}\mathbf{A}_{21}\mathbf{A}_{11}^{-1} & \mathbf{C}_{2}^{-1} \\
\end{array}\right)
\end{IEEEeqnarray}
where $\mathbf{C}_{1}=\mathbf{A}_{11}-\mathbf{A}_{12}\mathbf{A}_{
22}^{-1}\mathbf{A}_{21}$ and $\mathbf{C}_{2}=\mathbf{A}_{22}-\mathbf{A}_{21}\mathbf{A}_{
11}^{-1}\mathbf{A}_{12}$,
 \eqref{eq:equation_of_operator_value_cauchy_transform_1} can be split into
\begin{equation}
z\mathcal{G}_{\boldsymbol{\mathcal{B}}_N}^{\mathcal{M}_N}(z^2\mathbf{I}_N) = \left(z{\Tilde{{\boldsymbol{\Phi}}}}(z^2)
-  \overline{\mathbf{H}}\left(z{{\boldsymbol{\Phi}}}(z^2)\right)^{-1} \overline{\mathbf{H}}{}^H\right)^{-1}
\label{eq:Cauchy_transform_proof_temp1}
\end{equation}
and
\begin{equation}
z\mathcal{G}_k(z^2) = \left(\left(z{{\boldsymbol{\Phi}}}(z^2)
- \overline{\mathbf{H}}{}^H\left(z{\Tilde{{\boldsymbol{\Phi}}}}(z^2)\right)^{-1} \overline{\mathbf{H}}\right)^{-1}\right)_k
\label{eq:Cauchy_transform_proof_temp2}
\end{equation}
where
\begin{equation}
{{\boldsymbol{\Phi}}}(z^2) = {\rm{diag}}\left({{\boldsymbol{\Phi}}}_1(z^2),{{\boldsymbol{\Phi}}}_2(z^2),\cdots,{{\boldsymbol{\Phi}}}_K(z^2)\right).
\end{equation}
Furthermore, \eqref{eq:Cauchy_transform_proof_temp1} and \eqref{eq:Cauchy_transform_proof_temp2} are equivalent to
\begin{equation}
\mathcal{G}_{\boldsymbol{\mathcal{B}}_N}^{\mathcal{M}_N}(z\mathbf{I}_N) = \left(z{\Tilde{{\boldsymbol{\Phi}}}}(z)
-    \overline{\mathbf{H}}{{\boldsymbol{\Phi}}}(z)^{-1} \overline{\mathbf{H}}{}^H\right)^{-1}
\end{equation}
and
\begin{equation}
\mathcal{G}_k(z) = \left((z{{\boldsymbol{\Phi}}}(z)
-  \overline{\mathbf{H}}{}^H{\Tilde{{\boldsymbol{\Phi}}}}(z)^{-1} \overline{\mathbf{H}})^{-1}\right)_k.
\end{equation}
Finally, since the solution has the property $\Im(\mathcal{G}_{\boldsymbol{\mathcal{X}}^2}^{\mathcal{D}}(z\mathbf{I}_n)) \prec 0$ for $z \in \mathbb{C}^+$ and
$\mathcal{G}_{\boldsymbol{\mathcal{B}}_N}^{\mathcal{M}_N}(z\mathbf{I}_N)$ is a principal submatrix of $\mathcal{G}_{\boldsymbol{\mathcal{X}}^2}^{\mathcal{D}}(z\mathbf{I}_n)$, we have that $\Im(\mathcal{G}_{\boldsymbol{\mathcal{B}}_N}^{\mathcal{M}_N}(z\mathbf{I}_N)) \prec 0$ for $z \in \mathbb{C}^+$ by using Theorem $3.4$ of \cite{bapat2012linear}.

\section{Proof of Lemma \ref{lm:shannon_theorem_lemma_1}}
\label{sec:proof_of_shannon_theorem_lemma_1}
Recall that $\mathbf{E}_k(x)=-x\mathcal{G}_k(-x)$.
Let $\boldsymbol{\mathcal{E}}(x)$  denote
\begin{equation}
\left({{\boldsymbol{\Phi}}}(-x)+ x^{-1}\overline{\mathbf{H}}{}^H{\Tilde{{\boldsymbol{\Phi}}}}(-x)^{-1}\overline{\mathbf{H}}\right)^{-1}.
\nonumber \\
\end{equation}
Then, we have that
\begin{IEEEeqnarray}{Rl}
&\!\!\!\!\sum\limits_{k=1}^{K}{\rm{tr}}\left(\left({{\boldsymbol{\Phi}}}_k(-x)^{-1}-\mathbf{E}_k(x)\right)\frac{d{{\boldsymbol{\Phi}}}_k(-x)}{dx}\right)
 \nonumber \\
&~~~~~~~~={\rm{tr}}\left(\left({{\boldsymbol{\Phi}}}(-x)^{-1}-\boldsymbol{\mathcal{E}}(x)\right)\frac{d{{\boldsymbol{\Phi}}}(-x)}{dx}\right).
\end{IEEEeqnarray}
Recall that $\mathbf{A}(x)=({\Tilde{{\boldsymbol{\Phi}}}}(-x)
+ x^{-1}\overline{\mathbf{H}}{{\boldsymbol{\Phi}}}(-x)^{-1} \overline{\mathbf{H}}{}^H)^{-1}$.
Using the Woodbury identity \cite{higham2002accuracy}, we rewrite $\boldsymbol{\mathcal{E}}(x)$ as
\begin{equation}
\boldsymbol{\mathcal{E}}(x)=
\boldsymbol{\Phi}(-x)^{-1}-x^{-1}\boldsymbol{\Phi}(-x)^{-1}\overline{\mathbf{H}}{}^H\mathbf{A}(x)\overline{\mathbf{H}}\boldsymbol{\Phi}(-x)^{-1}
\end{equation}
which further leads to
\begin{IEEEeqnarray}{Rl}
&\!\!\!\!\sum\limits_{k=1}^{K}{\rm{tr}}\left(\left({{\boldsymbol{\Phi}}}_k(-x)^{-1}-\mathbf{E}_k(x)\right)\frac{d{{\boldsymbol{\Phi}}}_k(-x)}{dx}\right)
\nonumber \\
&={\rm{tr}}\left({{\boldsymbol{\Phi}}}(-x)^{-1}x^{-1}\overline{\mathbf{H}}{}^H\mathbf{A}(x)\overline{\mathbf{H}}{{\boldsymbol{\Phi}}}(-x)^{-1}\frac{d{{\boldsymbol{\Phi}}}(-x)}{dx}\right)  \nonumber \\
&={\rm{tr}}\left(x^{-1}\overline{\mathbf{H}}{}^H\mathbf{A}(x)\overline{\mathbf{H}}{{\boldsymbol{\Phi}}}(-x)^{-1}\frac{d{{\boldsymbol{\Phi}}}(-x)}{dx}{{\boldsymbol{\Phi}}}(-x)^{-1}\right)
\nonumber \\
&~~~~-{\rm{tr}}\left(x^{-1}\overline{\mathbf{H}}{}^H\mathbf{A}(x)\overline{\mathbf{H}}\frac{d{{\boldsymbol{\Phi}}}(-x)^{-1}}{dx}\right).
\end{IEEEeqnarray}

\section{Proof of Lemma \ref{lm:shannon_theorem_lemma_2}}
\label{sec:proof_of_shannon_theorem_lemma_2}
From
\begin{IEEEeqnarray}{Rl}
{{\boldsymbol{\Phi}}}_k(-x) - \mathbf{I}_{M_k}=&- \eta_{k} (\mathcal{G}_{\boldsymbol{\mathcal{B}}_N}^{\mathcal{M}_N}(-x\mathbf{I}_N)) \nonumber \\
 =&\eta_{k}(x^{-1}\mathbf{A}(x)) \label{eq:phi_k_Az_relation}
\end{IEEEeqnarray}
we have that
\begin{eqnarray}
\frac{d{{\boldsymbol{\Phi}}}_k(-x)}{dx}  =  \eta_{k} \left(\frac{dx^{-1}\mathbf{A}(x)}{dx}\right).
\label{eq:Phi_k_z_Az_relation}
\end{eqnarray}
From ${\Tilde{{\boldsymbol{\Phi}}}}(-x)- \mathbf{I}_N=\sum_{k=1}^{K}{\tilde{\eta}}_{k} (\mathcal{G}_k(-x))$, we then obtain that
\begin{IEEEeqnarray}{Rl}
  &\!\!\!\!\!\!\!\!\!\!\!\!{\rm{tr}}\left(\frac{dx^{-1}\mathbf{A}(x)}{dx}\left({\Tilde{{\boldsymbol{\Phi}}}}(-x)- \mathbf{I}_N\right)\right)
  \nonumber \\
  &= -{\rm{tr}}\left(\frac{dx^{-1}\mathbf{A}(x)}{dx}\sum\limits_{k=1}^{K}{\tilde{\eta}}_{k} (\mathcal{G}_k(-x))\right) \nonumber \\
  &= {\rm{tr}}\left(\frac{dx^{-1}\mathbf{A}(x)}{dx}\sum\limits_{k=1}^{K}{\tilde{\eta}}_{k} (x^{-1}\mathbf{E}_k(x))\right) \nonumber \\
  &= \sum\limits_{k=1}^{K}{\rm{tr}}\left({\eta}_k\left(\frac{dx^{-1}\mathbf{A}(x)}{dx}\right)x^{-1}\mathbf{E}_k(x)\right)
\end{IEEEeqnarray}
where the last equality is due to
\begin{IEEEeqnarray}{Rl}
{\rm{tr}}(\mathbf{A}_1{\tilde{\eta}}_{k} (\mathbf{A}_2))=& {\rm{tr}}({\mathbb E} \{\mathbf{A}_1 \widetilde{\mathbf{H}}_k \mathbf{A}_2 \widetilde{\mathbf{H}}_k^H\}) \IEEEnonumber \\
 =& {\rm{tr}}({\mathbb E}\{\widetilde{\mathbf{H}}_k^H\mathbf{A}_1\widetilde{\mathbf{H}}_k\mathbf{A}_2\})
  \IEEEnonumber \\
= &{\rm{tr}}(\eta_k({\mathbf{A}_1})\mathbf{A}_2). \nonumber
\end{IEEEeqnarray}
According to \eqref{eq:Phi_k_z_Az_relation}, we finally obtain
\begin{eqnarray}
  &&\!\!\!\!\!\!\!\!\!\!\!\!\!\!\!\!{\rm{tr}}\left(\frac{dx^{-1}\mathbf{A}(x)}{dx}\left({\Tilde{{\boldsymbol{\Phi}}}}(-x)- \mathbf{I}_N\right)\right)
   \nonumber \\
  &&=\sum\limits_{k=1}^{K}{\rm{tr}}\left(\frac{d{{\boldsymbol{\Phi}}}_k(-x)}{dx}x^{-1}\mathbf{E}_k(x)\right).
\end{eqnarray}

\section{Proof of Theorem \ref{th:shannon_theorem}}
\label{sec:proof_of_shannon_theorem}
We define ${J}(x)$  by
\begin{eqnarray}
{J}(x) = -x^{-1}-G_{\boldsymbol{\mathcal{B}}_N}(-x) = -x^{-1}{\rm{tr}}(\mathbf{A}(x)\mathbf{B}(x))
\end{eqnarray}
where  $\mathbf{B}(x)$ denotes ${\Tilde{{\boldsymbol{\Phi}}}}(-x)
+ x^{-1}\overline{\mathbf{H}}{{\boldsymbol{\Phi}}}(-x)^{-1} \overline{\mathbf{H}}{}^H - \mathbf{I}_N$.
For convenience, we rewrite ${J}(x)$ as
\begin{eqnarray}
{J}(x) = J_1(x) + J_2(x)
\end{eqnarray}
where $J_1(x)$ and $J_2(x)$ are defined by
\begin{eqnarray}
J_1(x) = -\frac{1}{x}{\rm{tr}}\left(\mathbf{A}(x)\left({\Tilde{{\boldsymbol{\Phi}}}}(-x)- \mathbf{I}_N\right)\right)
\end{eqnarray}
and
\begin{eqnarray}
 J_2(x)= -\frac{1}{x^2}{\rm{tr}}\left(\mathbf{A}(x)\overline{\mathbf{H}}{{\boldsymbol{\Phi}}}(-x)^{-1} \overline{\mathbf{H}}{}^H\right).
\end{eqnarray}
Differentiating ${\rm{tr}}(-\mathbf{A}(x)({\Tilde{{\boldsymbol{\Phi}}}}(-x)- \mathbf{I}_N))$ with respect to $x$,
we have that
\begin{IEEEeqnarray}{Rl}
&\!\!\!\!\!\!\!\!\!\!\frac{d}{dx}{\rm{tr}}\left(x\mathbf{I}_N-x^{-1}\mathbf{A}(x)\left({\Tilde{{\boldsymbol{\Phi}}}}(-x)- \mathbf{I}_N\right)\right)
\IEEEnonumber \\
&\!\!\!\!\!\!=  J_1(x) +K(x) -x{\rm{tr}}\left(\!\frac{dx^{-1}\mathbf{A}(x)}{dx}\left(\!{\Tilde{{\boldsymbol{\Phi}}}}(-x)- \mathbf{I}_N\!\right)\!\right)    \label{eq:shannon_derivate_1_1}
\end{IEEEeqnarray}
where $K(x)$ is defined as
\begin{eqnarray}
    K(x) = -{\rm{tr}}\left(\mathbf{A}(x)\frac{d{\Tilde{{\boldsymbol{\Phi}}}}(-x)}{dx}\right).
\end{eqnarray}
According to {Lemma \ref{lm:shannon_theorem_lemma_2}}, \eqref{eq:shannon_derivate_1_1} becomes
\begin{IEEEeqnarray}{Rl}
&\!\!\!\!\!\!\!\!\frac{d}{dx}{\rm{tr}}\left(-\mathbf{A}(x)\left({\Tilde{{\boldsymbol{\Phi}}}}(-x)- \mathbf{I}_N\right)\right)
\nonumber \\
&  =  J_1(x) + K(x)
    -\sum\limits_{k=1}^{K}{\rm{tr}}\left(\frac{d{{\boldsymbol{\Phi}}}_k(-x)}{dx}\mathbf{E}_k(x)\right).   \label{eq:shannon_derivate_1_2}
\end{IEEEeqnarray}
Defining $L(x)$ as
\begin{eqnarray}
  L(x) = -\sum\limits_{k=1}^{K}{\rm{tr}}\left(\frac{d{{\boldsymbol{\Phi}}}_k(-x)}{dx}\mathbf{E}_k(x)\right)
\end{eqnarray}
we obtain
\begin{eqnarray}
&&\!\!\!\!\!\!\!\!\!\!\!\!\!\!\!\!\!\!\!\!\!\!\!\!\!\!\!\!\!\!\frac{d}{dx}{\rm{tr}}\left(-\mathbf{A}(x)\left({\Tilde{{\boldsymbol{\Phi}}}}(-x)- \mathbf{I}_N\right)\right)
\nonumber \\
&&=  J_1(x) + K(x) + L(x).  \label{eq:shannon_derivate_1_3}
\end{eqnarray}
For a matrix-valued function $\mathbf{F}(x)$, we have that
\begin{equation}
\frac{d}{dx}\log\det(\mathbf{F}(x)) = {\rm{tr}}\left(\mathbf{F}(x)^{-1}\frac{d\mathbf{F}(x)}{dx}\right).
\end{equation}
When $\mathbf{F}(x)={\Tilde{{\boldsymbol{\Phi}}}}(-x)
+ x^{-1}\overline{\mathbf{H}}{{\boldsymbol{\Phi}}}(-x)^{-1} \overline{\mathbf{H}}{}^H$, we obtain
\begin{eqnarray}
  & &\!\!\!\!\!\!\!\!\!\!\frac{d}{dx}\log\det\left({\Tilde{{\boldsymbol{\Phi}}}}(-x)
+ x^{-1}\overline{\mathbf{H}}{{\boldsymbol{\Phi}}}(-x)^{-1} \overline{\mathbf{H}}{}^H\right)  \nonumber \\
  &&\!\!\!\!\!\!\!\!= {\rm{tr}}\left(\mathbf{A}(x)\frac{d\mathbf{B}(x)}{dx}\right)  \nonumber \\
  &&\!\!\!\!\!\!\!\!= {\rm{tr}}\left(\!\mathbf{A}(x)\frac{d{\Tilde{{\boldsymbol{\Phi}}}}(-x)}{dx}\!\right)
  + {\rm{tr}}\left(\!\mathbf{A}(x)\frac{dx^{-1}\overline{\mathbf{H}}{{\boldsymbol{\Phi}}}(-x)^{-1} \overline{\mathbf{H}}{}^H}{dx}\!\right)  \nonumber \\
  &&\!\!\!\!\!\!\!\!= -K(x) + J_2(x) + x^{-1}{\rm{tr}}\left(\!\mathbf{A}(x)\frac{d\overline{\mathbf{H}}{{\boldsymbol{\Phi}}}(-x)^{-1} \overline{\mathbf{H}}{}^H}{dx}\!\right).\nonumber \\  \label{eq:shannon_derivate_2_1}
\end{eqnarray}
According to {Lemma \ref{lm:shannon_theorem_lemma_1}},  \eqref{eq:shannon_derivate_2_1} becomes
\begin{eqnarray}
  & &\!\!\!\!\!\!\!\!\!\!\frac{d}{dx}\log\det\left({\Tilde{{\boldsymbol{\Phi}}}}(-x)
+ x^{-1}\overline{\mathbf{H}}{{\boldsymbol{\Phi}}}(-x)^{-1} \overline{\mathbf{H}}{}^H\right)  \nonumber \\
  &&=-K(x) + J_2(x)
  \nonumber \\
  &&~~~~~~- \sum\limits_{k=1}^{K}{\rm{tr}}\left(\left({{\boldsymbol{\Phi}}}_k(-x)^{-1}-\mathbf{E}_k(x)\right)\frac{d{{\boldsymbol{\Phi}}}_k(-x)}{dx}\right)  \nonumber \\
  &&=-K(x) + J_2(x) - L(x)
  \nonumber \\
  &&~~~~~~- \sum\limits_{k=1}^{K}{\rm{tr}}\left(\left({{\boldsymbol{\Phi}}}_k(-x)^{-1}\right)\frac{d{{\boldsymbol{\Phi}}}_k(-x)}{dx}\right).  \label{eq:shannon_derivate_2_2}
\end{eqnarray}
From \eqref{eq:shannon_derivate_1_3}, \eqref{eq:shannon_derivate_2_2} and
\begin{eqnarray}
  \frac{d}{dx}\log\det({{\boldsymbol{\Phi}}}(-x))
  =\sum\limits_{k=1}^{K}{\rm{tr}}\left({{\boldsymbol{\Phi}}}_k(-x)^{-1}\frac{d{{\boldsymbol{\Phi}}}_k(-x)}{dx}\right)
\end{eqnarray}
we obtain
 \begin{eqnarray}
J(x) \!\!\!\!&=&\!\!\!\!\frac{d}{dx}\log\det\left({\Tilde{{\boldsymbol{\Phi}}}}(-x)
+ x^{-1}\overline{\mathbf{H}}{{\boldsymbol{\Phi}}}(-x)^{-1} \overline{\mathbf{H}}{}^H\right)
\nonumber \\
&&+ \frac{d}{dx}\log\det({{\boldsymbol{\Phi}}}(-x))
\nonumber \\
&&~~
-\frac{d}{dx}{\rm{tr}}\left(\mathbf{A}(x)\left({\Tilde{{\boldsymbol{\Phi}}}}(-x)- \mathbf{I}_N\right)\right).
\end{eqnarray}
Since $\mathcal{V}_{\boldsymbol{\mathcal{B}}_N}(x)\rightarrow 0$ as $x \rightarrow \infty$, the Shannon transform $\mathcal{V}_{\boldsymbol{\mathcal{B}}_N}(x)$ can be obtained as
\begin{eqnarray}
\mathcal{V}_{\boldsymbol{\mathcal{B}}_N}(x)
 \!\!\!\!&=&\!\!\!\!   \log\det\left({\Tilde{{\boldsymbol{\Phi}}}}(-x)
+x^{-1}\overline{\mathbf{H}}{{\boldsymbol{\Phi}}}(-x)^{-1} \overline{\mathbf{H}}{}^H\right)
\nonumber \\
  &&\!\!+\log\det({{\boldsymbol{\Phi}}}(-x))
 \nonumber \\
 &&-{\rm{tr}}\left(\mathbf{A}(x)\left({\Tilde{{\boldsymbol{\Phi}}}}(-x)- \mathbf{I}_N\right)\right).
\end{eqnarray}
Furthermore, it is easy to verify that
\begin{eqnarray}
&&\!\!\!\!\!\!\!\!\!\!\!\!{\rm{tr}}\left(\mathbf{A}(x)\left({\Tilde{{\boldsymbol{\Phi}}}}(-x)- \mathbf{I}_N\right)\right)
\nonumber \\
&&= {\rm{tr}}\left(x\sum\limits_{k=1}^{K}{{\eta}}_{k}(\mathcal{G}_{\boldsymbol{\mathcal{B}}_N}^{\mathcal{M}_N}(-x\mathbf{I}_N))\mathcal{G}_k(-x) \right).
\end{eqnarray}
Finally, we obtain the Shannon transform $\mathcal{V}_{\boldsymbol{\mathcal{B}}_N}(x)$  as
\begin{IEEEeqnarray}{Rl}
\!\!\!\!\!\!\!\!\mathcal{V}_{\boldsymbol{\mathcal{B}}_N}(x) = & \log\det\left({\Tilde{{\boldsymbol{\Phi}}}}(-x)
+ x^{-1}\overline{\mathbf{H}}{{\boldsymbol{\Phi}}}(-x)^{-1} \overline{\mathbf{H}}{}^H\right)
\IEEEnonumber \\
&\!\!+\log\det({{\boldsymbol{\Phi}}}(-x))
\IEEEnonumber \\
&- {\rm{tr}}\left(x\sum\limits_{k=1}^{K}{{\eta}}_{k}(\mathcal{G}_{\boldsymbol{\mathcal{B}}_N}^{\mathcal{M}_N}(-x\mathbf{I}_N))\mathcal{G}_k(-x) \right).
\end{IEEEeqnarray}

\section{Proof of Theorem \ref{th:capaicty_achieving matrix_theorem}}
\label{sec:proof_of_capaicty_achieving matrix_theorem}
The way to show the strict convexity of $-\mathcal{V}_{\boldsymbol{\mathcal{B}}_N}(x)$ with respect to $\mathbf{Q}$ is similar to  Theorem $3$ of \cite{dupuy2011capacity} and Theorem $4$ of \cite{dumont2010capacity}, and thus omitted here.
Let the Lagrangian of the optimization problem \eqref{eq:optimization_problem_of_deterministic_equivalent} be defined as
\begin{IEEEeqnarray}{Rl}
\mathcal{L}(\mathbf{Q},\mathbf{\Upsilon},\boldsymbol{\mu})
=&  \mathcal{V}_{\boldsymbol{\mathcal{B}}_N}(x) + {\rm{tr}}\left(\sum\limits_{k=1}^{K}\mathbf{\Upsilon}_k\mathbf{Q}_k\right)
\nonumber \\
&~~~~+  \sum\limits_{k=1}^{K}\mu_k(M_k-{\rm{tr}}(\mathbf{Q}_K))
\end{IEEEeqnarray}
{where $\mathbf{\Upsilon}\triangleq\{\mathbf{\Upsilon}_k \succeq 0\}$ and $\boldsymbol{\mu} \triangleq\{ \mu_k  \geq 0 \}$ are the Lagrange multipliers associated with the problem constraints.}
In a similar manner to \cite{zhang2013capacity}, \cite{couillet2011deterministic}  and \cite{dumont2010capacity}, we write
the derivative of $\mathcal{V}_{\boldsymbol{\mathcal{B}}_N}(x)$ with respect to  $\mathbf{Q}_k$ as
\begin{IEEEeqnarray}{Rl}
&\!\!\!\!\!\!\frac{\partial\mathcal{V}_{\boldsymbol{\mathcal{B}}_N}(x)}{\partial \mathbf{Q}_k}
\nonumber \\
&=\frac{\partial\log\det(\mathbf{I}_M+\mathbf{\Gamma}\mathbf{Q})}{\partial \mathbf{Q}_k}
\nonumber \\
&~+\sum\limits_{ij}\frac{\partial\mathcal{V}_{\boldsymbol{\mathcal{B}}_N(x)}} {\partial\!\!\left[\mathcal{G}_{\boldsymbol{\mathcal{B}}_N}^{\mathcal{M}_N}(-x\mathbf{I}_N)\right]_{ij}} \frac{\partial\!\!\left[\mathcal{G}_{\boldsymbol{\mathcal{B}}_N}^{\mathcal{M}_N}(-x\mathbf{I}_N)\right]_{ij}}
{\partial\mathbf{Q}_k} \nonumber \\
&~~~+\sum\limits_{ij}\frac{\partial\mathcal{V}_{\boldsymbol{\mathcal{B}}_N}(x)}{\partial[{\tilde{\eta}}_{Q,k} (\mathcal{G}_k(-x))]_{ij}}\frac{\partial[{\tilde{\eta}}_{Q,k} (\mathcal{G}_k(-x))]_{ij}}{\partial\mathbf{Q}_k}
\end{IEEEeqnarray}
where
\begin{eqnarray}
\frac{\partial\log\det(\mathbf{I}_M+\mathbf{\Gamma}\mathbf{Q})}{\partial \mathbf{Q}_k}
=\left(\left(\mathbf{I}_M+\mathbf{\Gamma}\mathbf{Q}\right)^{-1}\mathbf{\Gamma}\right)_k.
\end{eqnarray}
Furthermore, we obtain equations \eqref{eq:derivative_of_VBN_tmp1} and \eqref{eq:derivative_of_VBN_tmp2} at the top of the following page.
\begin{figure*}[!t]
\normalsize
\setcounter{tempequationcounter}{\value{equation}}
\begin{IEEEeqnarray}{Rl}
\frac{\partial\mathcal{V}_{\boldsymbol{\mathcal{B}}_N}} {\partial\!\!\left[\mathcal{G}_{\boldsymbol{\mathcal{B}}_N}^{\mathcal{M}_N}(-x\mathbf{I}_N)\right]_{ij}}
=& {\rm{tr}}\left(\left({{\boldsymbol{\Phi}}}(-x)
+ x^{-1}\mathbf{Q}^{\frac{1}{2}}\overline{\mathbf{S}}{}^H{\Tilde{{\boldsymbol{\Phi}}}}(-x)^{-1}\overline{\mathbf{S}}\mathbf{Q}^{\frac{1}{2}} \right)^{-1}
\frac{\partial{{\boldsymbol{\Phi}}}(-x)} {\partial\!\!\left[\mathcal{G}_{\boldsymbol{\mathcal{B}}_N}^{\mathcal{M}_N}(-x\mathbf{I}_N)\right]_{ij}}\right)  \IEEEnonumber \\
  &~~- {\rm{tr}}\left(x\sum\limits_{k=1}^{K}{\tilde{\eta}}_{Q,k} (\mathcal{G}_k(-x))\frac{\partial\mathcal{G}_{\boldsymbol{\mathcal{B}}_N}^{\mathcal{M}_N}(-x\mathbf{I}_N)} {\partial\!\!\left[\mathcal{G}_{\boldsymbol{\mathcal{B}}_N}^{\mathcal{M}_N}(-x\mathbf{I}_N)\right]_{ij}}\right)  \IEEEnonumber \\
  =& 0
  \label{eq:derivative_of_VBN_tmp1} \\
\frac{\partial\mathcal{V}_{\boldsymbol{\mathcal{B}}_N}}{\partial[{\tilde{\eta}}_{Q,k} (\mathcal{G}_k(-x))]_{ij}}
=& {\rm{tr}}\left(
   \left({{\boldsymbol{\Phi}}}(-x)
+x^{-1}\mathbf{Q}^{\frac{1}{2}}\overline{\mathbf{S}}{}^H{\Tilde{{\boldsymbol{\Phi}}}}(-x)^{-1} \overline{\mathbf{S}}\mathbf{Q}^{\frac{1}{2}} \right)^{-1}\frac{\partial x^{-1}\mathbf{Q}^{\frac{1}{2}}\overline{\mathbf{S}}{}^H{\Tilde{{\boldsymbol{\Phi}}}}(-x)^{-1}\overline{\mathbf{S}} \mathbf{Q}^{\frac{1}{2}}}{\partial[{\tilde{\eta}}_{Q,k} (\mathcal{G}_k(-x))]_{ij}}\right)  \nonumber \\
&~~+ {\rm{tr}}\left({\Tilde{{\boldsymbol{\Phi}}}}(-x)^{-1}\frac{\partial{\Tilde{{\boldsymbol{\Phi}}}}(-x)}{{\partial[{\tilde{\eta}}_{Q,k} (\mathcal{G}_k(-x))]_{ij}}}\right) - {\rm{tr}}\left(x\frac{\partial{\tilde{\eta}}_{Q,k} (\mathcal{G}_k(-x))}{\partial[{\tilde{\eta}}_{Q,k} (\mathcal{G}_k(-x))]_{ij}}\mathcal{G}_{\boldsymbol{\mathcal{B}}_N}^{\mathcal{M}_N}(-x\mathbf{I}_N)\right)  \IEEEnonumber \\
=& 0  \label{eq:derivative_of_VBN_tmp2}
\end{IEEEeqnarray}
\addtocounter{tempequationcounter}{2}
\setcounter{equation}{\value{tempequationcounter}}
\hrulefill
\end{figure*}
The problem now becomes the same as that in \cite{zhang2013capacity}. Thus, the rest of the proof is omitted.

\section*{Acknowledgment}
We would like to thank the editor and the anonymous reviewers
for their helpful comments and suggestions.

\bibliographystyle{IEEEtran}
\bibliography{IEEEabrv,this_reference}

\end{document}